\documentclass[journal,transmag]{IEEEtran}
\usepackage{array}
\usepackage{multirow}
\usepackage[dvips]{graphicx}
\usepackage[english]{babel}
\usepackage[latin1]{inputenc}
\usepackage{amsfonts,amssymb,amsmath,dsfont,amscd,url,graphics,cite,calc,psfrag,theorem, epsfig}
\usepackage{url}
\usepackage{cite}
\usepackage{stfloats}
\usepackage{graphicx}
\usepackage[dvipsnames]{xcolor}
\usepackage{tikz}
\usepackage{pgf}
\usepackage{pgffor}
\usepackage{pgfpages}
\usepackage{wrapfig}
\usepackage{algorithm}
\usepackage{algorithmic}
\usepackage{datetime}
 \usetikzlibrary{arrows}
 \usetikzlibrary{snakes}
 \usetikzlibrary{shapes}
 \usetikzlibrary{automata}
 \usetikzlibrary{topaths} 
\usetikzlibrary[topaths] 
\usetikzlibrary{positioning}
\usetikzlibrary{decorations.pathmorphing}
\pgfdeclarelayer{background} \pgfdeclarelayer{foreground} %
\pgfsetlayers{background,main,foreground}
\usetikzlibrary{calc,3d}
\usetikzlibrary{shapes.gates.logic.US,trees,positioning,arrows}
\usetikzlibrary{positioning}
\usetikzlibrary{decorations.pathmorphing} 
\usetikzlibrary{fit}	
\usetikzlibrary{shapes,snakes}
 \usetikzlibrary{arrows}
 \usetikzlibrary{snakes}
 \usetikzlibrary{shapes}
 \usetikzlibrary{automata}
 \usetikzlibrary{topaths} 
\usetikzlibrary[topaths] 
\pgfdeclarelayer{background} \pgfdeclarelayer{foreground} %
\pgfsetlayers{background,main,foreground}
\usepackage{pgf}
\usepackage{pgfplots}
\pgfplotsset{compat=newest}
\pgfplotsset{plot coordinates/math parser=false}

\usepackage{wrapfig}\usepackage{subfig}
\newif\iffinal
\makeatletter
\def\input@path{{OSPAMT/}}
\makeatother

\graphicspath{{OSPAMT/Figs/}}
\usepackage{tikz}
\colorlet{drkblue}{blue!80}
\iffinal
  
\else
  
\fi
\makeatother

\newtheorem{definition}{Definition}

\newtheorem{rmk}{Remarks}

\usepackage{multicol}
\usepackage{trsym,trfsigns}
\usepackage{lipsum}
\usepackage{pdfpages}
\usepackage{epstopdf}
\pgfrealjobname{MainOSPAMT_SecondSubmitArxiv} 
\begin{document}
\title{Optimal Subpattern Assignment Metric for Multiple Tracks (OSPAMT Metric)
\author{Tuyet~Vu Member,~\IEEEmembership{IEEE} and~Rob~Evans, Fellow~\IEEEmembership{IEEE}
\thanks{Tuyet~Vu (email: Tuyet.Vu@dst.defence.gov.au is with the Department of Defence Science and Technology.
}
\thanks{Rob~Evans (email: robinje@unimelb.edu.au) is with the Department of Electrical and Electronic Engineering, the University of Melbourne, Vic. 3010, Australia.}
}
}
\maketitle
\begin{abstract}
In this paper, we propose a new metric which measures the distance between two finite sets of tracks (a track is a path of either a real or estimated target). This metric is based on the same principle as the Optimal Subpattern Assignment (OSPA) metric devised by Schuhmacher et al. Importantly however, the  new metric measures the distance between two finite sets of tracks whereas the OSPA metric measures the distance between two finite sets of target states.  By also considering the properties of false tracks, missed tracks and many tracks assigned to one track situations caused by missed detections and false alarms, the minimization of all distances between tracks across two finite sets of tracks employed by the new OSPAMT metric enables performance evaluation of multi-target tracking (MTT)  algorithms in a more comprehensive and accurate manner than existing metrics such as the  OSPA metric and the enhanced  OSPAT metric introduced by Ristic et al which measures the distance between two finite sets of labeled target states.
\end{abstract}
\selectlanguage{english}
\IEEEpeerreviewmaketitle
\section{Introduction}\label{S:intro}
In the multi-target tracking  problem, sequences of target states are estimated from sequences of noisy measurements and false measurements.Any track (i.e sequence of target states or loosely path of a target) may start and/or  end  during the scan time. As discussed in \cite{ristic2011metric}, for performance evaluation of MTT algorithms, \cite{Fridling1991,Ristic99TWSEval,rothrock2000performance},\cite[Ch.13]{Blackman99
}, and \cite{Colegrove96PerAss} defined suitable distance measures for their particular MTT algorithm. Computations of these distance measures can be difficult since they involves many aspects of the problem (e.g track initiation delay, velocity error, track label swaps, etc) and importantly it is often unclear which distance measure is most suitable for any particular MTT situation. There exist  mathematically rigorous metrics for measuring the distance between two finite sets. The Hausdorff metric measures the largest distance between two finite sets by finding the maximum of all distances from a point in one set to the closest point in another set. However, in \cite{HoffmanMahler02} it is shown that the Hausdorff distance is not suitable for measuring the performance of  MTT algorithms because it is insensitive to the difference between the number of elements in the two finite sets and a new metric is proposed, namely the $W_p$ distance which overcomes this limitation. As discovered in \cite{Schuhmacher2008}, the $W_p$ distance only partly fixes the undesirable cardinality error behavior of the Hausdorff metric because it is not defined if one of the two finite sets is empty and is inconsistent if the two finite sets have different numbers of elements. To overcome the shortcomings of the $W_p$ distance, Schuhmacher et al \cite{Schuhmacher2008} proposed a new metric called the optimal subpattern assignment (OSPA) metric which is based on a metric between the distributions of point process \cite{schuhmacher2008math}. It incorporates the spatial distance and the cardinality distance via a cut-off parameter $c$. This cut-off parameter is added to deal with any mismatch between the number of elements in the two finite sets (one set represents truth target and the other represent estimated target from an MTT algorithm). However in the MTT problem, a finite set do not represent a finite collection of states (generally not collection of finite vectors) instead it represents a finite collection of sequences of target states. This is because it includes more than one trajectories of targets where each trajectory of target is called track. Furthermore, in MTT problem the number of targets is unknown therefore a set of tracks is used to represent the randomness in number and in length of a track. Thus measuring the distance between the finite set of truth tracks and a finite set of estimated tracks from an MTT algorithm is important to find the reliable MTT algorithm. To find the short distance between these sets is not straight forward because the number of these two sets can be different, and the length of the match tracks across two track if applicable can be different. Furthermore, due to the essence of MTT problem, there is miss detection and imperfect sensor and hence several estimated tracks from an MTT algorithm can represent a truth track, etc. Thus, the OSPA metric does not fully resolve the problem.

To address the MTT problem, Ristic et al \cite{ristic2011metric} further extended the OSPA metric by adding the error between target labels (or target indices) to the  spatial distance. The extended metric is called the OSPAT metric (OSPA metric for track) between two finite set of tracks (one set represents a collection of truth tracks and the other represent a collection of estimated tracks from an MTT algorithm). This is an important advance, however it still does not fully address the MTT problem because it does not measure the distance between two finite sets of tracks (a track is a sequence of target states) instead it measures the distance between two finite sets of labeled states at each point in time. A labeled state is a vector whose components are the state of the target and target label. In order to add the labels to the states of the tracks, the ``optimal assignment'' between tracks across these two sets of tracks is defined and called an OSPAT assignment. While this work has gone a long way towards dealing with the MTT problem, there remain some important issues to be addressed. Firstly, in order to compute the OSPAT metric between two finite sets of tracks, the tracks in a set must be labeled or existing labels need to renamed in such a way that the sum of all spatial distances between pairs of tracks is minimized. If tracks across two sets are paired according the OSPAT assignment, they are labeled/relabeled the same. Otherwise they are labeled/relabeled differently. All labels in each set of tracks are distinguishable. A labeled track is used to indicate the track whose states are augmented with a  label.
 This step is questionable since the metric which is the sum of the spatial distance and the labeling errors will violate the triangle inequality. Furthermore, labels, which are arbitrary variables and are used to name targets in order to distinguish different targets, are added to the target states and hence they should play no role in the distance between two finite sets of tracks. As a result, the OSPAT metric does not always reliably evaluate the performance of an MTT algorithm. Secondly, the definition of the OSPAT assignment, which is ``optimal assignment'' of pairs of tracks in these two sets, is still vague and does not always achieve the optimal assignment between two finite sets of tracks. The optimal assignment is referred to the assignment which is sensible and rational; and most likely the correct assignment. This is because the spatial distance contributing to determining the OSPAT assignment ignores tracks which are not paired with other tracks in cases where this is applicable. Finally, the OSPAT assignment does not properly consider missed tracks, false tracks and track estimation error (caused by miss detection and/or clutter) as found in many MTT algorithms \cite{Reid1979,shalom1988tracking,Blackman2004,BlomHans2008Closespace,TuyetPMMHT14,Musicki04JIPDA,Oh2009}. Indeed, it only considers either missed tracks or false tracks but not both if the two sets have different numbers of tracks. In the case of the same number of tracks for both sets, the OSPAT assignment does not consider missed tracks or false tracks or both. In other words, the OSPAT assignment implicitly assumes there are no missing or false tracks.
  Recently \cite{Beard17}
  has introduced the new version of OSPAT metric, namely OSPA$^2$, which overcomes the limit of the OSPAT metric. However, one-to-one optimal assignment in the OSPA$^2$ is calculated for the whole duration of scanning. Hence, it does not considers the scenario where broken tracks exist.

In order to address these restrictions, we propose a new metric to measure the distance between two finite sets of tracks. We use the same underlying  principles as the OSPA metric, that is to incorporate the cardinality distance and the spatial distance at each time index. In addition, the new metric also makes use of the assignment between tracks across two finite sets of tracks. Here assignment is the optimal many-to-one assignment is calculated as easy as the optimal one-to-one assignment and is explained in Remark \ref{Rmk:Compute-Many-one Assignment} It also accounts for  MTT track maintenance (initiation, termination, false and missing tracks) by including these  into the computation of the spatial distance between a set of estimated tracks and the set of truth tracks. The formulae and performance results of the new metric were initially published in brief form in \cite{Tuyet14OSPAMT}. This paper properly details the new metric and studies its performance including comparison with other MTT metrics, on several example problems.

The structure of this paper is as follows. Section \ref{sec:finiterandomsets}  formulates the problem. The main contribution of this paper is contained in Section \ref{S:OSPA_metricTrks} where the OSPAMT metric is fully developed. Section \ref{SS:AnalysisMetrics} studies the performance of the OSPAMT metric via a number of illustrative example scenarios and compares OSPAMT with OSPA and OSPAT. For convenience brief summaries  of OSPA and OSPAT  are presented  in Appendix \ref{SS:Summary_OSPA} and \ref{SS:Summary_OSPAT}). Section \ref{S:Experiment} presents numerical studies to demonstrate the usefulness of the proposed metric and compares this with the other two metrics. Concluding remarks and suggestions for future work are given in section \ref{S:Conclusion}
\section{Problem Statement}\label{sec:finiterandomsets}

Let $\mathcal{T}=\{1,2,\ldots,T\}$ where $T$ is the number of scans. A track $i$ ($i\in\mathds{N}=\{1,2,\ldots\}$), denoted by $\tau_i$, is defined as a vector (of sets)
\begin{equation}\label{E:Trk}
\tau_i=(\tilde{x}_1,\tilde{x}_2,\ldots,\tilde{x}_T)
\end{equation}
where $\tilde{x}_t=\emptyset$ if target $i$ does not exist at time
$t\in\mathcal{T}$ or $\tilde{x}_t=\{x_t\}$ if target $i$ exists at time $t$ and generates state $x_t\in\mathcal{X}\subseteq\mathds{R}^{n_x}$ where $n_x$ is the dimension of the target state. Denote the state of track $i$ at time $t$ ($t\in\mathcal{T}$) as $\tau_i(t)=\tilde{x}_t$. If  target $i$ appears at time $t_0$ and disappear at time $t_1$ then the states of the track are $\tau_i(t)=\emptyset$ for all $t<t_0$ and $t\ge t_1$.
For the purpose of calculating the metric, all components of $x\in \mathds{R}^{n_x}$ are assumed to be appropriately scaled to deal with different units of measure. 

Let  $\mathcal{F}^T=\underbrace{\mathcal{F}\times\ldots\times\mathcal{F}}_{T \text{ times}}$ where $\mathcal{F}=\{\emptyset\}\cup\{\{x\}:x\in\mathcal{X}\}$ is a collection of all singletons and the empty set. Denote a finite collection of tracks by
\begin{align}\label{E:SetTrk}
\omega=\{&\tau_1,\ldots,\tau_n\in\mathcal{F}^T:\forall i\in\{1,\ldots,n\}, \tau_i\text{ is defined }\nonumber\\
& \text{ in \eqref{E:Trk} and }\exists t\in \mathcal{T}\text{ such that }\tau_i(t)\ne\emptyset\}.
\end{align}
%

Denote the space of finite sets of tracks defined in \eqref{E:SetTrk} and an empty set as
\begin{align*}
\Omega=\{\emptyset\}\cup\{\omega: \omega \text{ is defined in \eqref{E:SetTrk}}\}.
\end{align*}
For $\omega\in\Omega$, if $\omega=\{\tau_1,\ldots,\tau_n\}\ne\emptyset$ where $\tau_i$ is given in \eqref{E:Trk} for $i=1,\ldots,n$, then denote a track in  $\omega$ by
 $\tau^\omega_i=\tau_i$.

Our purpose is to define a metric on $\Omega$ which is a function $\mathbf{d}:\Omega\times\Omega\to\mathds{R}_+=[0,\infty)$ which satisfies the following conditions for all $\omega,\omega',\omega^*\in\Omega$
\begin{enumerate}
  \item (identity) $\mathbf{d}(\omega, \omega') = 0$  if and only if  $\omega=\omega'$
  \item (symmetry) $\mathbf{d}(\omega, \omega')=\mathbf{d}(\omega', \omega)$
  \item (triangle inequality) $\mathbf{d}(\omega, \omega')\le \mathbf{d}(\omega, \omega^*)+\mathbf{d}(\omega^*, \omega')$.
\end{enumerate}

Note that a function on $\Omega\times\Omega\to\mathds{R}_+=[0,\infty)$ that satisfies all axioms for a metric with the exception of the  symmetry property is called a quasimetric.

For the remainder of this paper we denote $\omega$, $\omega'$ as a set of truth tracks and estimated tracks respectively.
\section{Optimal subpattern assignment for multiple tracks metric (OSPAMT metric)}\label{S:OSPA_metricTrks}
\begin{figure}[htbp!]
\subfloat[$\omega=\{\tau_1\}$, $\omega_a=\{\tau'_1,\tau'_2\}$. 
]
{\label{Fig:ExamMotivateOSPMAT0}
\centering{\hspace{-2cm}
\scalebox{.9}
{\centering
\pgfdeclarelayer{background} \pgfdeclarelayer{foreground}
\pgfsetlayers{background,main,foreground}
\par
\par
\tikzset{cross/.style={cross out, draw=black, minimum size=2*(#1-\pgflinewidth), inner sep=0pt, outer sep=0pt},
cross/.default={3pt}}
\begin{tikzpicture}[y=.2cm, x=2cm,font=\sffamily]
 \draw [->,thick]
 |-(5.12,0) node (xaxis) [right]
  {\scriptsize Time};
\draw (1,0) -- coordinate (x axis mid) (5.1,0);
\foreach \x in {1,...,5}
     		\draw (\x,1pt) -- (\x,-3pt)
			node[anchor=north] {\x};
\draw (1,1.5) node[cross] (x1){};
\draw (2,1.5) node[cross] (x2){};
\draw (3,1.5) node[cross] (x3){};
\draw (4,1.5) node[cross] (x4){};
\draw (5,1.5) node[cross] (x5){};
\node(x11)[left=-0.12 of x1]{};
\node[circle,draw=black, fill=black, inner sep=0pt,minimum size=3pt] (y1) at (1,5.5) {};
\node(y11)[left=-0.11 of y1]{};
\node[circle,draw=black, fill=black, inner sep=0pt,minimum size=3pt] (y2) at (2,5.5){};
\node[circle,draw=black, fill=black, inner sep=0pt,minimum size=3pt] (y3) at (3,5.5){};
\node[circle,draw=black, fill=black, inner sep=0pt,minimum size=3pt] (y4) at (4,5.5){};
\node[circle,draw=black, fill=black, inner sep=0pt,minimum size=3pt] (y5) at (5,5.5){};
\node(y51)[right=-0.11 of y5]{};
%

 \draw[-,thin,black,dashed](x2)edge node[swap,near start]{}  (y2);
\draw[->,black](y1)edge (y2)(y2)edge (y3)(y4)edge (y5)
(x1)edge (x2)(x2)edge (x3)(x3)edge (x4)(x4)edge (x5)
;
\node(temdis1)  [below =.2cm of y1] {};
 \node (dis1) [right =-0.01cm of temdis1]{$\epsilon\ll 1$};
 \draw[-,thin,black,dashed](y1)edge node[swap,near start]{}
  (x1);
\node(temdis2) [below =.2cm of y2] {};
  \node (dis1) [right =-.5cm of temdis2]{$\epsilon$};
  \draw[-,thin,black,dashed](y2)edge node[swap,near start]{}
  (x2);
\node(temdis3) [below =.2cm of y3] {};
  \node (dis1) [right =-.5cm of temdis3]{$\epsilon$};
  \draw[-,thin,black,dashed](y3)edge node[midway]{}
  (x3);
\node(temdis4) [below =.2cm of y4] {};
  \node (dis1) [right =-.5cm of temdis4]{$\epsilon$};
  \draw[-,thin,black,dashed](y4)edge node[near start]{}
  (x4);
\node(temdis5) [below =.2cm of y5] {};
  \node (dis1) [right =-.5cm of temdis5]{$\epsilon$};
  \draw[-,thin,black,dashed](y5)edge node[near start]{}
  (x5);
 \begin{pgfonlayer}{background}
 \path (y5.east |-y5.east)+(.3,-.6) node(ii){$\tau'_2$};
  \path (y3.east |-y3.east)+(.3,-.6) node(aa){$\tau'_1$};
  \path (x5.east|- x5.east)+(.3,.4) node(bb){$\tau_1$};
\draw[->,black,dashed](bb) edge (x5)(aa)edge(y3)(ii) edge (y5);
\end{pgfonlayer}
\end{tikzpicture}
}}
}\\%
\subfloat[$\omega^*=\{\tau_1,\tau_2\}$, $\omega'=\{\tau'_1,\tau'_2\}$.
]
{\label{Fig:ExamMotivateOSPMAT1}
\centering{\hspace{-2cm}
\scalebox{.9}
{
\centering
\pgfdeclarelayer{background} \pgfdeclarelayer{foreground}
\pgfsetlayers{background,main,foreground}
\par
\par
\tikzset{cross/.style={cross out, draw=black, minimum size=2*(#1-\pgflinewidth), inner sep=0pt, outer sep=0pt},
cross/.default={3pt}}
\begin{tikzpicture}[y=.2cm, x=2cm,font=\sffamily]
 \draw [->,thick]
 |-(5.1,0) node (xaxis) [right]
  {\scriptsize Time};
\draw (1,0) -- coordinate (x axis mid) (5.1,0);
\foreach \x in {1,...,5}
     		\draw (\x,1pt) -- (\x,-3pt)
			node[anchor=north] {\x};
\draw (1,1.5) node[cross] (x1){};
\draw (2,1.5) node[cross] (x2){};
\draw (3,1.5) node[cross] (x3){};
\draw (4,1.5) node[cross] (x4){};
\draw (5,1.5) node[cross] (x5){};
\draw (4,16) node[cross] (x24){};
\draw (5,16) node[cross] (x25){};
\node(x11)[left=-0.12 of x1]{};
\node[circle,draw=black, fill=black, inner sep=0pt,minimum size=3pt] (y1) at (1,5.5) {};
\node(y11)[left=-0.11 of y1]{};
\node[circle,draw=black, fill=black, inner sep=0pt,minimum size=3pt] (y2) at (2,5.5){};
\node[circle,draw=black, fill=black, inner sep=0pt,minimum size=3pt] (y3) at (3,5.5){};
\node[circle,draw=black, fill=black, inner sep=0pt,minimum size=3pt] (y4) at (4,5.5){};
\node[circle,draw=black, fill=black, inner sep=0pt,minimum size=3pt] (y5) at (5,5.5){};
\node(y51)[right=-0.11 of y5]{};
 \draw[-,thin,black,dashed](x2)edge node[swap,near start]{}  (y2);
\draw[->,black](y1)edge (y2)(y2)edge (y3)(y4)edge (y5)
(x1)edge (x2)(x2)edge (x3)(x3)edge (x4)(x4)edge (x5)(x24)edge (x25);
\node(temdis1)  [below =.2cm of y1] {};
 \node (dis1) [right =-0.01cm of temdis1]{$\epsilon\ll 1$};
 \draw[-,thin,black,dashed](y1)edge node[swap,near start]{}
  (x1);
\node(temdis2) [below =.2cm of y2] {};
  \node (dis1) [right =-.5cm of temdis2]{$\epsilon$};
  \draw[-,thin,black,dashed](y2)edge node[swap,near start]{}
  (x2);
\node(temdis3) [below =.2cm of y3] {};
  \node (dis1) [right =-.5cm of temdis3]{$\epsilon$};
  \draw[-,thin,black,dashed](y3)edge node[midway]{}
  (x3);
\node(temdis4) [below =.2cm of y4] {};
  \node (dis1) [right =-.5cm of temdis4]{$\epsilon$};
  \draw[-,thin,black,dashed](y4)edge node[near start]{}
  (x4);
\node(temdis5) [below =.2cm of y5] {};
  \node (dis1) [right =-.5cm of temdis5]{$\epsilon$};
  \draw[-,thin,black,dashed](y5)edge node[near start]{}
  (x5);
\node(temdis6) [below =.2cm of x24] {};
  \node (dis1) [right =.01cm of temdis6]{$\beta\gg 1$};
  \draw[-,thin,black,dashed](x24)edge node[near start]{}(y4);
  \node(temdis7) [below =.2cm of x25] {};
  \node (dis1) [right =.01cm of temdis7]{};
  \draw[-,thin,black,dashed](x25)edge node[near start]{}(y5);
 \begin{pgfonlayer}{background}
\path (x24.west |- x24.west)+(-.3,-.6) node(dd){$\tau_3$};
 \path (y5.east |-y5.east)+(.3,-.6) node(ii){$\tau'_2$};
  \path (y3.east |-y3.east)+(.3,-.6) node(aa){$\tau'_1$};
  \path (x5.east|- x5.east)+(.3,.4) node(bb){$\tau_1$};
\draw[->,black,dashed](bb) edge (x5)(aa)edge(y3)(ii) edge (y5)(dd)edge (x24);
\end{pgfonlayer}
\end{tikzpicture}
}}
}\\
\subfloat[$\omega=\{\tau_1\}$, $\omega_b=\{\tau'_1,\tau'_2\}$. 
]
{\label{Fig:ExamMotivateOSPMAT2}
\centering{\hspace{-2cm}
\scalebox{.9}
{\centering
\pgfdeclarelayer{background} \pgfdeclarelayer{foreground}
\pgfsetlayers{background,main,foreground}
\par
\par
\tikzset{cross/.style={cross out, draw=black, minimum size=2*(#1-\pgflinewidth), inner sep=0pt, outer sep=0pt},
cross/.default={3pt}}
\begin{tikzpicture}[y=.2cm, x=2cm,font=\sffamily]
 \draw [->,thick]
 |-(5.1,0) node (xaxis) [right]
 {\scriptsize Time};
\draw (1,0) -- coordinate (x axis mid) (5.1,0);
\foreach \x in {1,...,5}
     		\draw (\x,1pt) -- (\x,-3pt)
			node[anchor=north] {\x};
\draw (1,1.5) node[cross] (x1){};
\draw (2,1.5) node[cross] (x2){};
\draw (3,1.5) node[cross] (x3){};
\draw (4,1.5) node[cross] (x4){};
\draw (5,1.5) node[cross] (x5){};
\node(x11)[left=-0.12 of x1]{};
\node[circle,draw=black, fill=black, inner sep=0pt,minimum size=3pt] (y1) at (1,10) {};
\node(y11)[left=-0.11 of y1]{};
\node[circle,draw=black, fill=black, inner sep=0pt,minimum size=3pt] (y2) at (2,10){};
\node[circle,draw=black, fill=black, inner sep=0pt,minimum size=3pt] (y3) at (3,10){};
\node[circle,draw=black, fill=black, inner sep=0pt,minimum size=3pt] (y4) at (4,10){};
\node[circle,draw=black, fill=black, inner sep=0pt,minimum size=3pt] (y5) at (5,10){};
\node(y51)[right=-0.11 of y5]{};
\draw[->,black](y1)edge (y2)(y2)edge(y3)(y3)edge (y4)(y4)edge (y5)
(x1)edge (x2)(x2)edge (x3)(x3)edge (x4)(x4)edge (x5)
;
\node(temdis1)  [below =.3cm of y1] {};
 \node (dis1) [right =-0.01cm of temdis1] {$\beta\gg 1$}; 
 \draw[-,thin,black,dashed](y1)edge node[swap,near start]{}
  (x1);
\node(temdis2)  [below =.3cm of y2] {};
  \node (dis1) [right =-.6cm of temdis2] {$\beta$};
  \draw[-,thin,black,dashed](y2)edge
   node[swap,near start]{}  (x2);
\node(temdis3)  [below =.3cm of y3] {};
  \node (dis1) [right =-.6cm of temdis3] {$\beta$};
  \draw[-,thin,black,dashed](y3)edge node[midway]{}
  (x3);
\node(temdis4)  [below =.3cm of y4] {};
  \node (dis1) [right =-.6cm of temdis4] {$\beta$};
  \draw[-,thin,black,dashed](y4)edge node[near start]{}
  (x4);
\node(temdis5)  [below =.3cm of y5] {};
  \node (dis1) [right =-.6cm of temdis5] {$\beta$};
  \draw[-,thin,black,dashed](y5)edge node[near start]{}
  (x5);

\begin{pgfonlayer}{background}
  \path (x5.east|- x5.east)+(.3,.4) node(bb){$\tau_1$};
  \path (y5.east |-y5.east)+(.3,-.7) node(aa){$\tau'_1$};
\draw[->,black,dashed](bb) edge (x5)(aa)edge(y5);
\end{pgfonlayer}
\end{tikzpicture}
}}}\\
\subfloat[$\omega=\{\tau_1\}$, $\omega_c=\{\tau'_1\}$. 
]
{\label{Fig:ExamMotivateOSPMAT3}
\centering{\hspace{-2cm}
\scalebox{.9}
{\centering
\pgfdeclarelayer{background} \pgfdeclarelayer{foreground}
\pgfsetlayers{background,main,foreground}
\par
\par
\tikzset{cross/.style={cross out, draw=black, minimum size=2*(#1-\pgflinewidth), inner sep=0pt, outer sep=0pt},
cross/.default={3pt}}
\begin{tikzpicture}[y=.2cm, x=2cm,font=\sffamily]
 \draw [->,thick]
 |-(5.1,0) node (xaxis) [right]
  {\scriptsize Time};
\draw (1,0) -- coordinate (x axis mid) (5.1,0);
\foreach \x in {1,...,5}
     		\draw (\x,1pt) -- (\x,-3pt)
			node[anchor=north] {\x};
\draw (1,1.5) node[cross] (x1){};
\draw (2,1.5) node[cross] (x2){};
\draw (3,1.5) node[cross] (x3){};
\draw (4,1.5) node[cross] (x4){};
\draw (5,1.5) node[cross] (x5){};
\node[circle,draw=black, fill=black, inner sep=0pt,minimum size=3pt] (y1) at (1,5.5) {};
\node(y11)[left=-0.11 of y1]{};
\node[circle,draw=black, fill=black, inner sep=0pt,minimum size=3pt] (y2) at (2,5.5){};
\node[circle,draw=black, fill=black, inner sep=0pt,minimum size=3pt] (y3) at (3,5.5){};
\node[circle,draw=black, fill=black, inner sep=0pt,minimum size=3pt] (y4) at (4,10){};
\node[circle,draw=black, fill=black, inner sep=0pt,minimum size=3pt] (y5) at (5,10){};
 \draw[-,thin,black,dashed](x2)edge node[swap,near start]{}  (y2);
\draw[->,black](y1)edge (y2)(y2)edge (y3)(y4)edge (y5)
(x1)edge (x2)(x2)edge (x3)(x3)edge (x4)(x4)edge (x5);
\node(temdis1)  [below =.2cm of y1] {};
 \node (dis1) [right =-0.01cm of temdis1]{$\epsilon\ll 1$};
 \draw[-,thin,black,dashed](y1)edge node[swap,near start]{}
  (x1);
\node(temdis2) [below =.2cm of y2] {};
  \node (dis1) [right =-.5cm of temdis2]{$\epsilon$};
  \draw[-,thin,black,dashed](y2)edge node[swap,near start]{}
  (x2);
\node(temdis3)  [below =.3cm of y3] {};
  \node (dis1) [right =-.5cm of temdis3] {$\epsilon$};
  \draw[-,thin,black,dashed](y3)edge node[midway]{}
  (x3);
\node(temdis4)  [below =.3cm of y4] {};
  \node (dis1) [right =-.6cm of temdis4] {$\beta$};
  \draw[-,thin,black,dashed](y4)edge node[near start]{}
  (x4);
\node(temdis5)  [below =.3cm of y5] {};
  \node (dis1) [right =-.6cm of temdis5] {$\beta$};
  \draw[-,thin,black,dashed](y5)edge node[near start]{}
  (x5);
 \begin{pgfonlayer}{background}
 \path (y5.east |-y5.east)+(.3,-.8) node(ii){$\tau'_2$};
  \path (y2.east |-y2.east)+(.3,-.6) node(aa){$\tau'_1$};
  \path (x5.east|- x5.east)+(.3,.6) node(bb){$\tau_1$};
\draw[->,black,dashed](bb) edge (x5)(aa)edge(y2)(ii) edge (y5);
\end{pgfonlayer}
\end{tikzpicture}
}}}
\caption{Intuitively, $\tau'_1,\tau'_2$ are estimated tracks of truth track $\tau_1$ in Figures \eqref{Fig:ExamMotivateOSPMAT0} and \eqref{Fig:ExamMotivateOSPMAT1} because distances between their states and the true target states at the corresponding time are very small. For Figure \eqref{Fig:ExamMotivateOSPMAT3}, only $\tau'_1$ is the estimated track of the truth track $\tau_1$ because the distances between the states of $\tau'_2$ and the truth track $\tau_1$ at corresponding time steps are very large. For Figure \eqref{Fig:ExamMotivateOSPMAT2}, $\tau'_1$ is a false track because the distances between the states of $\tau'_1$ and the truth track $\tau_1$ at corresponding time steps are very large.}
\label{Fig:ExamMotivateOSPMAT}
\end{figure}
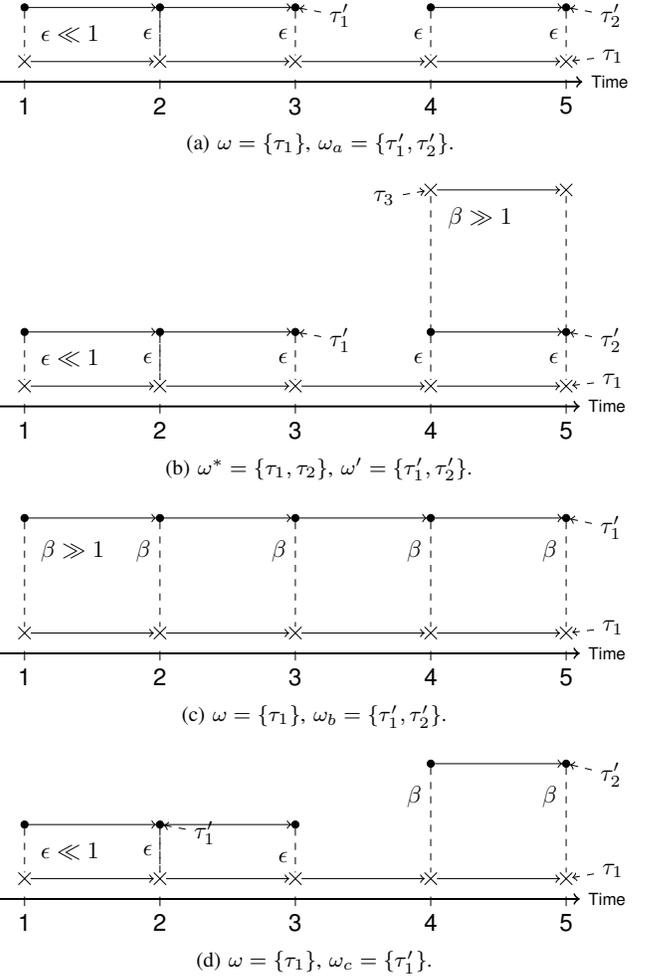

In this section, we illustrate, via examples,  the properties we would expect to see in a practically useful metric for MTT algorithms and show that the OSPAMT metric fulfills these requirements. The metric must measure  many situations which occur very often  in MTT problems including estimated tracks  broken into multiple disconnected sections as shown in Figure \eqref{Fig:ExamMotivateOSPMAT0}, missing tracks as shown in Figure \eqref{Fig:ExamMotivateOSPMAT1},  missing states and false tracks as shown in Figure \eqref{Fig:ExamMotivateOSPMAT3}, and both missing tracks and false tracks as in Figure \eqref{Fig:ExamMotivateOSPMAT2}. Note that a missing track is a truth track which is not close to any estimated track; a missing state is a state of a truth track which is not close to any state of any estimated tracks; and a false track is an estimated track (from a set of the estimated track) which does not close to any truth tracks. The metric also needs to be sensitive to the number of false tracks and missing tracks, which means the metric for MTT must tell us that the set of the estimated tracks in Figure \eqref{Fig:ExamMotivateOSPMAT3} is better than the set of the estimated tracks in Figure \eqref{Fig:ExamMotivateOSPMAT2} but worse than that in Figure \eqref{Fig:ExamMotivateOSPMAT0}.

The OSPAMT metric is a mathematical distance measure between an set of estimated tracks and a set of truth tracks and it successfully deals with all of the situations in Figure \ref{Fig:ExamMotivateOSPMAT} and provides an intuitively correct performance comparison (see detail in section \ref{SS:OSPAMT}). The process of calculating the OSPAMT metric also produces the matching assignments across these two sets, the identities of the missing tracks and the false tracks and the distance between these two sets at each point in time.

\subsection{Development of OSPAMT metric}\label{SS:OSPAMT}
Consider two finite sets of tracks $\omega$ and $\omega'$. An Assignments from $\omega'$ to $\omega$ are determined to be the sum of localization errors (localization error is capped at a preselected maximum or cut-off value $c$) and cardinality errors at each time index. The OSPAMT metric is the smallest sum over all assignments. The assignment here can be many to one so the localization error between each extra assigned track at each time index is charged with the preselected penalty $\Delta$ ($\Delta>0$) if there is more than one track in $\omega'$ assigned to a track in $\omega$. All unassigned tracks in $\omega$ and $\omega'$ and extra lengths of assigned tracks at each time index each are charged with $c$. Furthermore each extra length of assigned tracks is also charged with penalty $\Delta$ as an extra if that track is an extra track in $\omega'$ assigned to a track in $\omega$.

A many-to-one assignments from $\omega'$ to  $\omega$ described above can be explained as follows.

\begin{definition}[Assignment from $\omega'$ to $\omega$]\label{D:Assignment}

Given $\omega,\omega'\in\Omega$. Let $L^{\omega}=\{1,\ldots,|\omega|\}$ and $L^{\omega'}=\{1,\ldots,|\omega'|\}$ be the collections of all track indices in $\omega$ and in $\omega'$ respectively. Denote $L^{\omega}_0=\{0\}\cup L^{\omega}$. Then  $\mathcal{M}(L^{\omega'},L^{\omega}_0)$ is a collection of all functions from $L^{\omega'}$ to $L^{\omega}_0$ such that any track in $\omega$ and its associated tracks have at least one common time index, i.e. $\forall\lambda\in\mathcal{M}(L^{\omega'},L^{\omega}_0)$, $\forall i\in\lambda(L^{\omega'})\setminus\{0\}$, $\forall j\in\lambda^{-1}(i),\exists t\in\mathcal{T}$, we have

\begin{equation}\label{E:May-1Assign}
 \tau^{\omega'}_j(t)\ne\emptyset\text{ and }\tau^\omega_i(t)\ne\emptyset.
\end{equation}
Any element in $\mathcal{M}(L^{\omega'},L^{\omega}_0)$ is called an assignment from $\omega'$ to $\omega$ and denoted by $\lambda_{\overrightarrow{\omega'\omega}}$ or $\lambda$ if there is no ambiguity.
\end{definition}

Condition  \eqref{E:May-1Assign} means that an estimated target $j$ may be an estimated target of a true target $i$ if they both exists for at least one common time index.

Similarly, an element in $\lambda\in\mathcal{M}(L^{\omega},L^{\omega'}_0)$ is defined in the same way as Definition \ref{D:Assignment}.

Allowing many tracks in $\omega'$ to be assigned to a track in $\omega$ in each $\lambda\in \mathcal{M}(L^{\omega'},L^{\omega}_0)$ captures the situation where the MTT algorithm produces some separate estimated tracks instead of a track of a truth track due to miss-detection and/or clutter (e.g. scenarios in Figures \eqref{Fig:ExamMotivateOSPMAT0} and \eqref{Fig:ExamMotivateOSPMAT1}).

Element $0$ in set $L^\omega_0$ is introduced to capture false tracks in $\omega'$ if they exist by assigning to $0$ (e.g. scenario in Figure \eqref{Fig:ExamMotivateOSPMAT3}).

If a track in $\omega$ is not assigned to any tracks in $\omega'$, then that track is a missed track (e.g. scenario in Figure \eqref{Fig:ExamMotivateOSPMAT1}).

Similar to the OSPA metric (summary in Appendix \ref{SS:Summary_OSPA}) ,let $p$ ($1\le p<\infty$) be the 
order parameter
which penalizes the spatial distance between a track in
$\omega$ and any of its associated tracks in $\omega'$ at each time index. 
Let $c>0$ be the cutoff parameter. In order to distinguish between one-to-one
assignment and many-to-one assignment, we introduce $\Delta$ ($0<\Delta\le c$), called the assignment parameter. Note that $\Delta$ must be positive and 
is explained later.
This assignment parameter is usually chosen very small if there is miss-detection or low probability-of-detection
otherwise $\Delta$ is chosen large if there is no miss-detection or a high probability-of-detection.

Normalization of an MTT metric is an important term to ensure reliable measurement when one of the two sets $\omega$ or $\omega'$ is larger in cardinality. We follow a similar approach to that used in the OSPA metric. The normalization is the number of distances between two sets of tracks. In order to find this number, we need to sum all of the distances between these two sets at all time indices, as explained below.

Let $\mathbf{n}^{\omega}_t$ and $\mathbf{n}^{\omega'}_t$ be the number of existing targets at time 
$t$ in $\omega$ and $\omega'$ respectively.
Then the number of distances at time $t$ between $\omega$ and $\omega'$, denoted $\mathbf{n}_t$, is the maximum number of targets in $\omega$ and $\omega'$ at time $t$ (similar to the OSPA metric). Thus the number of distances between $\omega$ and $\omega'$, denoted by $\mathbf{n}$, is the sum of all numbers of distances between $\omega$ and $\omega'$ from time $1$ to $T$. Mathematically,

\begin{subequations}\label{E:NoTar_t_Sample1_2}
\begin{eqnarray}
\textstyle\mathbf{n}^{\omega}_t&=&
    \textstyle\sum_{\tau\in\omega}|\tau(t)|,\qquad\mathbf{n}^{\omega'}_t=\textstyle\sum_{\tau\in \omega'}|\tau(t)|,\label{E:NoTar_t_Sample1_2_1}\\
\mathbf{n}_t&=&\max\{\mathbf{n}^{\omega}_t,\mathbf{n}^{\omega'}_t\},\qquad
\mathbf{n}=\textstyle\sum_{t=1}^T\mathbf{n}_t.\label{E:NoTar_t_Sample1_22}
\end{eqnarray}
\end{subequations}

Note that $|\tau(t)|=1$ if track $\tau$ exists at time $t$ and $|\tau(t)|=0$ if track $\tau$ does not exists at time $t$.

Applying \eqref{E:NoTar_t_Sample1_2} to the example in Figure \ref{Fig:ExamMotivateOSPMAT}, 
$\mathbf{n}^{\omega}_t=\mathbf{n}^{\omega'}_t=\mathbf{n}_t=1$ for all $t=1,2,3$; for Figure \eqref{Fig:ExamMotivateOSPMAT1}, $\mathbf{n}^{\omega}_t=\mathbf{n}^{\omega'}_t=\mathbf{n}_t=2$ for all $t=4,5$; for Figures \eqref{Fig:ExamMotivateOSPMAT0}, \eqref{Fig:ExamMotivateOSPMAT2} and \eqref{Fig:ExamMotivateOSPMAT3}, $\mathbf{n}^{\omega}_t=\mathbf{n}^{\omega'}_t=\mathbf{n}_t=1$ for all $t=4,5$. Hence  $\mathbf{n}=7$ for Figure \eqref{Fig:ExamMotivateOSPMAT1} and $\mathbf{n}=5$ for Figures \eqref{Fig:ExamMotivateOSPMAT0}, \eqref{Fig:ExamMotivateOSPMAT2} and \eqref{Fig:ExamMotivateOSPMAT3}.

As mentioned above for a many-to-one assignment, $\Delta$ is added to each extra track in $\omega'$ which is assigned to a track in $\omega$. Consider the scenario in Figure \eqref{Fig:ExamMotivateOSPMAT1} where two tracks $\tau'_1$ and $\tau'_2$ are assigned to truth track $\tau_1$. If track $\tau'_1$ is an extra track assigned to truth track $\tau_1$, the length of this extra track is $3$. If  track $\tau'_1$ is an extra track assigned to truth track $\tau_1$, the length of this extra track is $2$. Thus we add $\Delta$ to each time an extra track exists and an order (defined below) between the many tracks in $\omega'$ assigned to a track in $\omega$ plays an important role. Indeed, without the order, the total distances in Figure  \eqref{Fig:ExamMotivateOSPMAT0} can be $3\epsilon+2(\epsilon+\Delta)$ or $2\epsilon+3(\epsilon+\Delta)$. If the orders are $(\tau'_1,\tau'_2)$ and $(\tau'_2,\tau'_1)$, the total distances are $3\epsilon+2(\epsilon+\Delta)$ and $2\epsilon+3(\epsilon+\Delta)$ respectively. Hence the notion of an order of assigned tracks at each time index is needed. The notions of the order of assigned tracks and the order of assigned track at each time index are now formally defined.
\begin{definition}[Track Order given $\lambda$ and $i$]\label{D:Tr_Ord}
Given $\lambda\in\mathcal{M}(L^{\omega'},L^{\omega}_0)$. For any $i\in L^{\omega}$. A \textbf{track order given $\lambda$ and $i$}, denoted by $\pi^\lambda_i$, is the order of the tracks in $\omega'$ which are assigned to track $i$ in $\omega$ (note that $\lambda$ roughly represents the mapping from $\omega'$ to $\omega$). Mathematically, it is defined as follows.

\begin{itemize}
  \item If $i\notin\lambda(L^{\omega'})$, then $\pi^\lambda_i$ is $0$.
  \item If $\lambda^{-1}(i)=\{k_1,\ldots,k_{l_i}\}$ where $k_1,\ldots,k_{l_i}\in L^{\omega'}$, then $\pi^\lambda_i=(k_{\sigma(1)},\ldots,k_{\sigma(l_i)})$ where $\sigma\in\Pi_{l_i}$ where $\Pi_k$ denotes the set of permutations on $\{1,2,...,k\}$  for any positive integer $k$.
\end{itemize}
\end{definition}

Note that if $\lambda\in\mathcal{M}(L^{\omega},L^{\omega'}_0)$ then $\pi^\lambda_i$ where $i\in L^{\omega'}$ is defined similarly.

To compute the distance between $\omega$ and $\omega'$, it is also important to know the  track order given $\lambda$ and $i$ at each time index. Hence the following definition will explain the meaning of track order given $\lambda$ and $i$ at each time index.  

\begin{definition}[Track Order given $\lambda$ and $i$ at $t$]\label{D:Tr_Ord_t}
Given $\lambda\in\mathcal{M}(L^{\omega'},L^{\omega}_0)$ and $\pi^\lambda_i$. Similar to the definition of a track order given $\lambda$ and $i$, a \textbf{track order given $\lambda$ and $i$ at time $t$}, denoted by $\pi_{i,t}^\lambda$, has the same order of the tracks given $\lambda$ and $i$ except from deleting the tracks which do not exist at time $t$. Mathematically, it is defined as follows.
\begin{itemize}
  \item If $i\notin\lambda(L^{\omega'})$, then $\pi_{i,t}^\lambda$ is $0$.
  \item Otherwise assume that $\pi^\lambda_i=(k_1,\ldots,k_{r_1},\ldots,k_{r_2},\ldots,k_{r_{i_t}},\ldots,k_{l_i})$.
  If $k_{r_1},k_{r_2},\ldots,k_{r_{i_t}}$ exist at time $t$, then $\pi_{i,t}^\lambda=(k_{r_1},k_{r_2},\ldots,k_{r_{i_t}})$.
\end{itemize}
\end{definition}

In order to denote a component of $\pi^\lambda_i$ or $\pi_{i,t}^\lambda$, we assume that
\begin{eqnarray}
  \pi^\lambda_i&=&(k_1,\ldots,k_{l_i}) \label{E:T:Asso1}\\
  \pi_{i,t}^\lambda&=&(k_{r_1},\ldots,k_{r_{i_t}}).\label{E:T:Asso2}
\end{eqnarray}
 Then the assigned track at the $s$th order in \eqref{E:T:Asso1} ($s=1,\ldots,l_i$), denoted by $\pi_i^\lambda(s)$, is $k_s$ (i.e. $\pi_i^\lambda(s)=k_s$). Similarly, if \eqref{E:T:Asso2} holds at time $t$, the assigned track at the $h$th order in \eqref{E:T:Asso2} ($h=1,\ldots,r_t$), denoted by $\pi_{i,t}^\lambda(h)$, is $k_{r_h}$ (i.e. $\pi_{i,t}^\lambda(h)=k_{r_h}$).

Denote the number of components in $\pi_{i,t}^\lambda$ as $\bar{\mathfrak{n}}^{\lambda}_{t,i}$. Indeed, $\bar{\mathfrak{n}}^{\lambda}_{t,i}$ is the number of existing targets in $\omega'$ assigned to target $i$ at time $t$ via mapping $\lambda$. If $i\notin\lambda(L^{\omega'})$, then $\bar{\mathfrak{n}}^{\lambda}_{t,i}=0$. If $\pi_{i,t}^\lambda=(k_{r_1},k_{r_2},\ldots,k_{r_{i_t}})$, then $\bar{\mathfrak{n}}^{\lambda}_{t,i}=i_t$. A collection of all $\pi^\lambda_i$ for $i \in L^{\omega}$ is defined as

\begin{definition}[Track Order given $\lambda$]\label{D:Tr_Ordlambda}
Given $\lambda\in\mathcal{M}(L^{\omega'},L^{\omega}_0)$. A \textbf{track order given $\lambda$}, denoted by $\pi^\lambda$, is a vector of $|\omega|$ ($\omega\ne\emptyset$) elements whose element is $\pi^\lambda_i$ for $i \in L^{\omega}$. Mathematically, it is defined as follows.
\begin{equation}\label{E:pilambda}
\pi^\lambda=(\pi^\lambda_1,\ldots,\pi^\lambda_{|\omega|}).
\end{equation}
\end{definition}
%

Hence we also define $$\Pi^\lambda_{|\omega|}=\{\pi^\lambda\text{ defined in \eqref{E:pilambda}: each }\pi^\lambda_i\text{ defined in Definition \ref{D:Tr_Ord}}\}$$ as a collection of all permutations of $\pi^\lambda_i$ ($i\in L^\omega$).

We are now in a position to formally define the distance from $\omega$ to  $\omega'(\omega'\ne\emptyset)$ conditioned on $\lambda\in\mathcal{M}(L^{\omega'},L^{\omega}_0)$ as follows.

\begin{definition}[Directional distance given $\lambda$]\label{D:OSPAMT Dist} Given the order parameter $p$ ($1\le p<\infty$), the cutoff parameter $c$, and $\lambda\in\mathcal{M}(L^{\omega'},L^{\omega}_0)$. Let $0<\Delta<c$. The distance from $\omega'$ to $\omega(\omega,\omega'\ne\emptyset)$ conditioned on $\lambda$ is the minimum distance of all distances of all permutations of track orders given $\lambda$ of many-to-one assignments and is defined as ($\omega,\omega'=\emptyset$ is discussed later).
 \begin{align}\label{E:lambdaMetric}
\textstyle\tilde{d}_{c,p}^{\Delta,\lambda}(\overrightarrow{\omega',\omega})=
\min_{\pi^{\lambda}\in \Pi^\lambda_{|\omega|}}\Big(\frac{1}{\mathbf{n}}\sum_{t=1}^T\tilde{d}_{c,p,t}^{\Delta,\lambda,\pi^\lambda}(\overrightarrow{\omega',\omega})\Big)^\frac{1}{p}
\end{align}
where
\begin{align}\label{E:lambdaMetric_pi}
&\textstyle\tilde{d}_{c,p,t}^{\Delta,\lambda,\pi^\lambda}(\overrightarrow{\omega',\omega})=
\sum_{i=1}^{|\omega|}d_{c,p,t}^{\Delta,\lambda,\pi^\lambda}\!\!(\tau^\omega_i,\omega')+s^{Card,\lambda}_{c,p,t,\Delta}(\omega)
\end{align}
and where $s^{Card,\lambda}_{c,p,t,\Delta}(\omega)$ is
\begin{align}\label{E:lambdaMetric_pi2}
&\textstyle\sum_{i=1}^{|\omega|}|\tau^\omega_i(t)|\max\{\bar{\mathfrak{n}}^{\lambda}_{t,i}-1,0\}(\Delta^p+c^p)\nonumber\\
+&\textstyle c^p(\!\mathbf{n}_t-\!\sum_{i=1}^{|\omega|}
\bar{\mathfrak{n}}^{\lambda}_{t,i}|\tau^\omega_i(t)|),
\end{align}
and for $j=\pi_{i,t}^\lambda(1)$, $k=\pi_i^\lambda(1)$, $d_{c,p,t}^{\Delta,\lambda,\pi^\lambda}\!\!(\tau^\omega_i,\omega')$ is
\begin{align}\label{E:lambdaMetric_pi1}
\!\!\left\{\!\!
   \begin{array}{ll}
     0,&\!\!\!\!\hbox{if $j|\tau^\omega_i(t)|\!=\!0$;} \\
     \!d^c(\tau^\omega_i(t),\tau^{\omega'}_j(t))^p\!+\!\bar{\delta}(j,k)\Delta^p, &\!\!\!\!\hbox{otherwise;}
   \end{array}
 \right.\!\!
\end{align}
where $\bar{\delta}(k,h)$ is the complement of a Kronecker delta (i.e $\bar{\delta}(k,h)$ is $1$ if $k\ne h$ or $0$ if $k=h$); and for $x,y\in\mathcal{X}$, $d(x,y)$ is typical $\|x-y\|_{p'}$, 
$p'\in[1,\infty)$
\begin{align}\label{E:OSPADist}
d^c(\tilde{x},\tilde{y})=
\left\{\!
  \begin{array}{ll}
    \min\{c,d(x,y)\}, & \hbox{if $\tilde{x}=\{x\},\tilde{y}=\{y\}$;} \\
    0, & \hbox{otherwise.}
  \end{array}
\right.
\end{align}
\end{definition}

\begin{rmk}\label{R:ExplainDef}
The arrow above the pair $(\omega',\omega)$ in Definition \ref{D:OSPAMT Dist}, $(\overrightarrow{\omega',\omega})$, represents a mapping from $\omega'$ to $\omega$, i.e. more than one track in $\omega'$ can be assigned to a track in $\omega$.

Equation \eqref{E:lambdaMetric_pi} represents the distance between two sets of tracks at a time index $t$. The first term on the
RHS of \eqref{E:lambdaMetric_pi} represents the total of all spatial distances (or local error) across two sets of tracks at time $t$. The symbol $d_{c,p,t}^{\Delta,\lambda,\pi^\lambda}(\tau^\omega_i,\omega')$ is a distance between a track $i$ in $\omega$ and a set of tracks $\omega'$ at time $t$ and is defined in \eqref{E:lambdaMetric_pi1}. The first line of \eqref{E:lambdaMetric_pi1} means that at time $t$ the distance between a track $i$ and a set of tracks $\omega'$, $d_{c,p,t}^{\Delta,\lambda,\pi^\lambda}(\tau^\omega_i,\omega')$, is $0$ if no track in $\omega'$ is assigned to track $i$ in $\omega$ or if a target $i$ does not exist. The first term on the second line of \eqref{E:lambdaMetric_pi1} is the distance between track $i$ and its first assigned track among the order of assigned tracks existing at time $t$ if both tracks exist (this distance is defined in the first line of \eqref{E:OSPADist}) and $0$ if either track $i$ does not exist at time $t$ or none of tracks assigned to $i$ exists at time $t$ (it is defined in the second line of \eqref{E:OSPADist}). The second term of the second line, $\Delta^p$, is included if the first track exists at time $t$ is not the first track among ordered tracks assigned to track $i$.

The second term of \eqref{E:lambdaMetric_pi}, $s^{Card,\lambda}_{c,p,t,\Delta}(\omega)$,  represents the cardinality error between these two sets of tracks at time $t$. Note that the parameter of $s^{Card,\lambda}_{c,p,t,\Delta}(\cdot)$ is in codomain $\omega$ in the mapping from $\omega'$ to $\omega$,  because the number of extra targets in $\omega'$ at time $t$ is explained in $\mathbf{n}_t$ and $\bar{\mathfrak{n}}^{\lambda}_{t,i}$.

The first term of \eqref{E:lambdaMetric_pi2} is the total of cardinality errors for extra states from extra tracks assigned to a truth track at time $t$. While the term under the sum, $|\tau^\omega_i(t)|\max\{\bar{\mathfrak{n}}^{\lambda}_{t,i}-1,0\}(\Delta^p+c^p)$, is the cardinality error at time $t$ for the extra states assigned to the state of target $i$. The term $\bar{\mathfrak{n}}^{\lambda}_{t,i}-1$ represent the number of tracks in $\omega'$ assigned to track $i$ minus the first track which is used to calculate the spatial distance between that first track and track $i$ at time $t$. The cardinality error,$\Delta^p+c^p$, is the sum of $\Delta^p$ and $c^p$ if there is more than one track in $\omega'$ assigned to track $i$ at time $t$. That is, the number of the existing assigned targets at time $t$ is larger than one, i.e. $\bar{\mathfrak{n}}^{\lambda}_{t,i}>1$.

The second term of \eqref{E:lambdaMetric_pi2} represents the cardinality errors of missing tracks or false tracks or missing states or false states. The term $
\bar{\mathfrak{n}}^{\lambda}_{t,i}|\tau^\omega_i(t)|$ represents the number of distances at time $t$ that are used in \eqref{E:lambdaMetric_pi1} and in the first term of \eqref{E:lambdaMetric_pi2}
\end{rmk}

\begin{rmk}\label{Rm:OSPAt}By \eqref{E:lambdaMetric_pi2}-\eqref{E:lambdaMetric_pi1},
 there are $\mathbf{n}_t$ terms on the right hand side of \eqref{E:lambdaMetric_pi} and all terms are smaller than or equal to $c^p+\Delta^p$.
\end{rmk}

\underline{\textbf{Example 1:}} 
Applying
\eqref{E:lambdaMetric_pi} to the example in Figure \ref{Fig:ExamMotivateOSPMAT}. Choose $c<\beta$.

\begin{itemize}
  \item For Figure \eqref{Fig:ExamMotivateOSPMAT0} and \eqref{Fig:ExamMotivateOSPMAT1}. Let $\lambda\in\mathcal{M}(L^{\omega'},L^\omega_0)$ such that $\lambda(1)=\lambda(2)=1$, $\pi^{\lambda}(1)=(1,2)$. Then for $t=4,5$ and $j=1,2,3$, $\tilde{d}_{c,p,j}^{\Delta,\lambda,\pi^\lambda}(\overrightarrow{\omega',\omega})=\epsilon^p$,
      \begin{align*}
\tilde{d}_{c,p,t}^{\Delta,\lambda,\pi^\lambda}(\overrightarrow{\omega',\omega})&
=\epsilon^p+\Delta^p\text{ for Fig.\eqref{Fig:ExamMotivateOSPMAT0}}\\
\tilde{d}_{c,p,t}^{\Delta,\lambda,\pi^\lambda}(\overrightarrow{\omega',\omega})&
=\epsilon^p+\Delta^p+c^p\text{ for Fig.\eqref{Fig:ExamMotivateOSPMAT1}}.
\end{align*}
  \item For Figure \eqref{Fig:ExamMotivateOSPMAT2}. Let $\lambda\in\mathcal{M}(L^{\omega'},L^\omega_0)$ such that $\lambda(1)=0$. Then for $t=1,\ldots,5$, $\tilde{d}_{c,p,t}^{\Delta,\lambda,\pi^\lambda}(\overrightarrow{\omega',\omega})=c^p$.
  \item For Figure \eqref{Fig:ExamMotivateOSPMAT3}. Let $\lambda\in\mathcal{M}(L^{\omega'},L^\omega_0)$ such that $\lambda(1)=1$, $\pi^{\lambda}(1)=1$. Then for $t=4,5$ and $j=1,2,3$
      \begin{align*}
\tilde{d}_{c,p,j}^{\Delta,\lambda,\pi^\lambda}(\overrightarrow{\omega',\omega})&=\epsilon^p,\qquad \tilde{d}_{c,p,t}^{\Delta,\lambda,\pi^\lambda}(\overrightarrow{\omega',\omega})
=c^p.
\end{align*}
\end{itemize}
Now we want to find a particular $\lambda\in\mathcal{M}(L^{\omega'},L^\omega_0)$ that minimizes the distance $\tilde{d}_{c,p}^{\Delta,\lambda}(\overrightarrow{\omega',\omega})$ in \eqref{E:lambdaMetric}
\begin{definition}[Quasi-OSPAMT Metric]\label{D:OSPAMT_QuaMetric} Given the order parameter $p$ ($1\le p<\infty$) and the cutoff parameter $c$. Let $0<\Delta<c$. A distance from $\omega'$ to $\omega$, denoted by $d_{c,p}^{\Delta}(\overrightarrow{\omega',\omega})$,  is defined as follows
\begin{align}\label{E:QuaMetric_OSPATrks}
\!\!d_{c,p}^{\Delta}(\overrightarrow{\omega',\omega})=\left\{\!\!\!
  \begin{array}{ll}
 \displaystyle\min_{\lambda\in\mathcal{M}(L^{\omega'},L^{\omega}_0)}
\!\!\tilde{d}_{c,p}^{\Delta,\lambda}(\overrightarrow{\omega',\omega}),&\!\! \hbox{\!\!if $\omega,\omega'\!\!\ne\!\emptyset$;} \\
    0, &\!\! \hbox{\!\!if $\omega=\omega'\!\!=\!\emptyset$;} \\
    c, &\!\! \hbox{otherwise.}
  \end{array}
\right.\!\!
\end{align}
\end{definition}
Note that if $\Delta=0$ then the distance \eqref{E:quasi-OSPAMTAss} violates the identity property.

\begin{rmk}$d_{c,p}^{\Delta}(\overrightarrow{\omega',\omega})$ in \eqref{E:QuaMetric_OSPATrks} is less than or equal to the cutoff parameter $c$.
\end{rmk}
The distance from $\omega'$ to $\omega$ in Definition \ref{D:OSPAMT_QuaMetric} is called the Quasi-OSPAMT because it is a quasimetric. It can be useful for MTT problems when the tracker estimates a truth track as many broken tracks  (see Figure \eqref{Fig:ExamMotivateOSPMAT0} and \eqref{Fig:ExamMotivateOSPMAT1}).
\begin{figure}[htbp!]
\centering{\hspace{-2cm}
\scalebox{0.9}
{\centering
\pgfdeclarelayer{background} \pgfdeclarelayer{foreground}
\pgfsetlayers{background,main,foreground}
\par
\par
\tikzset{cross/.style={cross out, draw=black, minimum size=2*(#1-\pgflinewidth), inner sep=0pt, outer sep=0pt},
cross/.default={3pt}}
\begin{tikzpicture}[y=.2cm, x=2cm,font=\sffamily]
 \draw [->,thick]
 |-(5.12,0) node (xaxis) [right]
  {\scriptsize Time};
\draw (1,0) -- coordinate (x axis mid) (5.1,0);
\foreach \x in {1,...,5}
     		\draw (\x,1pt) -- (\x,-3pt)
			node[anchor=north] {\x};
\draw (1,1.5) node[cross] (x1){};
\draw (2,1.5) node[cross] (x2){};
\draw (3,1.5) node[cross] (x3){};
\draw (4,1.5) node[cross] (x4){};
\draw (5,1.5) node[cross] (x5){};
\node(x11)[left=-0.12 of x1]{};
\node[circle,draw=black, fill=black, inner sep=0pt,minimum size=3pt] (y1) at (1,5.5) {};
\node(y11)[left=-0.11 of y1]{};
\node[circle,draw=black, fill=black, inner sep=0pt,minimum size=3pt] (y2) at (2,5.5){};
\node[circle,draw=black, fill=black, inner sep=0pt,minimum size=3pt] (y3) at (3,5.5){};
\node[circle,draw=black, fill=black, inner sep=0pt,minimum size=3pt] (y4) at (4,5.5){};
\node[circle,draw=black, fill=black, inner sep=0pt,minimum size=3pt] (y5) at (5,5.5){};
\node(y51)[right=-0.11 of y5]{};
%

 \draw[-,thin,black,dashed](x2)edge node[swap,near start]{}  (y2);
\draw[->,black](y1)edge (y2)(y2)edge (y3)(y3)edge(y4)(y4)edge (y5)
(x1)edge (x2)(x2)edge (x3)(x4)edge (x5)
;
\node(temdis1)  [below =.2cm of y1] {};
 \node (dis1) [right =-0.01cm of temdis1]{$\epsilon\ll 1$};
 \draw[-,thin,black,dashed](y1)edge node[swap,near start]{}
  (x1);
\node(temdis2) [below =.2cm of y2] {};
  \node (dis1) [right =-.5cm of temdis2]{$\epsilon$};
  \draw[-,thin,black,dashed](y2)edge node[swap,near start]{}
  (x2);
\node(temdis3) [below =.2cm of y3] {};
  \node (dis1) [right =-.5cm of temdis3]{$\epsilon$};
  \draw[-,thin,black,dashed](y3)edge node[midway]{}
  (x3);
\node(temdis4) [below =.2cm of y4] {};
  \node (dis1) [right =-.5cm of temdis4]{$\epsilon$};
  \draw[-,thin,black,dashed](y4)edge node[near start]{}
  (x4);
\node(temdis5) [below =.2cm of y5] {};
  \node (dis1) [right =-.5cm of temdis5]{$\epsilon$};
  \draw[-,thin,black,dashed](y5)edge node[near start]{}
  (x5);
 \begin{pgfonlayer}{background}
 \path (y5.east |-y5.east)+(.3,-.8) node(ii){$\tau'_1$};
  \path (x3.east |-x3.east)+(.3,.6) node(a){$\tau_1$};
  \path (x5.east|- x5.east)+(.3,.4) node(b){$\tau_2$};
\draw[->,black,dashed](b) edge (x5)(a)edge(x3)(ii) edge (y5);
\end{pgfonlayer}
\end{tikzpicture}
}}
\caption{$\omega=\{\tau_1,\tau_2\},\omega'=\{\tau'_1\}$ where $\epsilon\ll c$.}
\label{Fig:ManyTruthsEstAs1}
\end{figure}
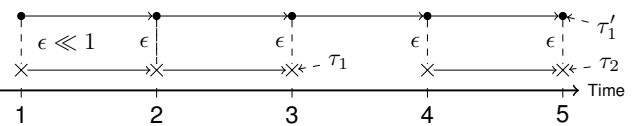
Identifying pairs of tracks across two sets of tracks is also important because it tells us which one is assigned to which. Similarly, the tracks, which are not assigned to any track, can be identified as missing tracks if they are from the set of truth track ($\omega$) and as false tracks if they are from the set of the estimated track ($\omega'$). In order to identify them, the quasi-OSPAMT assignment is defined as follows
\begin{definition}[Quasi-OSPAMT Assignment]Given the order parameter $p$ ($1\le p<\infty$) and the cutoff parameter $c$. Let $0<\Delta<c$. For $\omega,\omega'\ne\emptyset$, the quasi-OSPAMT assignment is a mapping $\lambda_*$ (from tracks in $\omega'$ to tracks in $\omega$ if $\omega,\omega'\!\!\ne\!\emptyset$) which is defined as
\begin{align}\label{E:quasi-OSPAMTAss}
&\lambda_*=\arg\min\min_{\lambda\in\mathcal{M}(L^{\omega'},L^{\omega}_0)}
\!\!\tilde{d}_{c,p}^{\Delta,\lambda}(\overrightarrow{\omega',\omega}).
\end{align}
\end{definition}

\begin{rmk}If $d_{c,p}^{\Delta}(\overrightarrow{\omega',\omega})$ in \eqref{E:QuaMetric_OSPATrks} is equal to the cutoff parameter and $\omega,\omega'\!\!\ne\!\emptyset$, all tracks in $\omega'$ are false tracks and all tracks in $\omega$ are missing track. In this case, the user can choose the new cutoff parameter which is larger than the previous cutoff parameter.
\end{rmk}
The quasi-OSPAMT metric will not be reliable for evaluating MTT algorithms. Assume that two different trackers have two different results in Figures \ref{Fig:ManyTruthsEstAs1} and \ref{Fig:MotiExaOSPAMTMetric}. The quasi-OSPAMT metric from $\omega'$ to $\omega$ for the scenarios in Figures \ref{Fig:ManyTruthsEstAs1} and  \ref{Fig:MotiExaOSPAMTMetric} is  $(3\epsilon+2c)/5$ even though the scenario in Figure \ref{Fig:MotiExaOSPAMTMetric} is clearly much worse than scenario in Figure \ref{Fig:ManyTruthsEstAs1}. Hence a distance which also satisfies the symmetric property of metrics is necessary for reliable performance evaluation  of  MTT algorithms. When symmetry is included, the distance between $\omega$ and $\omega'$ is called the OSPAMT metric and is defined as follows.
\begin{figure}[htbp!]
\centering{\hspace{-2cm}
\scalebox{0.9}
{\centering
\pgfdeclarelayer{background} \pgfdeclarelayer{foreground}
\pgfsetlayers{background,main,foreground}
\par
\par
\tikzset{cross/.style={cross out, draw=black, minimum size=2*(#1-\pgflinewidth), inner sep=0pt, outer sep=0pt},
cross/.default={3pt}}
\begin{tikzpicture}[y=.2cm, x=2cm,font=\sffamily]
 \draw [->,thick]
 |-(5.1,0) node (xaxis) [right]
  {\scriptsize Time};
\draw (1,0) -- coordinate (x axis mid) (5.1,0);
\foreach \x in {1,...,5}
     		\draw (\x,1pt) -- (\x,-3pt)
			node[anchor=north] {\x};
\draw (1,1.5) node[cross] (x1){};
\draw (2,1.5) node[cross] (x2){};
\draw (3,1.5) node[cross] (x3){};
\draw (4,1.5) node[cross] (x4){};
\draw (5,1.5) node[cross] (x5){};
\node[circle,draw=black, fill=black, inner sep=0pt,minimum size=3pt] (y1) at (1,5.5) {};
\node(y11)[left=-0.11 of y1]{};
\node[circle,draw=black, fill=black, inner sep=0pt,minimum size=3pt] (y2) at (2,5.5){};
\node[circle,draw=black, fill=black, inner sep=0pt,minimum size=3pt] (y3) at (3,5.5){};
\node[circle,draw=black, fill=black, inner sep=0pt,minimum size=3pt] (y4) at (4,10){};
\node[circle,draw=black, fill=black, inner sep=0pt,minimum size=3pt] (y5) at (5,10){};
 \draw[-,thin,black,dashed](x2)edge node[swap,near start]{}  (y2);
\draw[->,black](y1)edge (y2)(y2)edge (y3)(y3)edge (y4)(y4)edge (y5)
(x1)edge (x2)(x2)edge (x3)(x4)edge (x5);
\node(temdis1)  [below =.2cm of y1] {};
 \node (dis1) [right =-0.01cm of temdis1]{$\epsilon\ll 1$};
 \draw[-,thin,black,dashed](y1)edge node[swap,near start]{}
  (x1);
\node(temdis2) [below =.2cm of y2] {};
  \node (dis1) [right =-.5cm of temdis2]{$\epsilon$};
  \draw[-,thin,black,dashed](y2)edge node[swap,near start]{}
  (x2);
\node(temdis3)  [below =.3cm of y3] {};
  \node (dis1) [right =-.5cm of temdis3] {$\epsilon$};
  \draw[-,thin,black,dashed](y3)edge node[midway]{}
  (x3);
\node(temdis4)  [below =.3cm of y4] {};
  \node (dis1) [right =-.6cm of temdis4] {$\beta$};
  \draw[-,thin,black,dashed](y4)edge node[near start]{}
  (x4);
\node(temdis5)  [below =.3cm of y5] {};
  \node (dis1) [right =-.6cm of temdis5] {$\beta$};
  \draw[-,thin,black,dashed](y5)edge node[near start]{}
  (x5);
 \begin{pgfonlayer}{background}
 \path (y5.east |-y5.east)+(.3,-.8) node(ii){$\tau'_1$};
  \path (x3.east |-x3.east)+(.3,.6) node(a){$\tau_1$};
  \path (x5.east|- x5.east)+(.3,.4) node(b){$\tau_2$};
\draw[->,black,dashed](b) edge (x5)(a)edge(x3)(ii) edge (y5);
\end{pgfonlayer}
\end{tikzpicture}
}}
\caption{$\omega=\{\tau_1,\tau_2\},\omega'=\{\tau'_1\}$ where $\epsilon\ll c<\beta$}
\label{Fig:MotiExaOSPAMTMetric}
\end{figure}
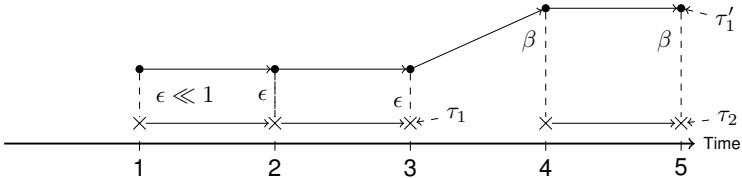
\begin{definition}[OSPAMT metric]\label{D:OSPAMTM metric} Given the order parameter $p$ ($1\le p<\infty$) and the cutoff parameter $c$. Let $0<\Delta<c$. The OSPAMT distance between $\omega$ and $\omega'$, denoted by $d_{c,p}^{\Delta}(\omega,\omega')$, is given by
\begin{align}\label{E:dis_OSPATrks}
d_{c,p}^{\Delta}(\omega,\omega')=\min\left\{d_{c,p}^{\Delta}(\overrightarrow{\omega',\omega}),d_{c,p}^{\Delta}(\overrightarrow{\omega,\omega'})
\right\}
\end{align}
\end{definition}

The distance in \eqref{E:dis_OSPATrks} is a metric.
The proof is presented in Appendix \ref{A:OSPAMT-metric}.
The OSPA metric in \cite{Schuhmacher2008} is \eqref{E:dis_OSPATrks}  when $T=1$ because by \eqref{E:lambdaMetric_pi} a many-to-one mapping gives a larger distance than a one-to-one mapping. Using the OSPAMT metric in Definition \ref{D:OSPAMTM metric} for the scenarios in Figures \ref{Fig:ManyTruthsEstAs1} and  \ref{Fig:MotiExaOSPAMTMetric} with $\epsilon+\Delta<c$, we can conclude that the scenario in Figure \ref{Fig:MotiExaOSPAMTMetric} is much worse than the scenario in Figure \ref{Fig:ManyTruthsEstAs1} because the  OSPAMT metric for Figures \ref{Fig:ManyTruthsEstAs1} and  \ref{Fig:MotiExaOSPAMTMetric} are $(5\epsilon+2\Delta)/5$ and $(3\epsilon+2c)/5$ respectively.

In MTT applications the situation of more than one truth track being assigned to an estimated track is not practical. In such a scenario, the estimated track must have been considered as more than two estimated tracks. In order to avoid this scenario, we must find a way to check whether the estimated track can be potentially a combination of many estimated tracks (see Figure \ref{Fig:ManyTruthsEstAs1}). If this happens, a method of identify and solving this situation is needed for reliable evaluation. Take the examples in Figures \ref{Fig:MotiExaOSPAMTMetric}. If we can identify the estimated track $\tau'_1$ which is assigned to many truth tracks, we can break this estimated into many tracks. In order to do this, the optimal assignment obtained from the process of calculating the OSPAMT metric (i.e.$\lambda$) is the OSPAMT assignment which is defined below.

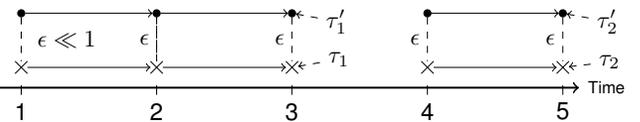
\begin{figure}[htbp!]
\centering{\hspace{-2cm}
\scalebox{0.9}
{\centering
\pgfdeclarelayer{background} \pgfdeclarelayer{foreground}
\pgfsetlayers{background,main,foreground}
\par
\par
\tikzset{cross/.style={cross out, draw=black, minimum size=2*(#1-\pgflinewidth), inner sep=0pt, outer sep=0pt},
cross/.default={3pt}}
\begin{tikzpicture}[y=.2cm, x=2cm,font=\sffamily]
 \draw [->,thick]
 |-(5.12,0) node (xaxis) [right]
  {\scriptsize Time};
\draw (1,0) -- coordinate (x axis mid) (5.1,0);
\foreach \x in {1,...,5}
     		\draw (\x,1pt) -- (\x,-3pt)
			node[anchor=north] {\x};
\draw (1,1.5) node[cross] (x1){};
\draw (2,1.5) node[cross] (x2){};
\draw (3,1.5) node[cross] (x3){};
\draw (4,1.5) node[cross] (x4){};
\draw (5,1.5) node[cross] (x5){};
\node(x11)[left=-0.12 of x1]{};
\node[circle,draw=black, fill=black, inner sep=0pt,minimum size=3pt] (y1) at (1,5.5) {};
\node(y11)[left=-0.11 of y1]{};
\node[circle,draw=black, fill=black, inner sep=0pt,minimum size=3pt] (y2) at (2,5.5){};
\node[circle,draw=black, fill=black, inner sep=0pt,minimum size=3pt] (y3) at (3,5.5){};
\node[circle,draw=black, fill=black, inner sep=0pt,minimum size=3pt] (y4) at (4,5.5){};
\node[circle,draw=black, fill=black, inner sep=0pt,minimum size=3pt] (y5) at (5,5.5){};
\node(y51)[right=-0.11 of y5]{};
%

 \draw[-,thin,black,dashed](x2)edge node[swap,near start]{}  (y2);
\draw[->,black](y1)edge (y2)(y2)edge (y3)(y4)edge (y5)
(x1)edge (x2)(x2)edge (x3)(x4)edge (x5)
;
\node(temdis1)  [below =.2cm of y1] {};
 \node (dis1) [right =-0.01cm of temdis1]{$\epsilon\ll 1$};
 \draw[-,thin,black,dashed](y1)edge node[swap,near start]{}
  (x1);
\node(temdis2) [below =.2cm of y2] {};
  \node (dis1) [right =-.5cm of temdis2]{$\epsilon$};
  \draw[-,thin,black,dashed](y2)edge node[swap,near start]{}
  (x2);
\node(temdis3) [below =.2cm of y3] {};
  \node (dis1) [right =-.5cm of temdis3]{$\epsilon$};
  \draw[-,thin,black,dashed](y3)edge node[midway]{}
  (x3);
\node(temdis4) [below =.2cm of y4] {};
  \node (dis1) [right =-.5cm of temdis4]{$\epsilon$};
  \draw[-,thin,black,dashed](y4)edge node[near start]{}
  (x4);
\node(temdis5) [below =.2cm of y5] {};
  \node (dis1) [right =-.5cm of temdis5]{$\epsilon$};
  \draw[-,thin,black,dashed](y5)edge node[near start]{}
  (x5);
 \begin{pgfonlayer}{background}
 \path (y5.east |-y5.east)+(.3,-.6) node(bb){$\tau'_2$};
 \path (y3.east |-y3.east)+(.3,-.6) node(aa){$\tau'_1$};
  \path (x3.east |-x3.east)+(.3,.6) node(a){$\tau_1$};
  \path (x5.east|- x5.east)+(.3,.4) node(b){$\tau_2$};
\draw[->,black,dashed](b) edge (x5)(a)edge(x3)(bb) edge (y5)(aa)edge(y3);
\end{pgfonlayer}
\end{tikzpicture}
}}
\caption{$\omega=\{\tau_1,\tau_2\},\omega^{\prime\prime}=\{\tau^{\prime\prime}_1,\tau^{\prime\prime}_2\}$ where $<\epsilon<\ll c$.}
\label{Fig:ManyTruthsEstAs1_Same}
\end{figure}

\begin{definition}[OSPAMT Assignment] Given the order parameter $p$ ($1\le p<\infty$) and the cutoff parameter $c$. Let $0<\Delta<c$. For $\omega,\omega'\ne\emptyset$, the OSPAMT assignment is a mapping $\lambda_*$ (between tracks in $\omega$ and tracks in $\omega'$) which is defined as
\begin{align}\label{E:OSPATrks}
\arg\min\left\{\min_{\lambda\in\mathcal{M}(L^{\omega'},L^{\omega}_0)}
\tilde{d}_{c,p}^{\Delta,\lambda}(\overrightarrow{\omega',\omega}),\min_{\lambda\in\mathcal{M}(L^{\omega},L^{\omega'}_0)}
\tilde{d}_{c,p}^{\Delta,\lambda}(\overrightarrow{\omega,\omega'})
\!\!\right\}.
\end{align}
\end{definition}
As discussed above, the OSPAMT distances for Figures \ref{Fig:ManyTruthsEstAs1} and \ref{Fig:MotiExaOSPAMTMetric} are $(5\epsilon+2\Delta)/5$ and $(3\epsilon+2c)/5$ respectively. Then the OSPAMT assignment for Figures \ref{Fig:ManyTruthsEstAs1} and \ref{Fig:MotiExaOSPAMTMetric} are $\lambda_{*,1}\in\mathcal{M}(L^{\omega},L^{\omega'}_0)$ and $\lambda_{*,2}\in\mathcal{M}(L^{\omega'},L^\omega_0)$ respectively where $\lambda_{*,1}^{-1}(1)=\{1,2\}$ and $\lambda_{*,2}^{-1}(1)=1$. Hence, the estimated track $\tau'_1$ in Figure \ref{Fig:ManyTruthsEstAs1} is assigned to both truth tracks $\tau_1$ and $\tau_2$. Since $\tau_1$ exists from time index $t=1$ to time index $t=3$ and $\tau_2$ exists from time index $t=4$ to time index $t=5$, we can divide $\tau'_1$ into two estimated tracks $\tau^{\prime\prime}_1$ and $^{\prime\prime}_2$ as in Figure \ref{Fig:ManyTruthsEstAs1_Same}. Then the OSPAMT metric after splitting for Figure \ref{Fig:ManyTruthsEstAs1} ($\epsilon$) is smaller than the OSPAMT metric for the original Figure \ref{Fig:ManyTruthsEstAs1} ($(5\epsilon+2\Delta)/5$).

\begin{rmk}\label{R:NoMatchTrBig1}
For improved performance evaluation we calculate the quasi-OSPAMT distances from $\omega$ to $\omega'$. If the quasi-OSPAMT assignment from $\omega$ to $\omega'$ is many-to-one, then we can split all estimated tracks which are assigned to more than one truth track, until the quasi-OSPAMT assignment (after splitting) is one-to-one. Then the OSPAMT metric and assignment will be optimal (see Figure \ref{Fig:ManyTruthsEstAs1} and Figure \ref{Fig:ManyTruthsEstAs1_Same} after the scenario in Figure \ref{Fig:ManyTruthsEstAs1} is split).
\end{rmk}

Computation of the OSPAMT assignment can be carried out using the following (not necessarily the most computationally efficient) algorithm. A The code can be found in \url{https://www.researchgate.net/publication/318922031_OSPAMT_Matlab_Code}
\begin{rmk}[Steps for computing the OSPAMT Metric]{\underline{\textbf{Steps for computing the OSPAMT Metric:}}}\label{Rmk:Compute-Many-one Assignment}
\begin{itemize}
  \item Calculate all distances of combinations between two sets of finite tracks with the cost function \eqref{E:QuaMetric_OSPATrks}. For example, if $m=|\omega|$ and $n=|\omega'|$. Then the matrix $m\times\!n$ $D$ whose elements are distance between all combination of two sets of tracks.
  \item For each row of $D$, a column which has the smallest distance of all other columns keeps its value in  matrix $m\times\!n$ $D_1$ and other columns will have value $\infty$ in matrix $m\times\!n$ $D_1$. Similarly,  we obtain a matrix $m\times\!n$ $D_2$ such that for each column of $D_2$, only row with minimum distance in matrix $D$ has its value of $D$ the rest is  $\infty$.  For example if
   \begin{align*}
   D=\left[ {\begin{array}{cccc}
   	70& 80& 80 &80\\
   	79& 80 &29 &80 \\
   	80 &50 &80& 55\end{array} } \right]
   \end{align*}
   then  
     \begin{align}\label{E:ComputeOSPAMT1}
   \!\!D_1\!=\!\!\left[\!\! {\begin{array}{cccc}
   	70& \infty& \infty &\infty\\
   	\infty& \infty &29 &\infty \\
   	\infty &50 &\infty& \infty\end{array} } \!\!\right],
   \end{align}
    \begin{align}\label{E:ComputeOSPAMT2}
   D_2=\left[ {\begin{array}{cccc}
   	70& \infty&\infty &\infty\\
   	\infty    & \infty&29     &\infty \\
   	\infty&50 	  &\infty & 55\end{array} } \right]
   \end{align}
  \item 
   Combine 2 matrices $D_1$ and $D_2$ as follows. For each row of $D_1$, fill in the non $\infty$ value of $D_2$ into the corresponding element of matrix $D_1$ whose value is $\infty$ and produce matrix $D_3$. From example of $D_1$ and $D_2$ in \eqref{E:ComputeOSPAMT1}-\eqref{E:ComputeOSPAMT2},
   \begin{align}\label{E:ComputeOSPAMT12}
  D_3=\!\!\left[\!\! {\begin{array}{cccc}
  	70& \infty& \infty&\infty\\
  	\infty& \infty &\infty &\infty \\
  	\infty &50 &\infty& 55\end{array} } \!\!\right]
  \end{align}
    \item {\bf{Find Optimal Assignment}}:
  Find the minimum value of $D_3$ denoted by $minD3$, if row and column of that value has more than one value smaller than $\infty$, the row or column that has the smallest value which is larger than $minD3$ will be ordered according to the value.The entry of $D_3$ which has value $minD3$ will become $\infty$. We continue until all the value of $D_3$ is $\infty$. As a result, we got a matrix $D_4$ is the same size as matrix $D$ whose values is non-negative integer. Elements of $D_4$ is $0$ if their values in $D_3$ are $\infty$.  The following steps will demonstrate how to find the optimal many-to-one assignment of $D_3$
  \begin{equation*}
 \!\! D'_3=\!\!\left[\!\! {\begin{array}{cccc}
 	70& \infty& \infty&\infty\\
 	\infty& \infty &\infty &\infty \\
 	\infty &50 &\infty& 55\end{array} } \!\!\right],\qquad
 \!\!D_4\!=\!\!\left[\!\! {\begin{array}{cccc}
 	0& 0&0 &0\\
 	0    & 0&1     &0 \\
 	0&50 	  &0 & 55\end{array} }\!\! \right]
 \end{equation*}
 the minimum value of $D'_3=D_3$ is 29 at $2nd$ row and $3rd$ column. There is no other value larger than $29$ and smaller the $\infty$ in  $2nd$ row and $3rd$ column. Hence the entry in $2nd$ row and $3rd$ column of $D_4$ is $1$. Then the entry in $2nd$ row and $3rd$ column of $D'_3$ become $\infty$.  The next minimum value of $D'_3$ is $50$ at $3rd$ row and $2nd$ column then the entry of $3rd$ row and $2nd$ column of $D_4$ is $1$. The $3rd$ row has other value $50<55<\infty$ at $4th$ column then  the entry of $3rd$ row  and $4th$ column  of $D_4$ is $2$ and of $D'_3$ is $\infty$. We continue until all the value of $D'_3$ is $\infty$.
 \begin{equation*}
 \!\! D'_3=\!\!\left[\!\! {\begin{array}{cccc}
 	70& \infty& \infty&\infty\\
 	\infty& \infty &\infty &\infty \\
 	\infty &\infty &\infty& \infty\end{array} } \!\!\right],\qquad
 \!\!D_4\!=\!\!\left[\!\! {\begin{array}{cccc}
 	0& 0&0 &0\\
 	0    & 0&1     &0 \\
 	0&1 	  &0 & 2\end{array} }\!\! \right]
 \end{equation*}
  \begin{equation*}
  \!\! D'_3=\!\!\left[\!\! {\begin{array}{cccc}
  	\infty& \infty& \infty&\infty\\
  	\infty& \infty &\infty &\infty \\
  	\infty &\infty &\infty& \infty\end{array} } \!\!\right],\qquad
  \!\!D_4\!=\!\!\left[\!\! {\begin{array}{cccc}
  1& 0&0 &0\\
  	0    & 0&1     &0 \\
  	0&1 	  &0 & 2\end{array} }\!\! \right]
  \end{equation*}
  \item $D_4$ represent the assignment between $\omega$ and $\omega'$.
  \item Note that the missing track  appears if its distance to other track in $\omega'$ is $c$ where $c=80$. Hence if $D$ has the entire row or column whose entries are $c$, remove it before finding the optimal asignment.
\end{itemize}
\end{rmk}

In the remaining of this section, we will find the OSPAMT metric at each time index; the OSPAMT cardinality error and localization error; and the OSPAMT cardinality error and localization error at each time index. Assume that we find OSPAMT assignment $\lambda_*$ from \eqref{E:OSPATrks}. To avoid to distinguish whether $\lambda$ is a mapping from $\omega$ to $\omega'$ or reverse. We denote
%
$\omega^\bullet=\omega'$ and $\omega^*=\omega$
if $\lambda_*\in\mathcal{M}(L^{\omega'},L^{\omega}_0)$, otherwise $\omega^\bullet=\omega$ and $\omega^*=\omega'$ if $\lambda_*\in\mathcal{M}(L^{\omega},L^{\omega'}_0)$. Thus the optimal order $\pi^{*\lambda_*}$ of $\lambda_*\in\mathcal{M}(L^{\omega^\bullet},L^{\omega^*}_0)$ is
 \begin{align}\label{E:lambdaMetric_opt_pi}
&\pi^{*\lambda_*}=\textstyle
\arg\min_{\pi^{\lambda_*}}\Big(\frac{1}{\mathbf{n}}\sum_{t=1}^T
\tilde{d}_{c,p,t}^{\Delta,\lambda_*,\pi^{\lambda_*}}(\overrightarrow{\omega^\bullet,\omega^*})\Big)^\frac{1}{p}
\end{align}
where $\pi^{\lambda_*}$ is given in \eqref{E:pilambda} and $\bar{\mathfrak{n}}^{\lambda_*}_{t,i}$ is the number of targets at time $t$ in $\omega^\bullet$ assigned to target $i$ in $\omega^*$.

The OSPAMT metric also gives the distance between the two sets of tracks $\omega$ and $\omega'$ at any time
\begin{definition}[OSPAMT distance at time $t$]\label{D:OSPAMT_T} Given the order parameter $p$ ($1\le p<\infty$) and the cutoff parameter $c$. Let $0<\Delta<c$. The distance between $\omega$ and $\omega'$ at time $t$, denoted by $d_{c,p}^{\Delta,t}(\omega,\omega')$, is
\begin{align}\label{E:OSPAMT_t}
\left\{
  \begin{array}{ll}
    0, & \hbox{if $\mathbf{n}_t=0$;} \\
    \left[\frac{1}{\mathbf{n}_t}\tilde{d}_{c,p,t}^{\Delta,\lambda_*,\pi^{*\lambda_*}}(\overrightarrow{\omega^{\bullet},\omega^*})
    \right]^{\frac{1}{p}},& \hbox{otherwise}
  \end{array}
\right.
\end{align}
where $\tilde{d}_{c,p,t}^{\Delta,\lambda_*,\pi^{*\lambda_*}}(\overrightarrow{\omega^{\bullet},\omega^*})$ is given in \eqref{E:lambdaMetric_pi}.
\end{definition}
Note that distance in  \eqref{E:OSPAMT_t} is not a metric.

The OSPAMT metric can be broken into OSPAMT cardinality error and localization error as described in the following definition. OSPAMT localization error also describes the closeness between estimated tracks and their truth tracks in the matched associations.
\begin{definition}[OSPAMT Card and Loc error]Given the order parameter $p$ ($1\le p<\infty$) and the cutoff parameter $c$. Let $0<\Delta<c$. The OSPAMT cardinality error and the OSPAMT localization error between $\omega$ and $\omega'$ are
\begin{align}
\textstyle
d_{c,p}^{\Delta,Card}(\omega,\omega')&=\textstyle
\bigg[\frac{1}{\mathbf{n}}\sum_{t=1}^Ts^{Card,\lambda_*}_{c,p,t,\Delta}(\omega^*)\bigg]^{\frac{1}{p}}
\label{E:CardDis}\\
\textstyle
d_{c,p}^{\Delta,Loc}(\omega,\omega')&=\textstyle
\left[\frac{1}{\mathbf{n}}\sum_{t=1}^T\sum_{i=1}^{|\omega^*|}
d_{c,p,t}^{\Delta,\lambda_*,\pi^{*\lambda_*}}\!\!(\tau^{\omega^*}_i\!\!,\omega^\bullet)\right]^{\frac{1}{p}}
\label{E:LocDis}
\end{align}
respectively where $d_{c,p,t}^{\Delta,\lambda_*,\pi^{*\lambda_*}}\!\!(\tau^{\omega^*}_i,\omega^\bullet)$ is given in \eqref{E:lambdaMetric_pi1} and $s^{Card,\lambda_*}_{c,p,t,\Delta}(\omega^*)$ is given in \eqref{E:lambdaMetric_pi2}.
\end{definition}
OSPAMT cardinality error can be calculated at a particular time index. Similarly, the localization error can be calculated at a particular time index. These errors are easy to plot for comparison at each time index.
\begin{definition}[OSPAMT Card and Loc error at time $t$]
The OSPAMT cardinality error and the OSPAMT localization error between $\omega$ and $\omega'$ at time $t$ are
\begin{align}
\textstyle d_{c,p,t}^{\Delta,Card}(\omega,\omega')&=\textstyle
\left[\frac{1}{\mathbf{n}_t}s^{Card,\lambda_*}_{c,p,t,\Delta}(\omega^*)\right]^{\frac{1}{p}}
\label{E:CardDist}\\
\textstyle d_{c,p,t}^{\Delta,Loc}(\omega,\omega')&=\textstyle
\left[\frac{1}{\mathbf{n}_t}\sum_{i=1}^{|\omega^*|}
d_{c,p,t}^{\Delta,\lambda_*,\pi^{*\lambda_*}}\!\!(\tau^{\omega^*}_i,\omega^\bullet)\right]^{\frac{1}{p}}
\label{E:LocDist}
\end{align}
respectively.
\end{definition}

Note that the localization error in \eqref{E:LocDist} differs from the one in the OSPA metric in the sense that it considers many-to-one assignments. Thus the localization error in \eqref{E:LocDist} may be either larger than the one in the OSPA metric because of the term $\Delta^p$ in \eqref{E:lambdaMetric_pi1}; or smaller than the one in the OSPA metric because of \eqref{E:lambdaMetric_pi1}, $d_{c,p,t}^{\Delta,\lambda_*,\pi^{*\lambda_*}}\!\!(\tau^{\omega^*}_i,\omega^\bullet)=0$ if at time $t$ target $i$ is not assigned to any target in $\omega^\bullet$ even if the distance between target $i$ and a target $j$ in $\omega^\bullet$ is the smallest. This is because the cardinality error will be calculated in \eqref{E:lambdaMetric_pi2} for this case at those time indices. By \eqref{E:lambdaMetric_pi2}, the cardinality error \eqref{E:CardDist} is always larger than or equal to the cardinality error in the OSPA metric (See Figure \ref{Simul2_CardLoc} for explanation).

We now calculate the OSPAMT metric for the scenario in Figure \eqref{Fig:ExamMotivateOSPMAT0}.

\vspace{0.5cm}
\underline{\textbf{Example 2:}} Consider  Example $1$ for Figure \eqref{Fig:ExamMotivateOSPMAT0}. Choose $\epsilon\ll c$. We have four different possibilities which contribute to the calculation of the OSPAMT distance between $\omega$ and $\omega'$ such as  $\lambda,\lambda',\lambda^*\in\mathcal{M}(L^{\omega'},L^{\omega}_0)$ where $\lambda(1)=\lambda(2)=1$, $\pi^{\lambda}(1)=(1,2)$,
$\gamma^{\lambda}(1)=(2,1)$, $\lambda'(2)=1,\lambda'(1)=0$ and $\lambda^*(1)=1,\lambda^*(2)=0$ where $\pi^{\lambda}_1$ and $\gamma^{\lambda}_1$ are different order of track indices in $\omega'$ assigned to track $1$ in $\omega$. By \eqref{E:lambdaMetric}, we have 
$\tilde{d}_{c,p}^{\Delta,\lambda}(\overrightarrow{\omega',\omega})=\min\{A_1,A_2\}$,
$\tilde{d}_{c,p}^{\Delta,\lambda'}\!(\overrightarrow{\omega',\omega})=A_3$, $
\tilde{d}_{c,p}^{\Delta,\lambda^*}\!(\overrightarrow{\omega',\omega})=A_4$ where
\begin{align*}
A_1&=\!\!\sqrt[p]{(5\epsilon^p+2\Delta^p)/5},\:A_3=\!\!\sqrt[p]{(2\epsilon^p+3c^p)/5},\\
A_2&=\!\!\sqrt[p]{(5\epsilon^p+3\Delta^p)/5},\:
A_4=\!\!\sqrt[p]{(3\epsilon^p+2c^p)/5}.
\end{align*}

It is clear that $A_4<A_3;$ $A_1<A_2$. Hence  $\tilde{d}_{c,p}^{\Delta,\lambda}(\overrightarrow{\omega',\omega})=A_1$. Thus
$\lambda^*$ is the OSPAMT assignment and $d_{c,p}^{\Delta}(\omega,\omega')=A_4$ if
$c^p\le\epsilon^p+\Delta^p$; else $\lambda$ is the OSPAMT assignment and $d_{c,p}^{\Delta}(\omega,\omega')=A_1$ if $\epsilon^p+\Delta^p\le c^p$.

From this example, we see that different choices of $\Delta$ result in different OSPAMT assignments. The following section will discuss the choice of $c$ and $\Delta$.
\subsection{Choice of $c$ and $\Delta$}
In reality, the choice of $c$ and $\Delta$ may result in different OSPAMT assignment (see Figure \eqref{Fig:ExamMotivateOSPMAT0})
 and the discussion in Example $2$) and also depends on the problem. Thus checking the OSPAMT assignment is advisable to ensure that the OSPAMT assignment accords with the optimal assignment. This situation is similar to other metrics.
\section{Evaluation of OSPAMT metric}\label{S:Eval_Metrics}
In this section, the properties of the OSPAMT metric will be evaluated on several examples. We also discuss the OSPA (summarized in Appendix \ref{SS:Summary_OSPA}) and OSPAT metric (summarized in Appendix \ref{SS:Summary_OSPAT}) on the same examples.
\subsection{Simplistic interpretation of OSPA, OSPAT and OSPAMT metrics between two finite sets of tracks}\label{SS:Dis_metrics}
 \begin{itemize}
   \item The OSPA metric is the minimum sum of distance between two tracks across two finite sets of tracks at each time, adding the penalty $c$ for each extra track at each time where each distance between tracks is penalized with the same $c$ if this distance is larger than $c$.
   \item The OSPAT metric is the OSPA metric at each time with the requirement that the tracks are augmented with  labels and an added penalty $\alpha$ for each distance between two labeled tracks with the different labels.
   \item The OSPAMT metric is the OSPA metric with  additions that penalizes changes in assignments of tracks to truth over time without having to add any label distance. As the OSPAMT allows multiple tracks in one set to be assigned to a track in another set, the penalty applies to assigned tracks except the closest assigned track at each time.
 \end{itemize}
\subsection{Analysis of OSPAMT metric via scenarios}\label{SS:AnalysisMetrics}
Let $p=p'$ and $0<\Delta=\alpha<c$ where $\alpha$ is in \eqref{E:dis-label}. From now on,  $[\tau_1\leftrightarrow(\tau'_1,\tau'_2)]$ means that tracks $\tau'_1$ and $\tau'_2$ in $\omega'$ are assigned to track $\tau_1$. Similarly, $[(\tau_1,\tau_2)\leftrightarrow\emptyset]$ means that $\tau_1$ and $\tau_2$ are assigned to $0$ or are not assigned to any track in $\omega'$.
\subsubsection{Reliability of OSPAMT metric}
\begin{figure}[htbp!]
\hspace{-.1cm}%
\subfloat[$\omega=\{\tau_1,\tau_2\}$, $\omega'=\{\tau'_1,\tau'_2\}$.
]
{\label{Fig:Inc_Sol1}
\centering{\hspace{-1.6cm}
\scalebox{0.8}
{
\pgfdeclarelayer{background} \pgfdeclarelayer{foreground}
\pgfsetlayers{background,main,foreground}
\par
\par
\tikzset{cross/.style={cross out, draw=black, minimum size=2*(#1-\pgflinewidth), inner sep=0pt, outer sep=0pt},
cross/.default={3pt}}
\begin{tikzpicture}[y=.2cm, x=2cm,font=\sffamily,bend angle=50]
 \draw [->,thick]
 |-(3.12,0) node (xaxis) [right]
  {\scriptsize Time};
\draw (1,0) -- coordinate (x axis mid) (3.1,0);
\foreach \x in {1,...,3}
     		\draw (\x,1pt) -- (\x,-3pt)
			node[anchor=north] {\x};
\draw (1,1.5) node[cross] (x1){};
\draw (2,1.5) node[cross] (x2){};
\draw (3,1.5) node[cross] (x3){};
\draw (1,15) node[cross] (2x1){};
\draw (2,15) node[cross] (2x2){};
\draw (3,15) node[cross] (2x3){};
\node[circle,draw=black, fill=black, inner sep=0pt,minimum size=3pt] (y1) at (1,5.5) {};
\node[circle,draw=black, fill=black, inner sep=0pt,minimum size=3pt] (y2) at (2,5.5){};
\node[circle,draw=black, fill=black, inner sep=0pt,minimum size=3pt] (y3) at (3,5.5){};

\node[circle,draw=black, fill=black, inner sep=0pt,minimum size=3pt] (2y1) at (1,11) {};
\node[circle,draw=black, fill=black, inner sep=0pt,minimum size=3pt] (2y2) at (2,11){};
\node[circle,draw=black, fill=black, inner sep=0pt,minimum size=3pt] (2y3) at (3,11){};
\draw[-,thin,black,dashed](x2)edge node[swap,near start]{}  (y2);
\draw[->,black](2y1)edge (y2)(y2)edge (y3)(y1)edge (2y2)(2y2)edge (2y3)
(x1)edge (x2)(x2)edge (x3)(2x1)edge (2x2)(2x2)edge (2x3)
;
\node(temdis1)  [below =.2cm of y1] {};
 \node (dis1) [left =-.5cm of temdis1]{$\epsilon$};
 \draw[-,thin,black,dashed](y1)edge node[swap,near start]{}
  (x1);
\node(temdis2) [below =.2cm of y2] {};
  \node (dis1) [left =-1.2cm of temdis2]{$\epsilon\ll 1$};
  \draw[-,thin,black,dashed](y2)edge node[swap,near start]{}
  (x2);
\node(temdis3) [below =.2cm of y3] {};
  \node (dis1) [right =-.5cm of temdis3]{$\epsilon$};
  \draw[-,thin,black,dashed](y3)edge node[midway]{}
  (x3);
\node(temdis4) [below =.2cm of 2x1] {};
  \node (dis1) [left =-.5cm of temdis4]{$\epsilon$};
  \draw[-,thin,black,dashed](2x1)edge node[near start]{}
  (2y1);
\node(temdis5) [below =.2cm of 2x2] {};
  \node (dis1) [right =-.5cm of temdis5]{$\epsilon$};
  \draw[-,thin,black,dashed](2x2)edge node[near start]{}
  (2y2);
\node(temdis6) [below =.2cm of 2x3] {};
  \node (dis1) [right =-.5cm of temdis6]{$\epsilon$};
  \draw[-,thin,black,dashed](2x3)edge node[near start]{}
  (2y3);
\node(temdis7) [below =.7cm of 2x1] {};
  \node (dis1) [right =.3cm of temdis7]{$\beta$};
  \draw[-,thin,black,dashed](2x1)edge[bend left] node[near start]{}
  (y1);
\node(temdis8) [below =.7cm of 2y1] {};
  \node (dis1) [right =.3cm of temdis8]{$\beta$};
  \draw[-,thin,black,dashed](2y1)edge[bend left] node[near start]{}
  (x1);    
 \begin{pgfonlayer}{background}
 \path (2y3.east |-2y3.east)+(.3,-.6) node(ii){$\tau'_2$};
  \path (y3.east |-y3.east)+(.3,-.6) node(aa){$\tau'_1$};
  \path (x3.east|- x3.east)+(.3,.4) node(b){$\tau_1$};
  \path (2x3.east|- 2x3.east)+(.3,.4) node(a){$\tau_2$};
\draw[->,black,dashed](b) edge (x3)(a) edge (2x3)(aa)edge(y3)(ii) edge (2y3);
\end{pgfonlayer}
\end{tikzpicture}
}}
}
\hspace{-0.5cm}
\subfloat[$\omega=\{\tau_1,\tau_2\}$, $\omega^{\prime\prime}=\{\tau'_1,\tau'_2\}$.
]
{\label{Fig:Inc_Sol2}
\centering{\hspace{-1.8cm}
\scalebox{.8}
{\centering
\pgfdeclarelayer{background} \pgfdeclarelayer{foreground}
\pgfsetlayers{background,main,foreground}
\par
\par
\tikzset{cross/.style={cross out, draw=black, minimum size=2*(#1-\pgflinewidth), inner sep=0pt, outer sep=0pt},
cross/.default={3pt}}
\begin{tikzpicture}[y=.2cm, x=2cm,font=\sffamily]
 \draw [->,thick]
 |-(3.12,0) node (xaxis) [right]
  {\scriptsize Time};
\draw (1,0) -- coordinate (x axis mid) (3.1,0);
\foreach \x in {1,...,3}
     		\draw (\x,1pt) -- (\x,-3pt)
			node[anchor=north] {\x};
\draw (1,1.5) node[cross] (x1){};
\draw (2,1.5) node[cross] (x2){};
\draw (3,1.5) node[cross] (x3){};
\draw (1,15) node[cross] (2x1){};
\draw (2,15) node[cross] (2x2){};
\draw (3,15) node[cross] (2x3){};
\node[circle,draw=black, fill=black, inner sep=0pt,minimum size=3pt] (y1) at (1,7) {};
\node[circle,draw=black, fill=black, inner sep=0pt,minimum size=3pt] (y2) at (2,5.5){};
\node[circle,draw=black, fill=black, inner sep=0pt,minimum size=3pt] (y3) at (3,5.5){};

\node[circle,draw=black, fill=black, inner sep=0pt,minimum size=3pt] (2y1) at (1,9.5) {};
\node[circle,draw=black, fill=black, inner sep=0pt,minimum size=3pt] (2y2) at (2,11){};
\node[circle,draw=black, fill=black, inner sep=0pt,minimum size=3pt] (2y3) at (3,11){};
\draw[-,thin,black,dashed](x2)edge node[swap,near start]{}  (y2);
\draw[->,black](y1)edge (y2)(y2)edge (y3)(2y1)edge (2y2)(2y2)edge (2y3)
(x1)edge (x2)(x2)edge (x3)(2x1)edge (2x2)(2x2)edge (2x3)
;
\node(temdis1)  [below =.2cm of y1] {};
 \node (dis1) [left =-.5cm of temdis1]{$\eta$};
 \draw[-,thin,black,dashed](y1)edge node[swap,near start]{}
  (x1);
\node(temdis2) [below =.2cm of y2] {};
  \node (dis1) [left =-1.2cm of temdis2]{$\epsilon\ll 1$};
  \draw[-,thin,black,dashed](y2)edge node[swap,near start]{}
  (x2);
\node(temdis3) [below =.2cm of y3] {};
  \node (dis1) [right =-.5cm of temdis3]{$\epsilon$};
  \draw[-,thin,black,dashed](y3)edge node[midway]{}
  (x3);
\node(temdis4) [below =.2cm of 2x1] {};
  \node (dis1) [left =-.5cm of temdis4]{$\eta$};
  \draw[-,thin,black,dashed](2x1)edge node[near start]{}
  (2y1);
\node(temdis5) [below =.2cm of 2x2] {};
  \node (dis1) [right =-.5cm of temdis5]{$\epsilon$};
  \draw[-,thin,black,dashed](2x2)edge node[near start]{}
  (2y2);
\node(temdis6) [below =.2cm of 2x3] {};
  \node (dis1) [right =-.5cm of temdis6]{$\epsilon$};
  \draw[-,thin,black,dashed](2x3)edge node[near start]{}
  (2y3);
 \begin{pgfonlayer}{background}
 \path (2y3.east |-2y3.east)+(.3,-.6) node(ii){$\tau'_2$};
  \path (y3.east |-y3.east)+(.3,-.6) node(aa){$\tau'_1$};
  \path (x3.east|- x3.east)+(.3,.4) node(b){$\tau_1$};
  \path (2x3.east|- 2x3.east)+(.3,.4) node(a){$\tau_2$};
\draw[->,black,dashed](b) edge (x3)(a) edge (2x3)(aa)edge(y3)(ii) edge (2y3);
\end{pgfonlayer}
\end{tikzpicture}
}}}
\caption{Let 
$\epsilon<\eta<c\ll\beta$. Clearly, $\omega^{\prime\prime}$ is a better estimate of $\omega$ than $\omega'$. This is also the conclusion of OSPAMT distance since $d_{c,p}^{\Delta}(\omega,\omega^{\prime\prime})=\sqrt[p]{(2\epsilon^p+\eta^p)/3}<\sqrt[p]{(2\epsilon^p+c^p)/3}=d_{c,p}^{\Delta}(\omega,\omega')$. }
\label{Fig:Inc_Sol}
\end{figure}
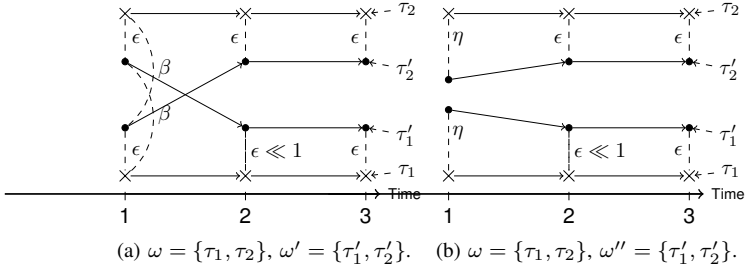

Consider the scenarios in Figure \ref{Fig:Inc_Sol}
where $\epsilon$ and $\eta$ ($\epsilon<\eta\ll \beta$) are the distances between the indicated target states. Without loss of generality, let $\epsilon,\eta<c$. Obviously, the algorithm that produced $\omega^{\prime\prime}$ (i.e. the scenario in Figure \eqref{Fig:Inc_Sol2}) is better than the algorithm that produced $\omega'$ (i.e. the scenario in Figure \eqref{Fig:Inc_Sol1}). Although the OSPAT and OSPAMT metrics have the same optimal assignment, only the OSPAMT metric comes to this conclusion, while the OSPAT metric comes to various conclusions depending on the value of $\alpha$. That is, the algorithm that produced $\omega^{\prime\prime}$ is better than the algorithm that produced $\omega'$ if $\sqrt[p]{\alpha^{p}+\epsilon^{p}}\le\eta$ otherwise it is worse than the algorithm that  produced $\omega'$ (see Table \ref{table:FalseSelGdAlg}). The OSPA metric comes to a different conclusion, that is, the algorithm that produced $\omega'$ is always better than the algorithm that produced $\omega^{\prime\prime}$ (see Table \ref{table:FalseSelGdAlg}).

Note that the OSPA metric is generally not applicable for the comparison between two finite sets of tracks because it measures the distance between two finite sets of states at each time regardless of which track a state belongs to.

\renewcommand{\arraystretch}{1.25}
\begin{center}
\begin{table}[htpb!]%
\caption{\textbf{Distances and Optimal Assignment at $t=1$ of Figure \ref{Fig:Inc_Sol}}}
\begin{center}
\begin{tabular}
{|l|l|l|l|}
\hline
\!Metric\!&\!Figure \eqref{Fig:Inc_Sol1}&\!Figure \eqref{Fig:Inc_Sol2}\!&\!Assignment of Figure \ref{Fig:Inc_Sol}\!\\ \hline
\!OSPA &\!$\epsilon$ & \!$\eta$ &\!$[\tau_1\leftrightarrow\tau'_2;\tau_2\leftrightarrow\tau'_1]$\\ \hline
\!OSPAT&\!$\min\{\sqrt[p]{\alpha^{p}+\epsilon^{p}},c\}$\!& \!$\eta$&\!$[\tau_1\leftrightarrow\tau'_2;\tau_2\leftrightarrow\tau'_1]$\\\hline
\!OSPAMT\!&\!$c$ &\!$\eta$&\!$[\tau_1\leftrightarrow\tau'_1;\tau_2\leftrightarrow\tau'_2]$\\\hline
\end{tabular}
\label{table:FalseSelGdAlg}
\end{center}
\end{table}
\end{center}
\renewcommand{\arraystretch}{1}

The metric producing the smallest distance may not be the best metric. Indeed, by Table \ref{table:FalseSelGdAlg} for the scenario in Figure \eqref{Fig:Inc_Sol1}, the OSPA metric has the smallest distance when compared to the other two metrics but it does not evaluate the scenario in Figures \eqref{Fig:Inc_Sol1} and \eqref{Fig:Inc_Sol2} correctly because it does not consider which state belongs to which track. On the other hand, the metric producing the largest distance may not be the worst metric. Indeed, by Table \ref{table:FalseSelGdAlg} for the scenario in Figure \eqref{Fig:Inc_Sol1}, the OSPAMT metric at time $1$ has the largest distance compared to the other two metrics but its OSPAMT metric interprets the two scenarios in Figure \ref{Fig:Inc_Sol} correctly as discussed earlier.
\subsubsection{Coherence of OSPAMT assignment}\label{SS:Co_OSPAMT_Ass}
The OSPAMT assignment contributes to the computation of the OSPAMT metric, whereas the OSPAT assignment plays no role in the computation of the OSPAT metric by \eqref{E:lambdaMetric} and \eqref{E:dis_OSPATrks}. By Table \ref{table:FalseSelGdAlg}, the OSPAMT assignment is also the OSPAMT assignment at any time $t=1,2,3$ while the OSPAT assignment is not the OSPAT assignment at $t=1$ for computation of the OSPAT metric at $t=1$. This can also be seen in Table \ref{table:FalseOptAss} for the scenario in Figure \eqref{Fig:ExamMotivateOSPMAT0} where at time $1$ and $5$ the OSPAT assignments for computation of the OSPAT metrics differ.

\renewcommand{\arraystretch}{1.25}
\begin{center}
\begin{table}[htpb!]%
\caption{\textbf{Optimal Assignment at $t=1,5$ of Figure \eqref{Fig:ExamMotivateOSPMAT0}}}
\begin{center}
\begin{tabular}
{|l|l|l|l|}
\hline
\!\!Metric\!\!&\!\!$t=1$\!\!&\!\!$t=5$\!\!\\ \hline
\!\!OSPA\!\!&\!\!$[\tau_1\leftrightarrow\tau'_1]$\!\!&\!\!$[\tau_1\leftrightarrow\tau'_2]$\!\!\\ \hline
\!\!OSPAT\!\!&\!\!$[\tau_1\leftrightarrow\tau'_1]$\!\! &\!\!$[\tau_1\leftrightarrow\tau'_2]$\!\!\\\hline
\!\!OSPAMT\!\!&\!\!$[\tau_1\leftrightarrow(\tau'_1,\tau'_2)]$\!\! &\!\!$[\tau_1\leftrightarrow(\tau'_1,\tau'_2)]$\!\!\\\hline
\end{tabular}
\label{table:FalseOptAss}
\end{center}
\end{table}
\end{center}
\renewcommand{\arraystretch}{1}

Consideration of many-to-one assignments helps the OSPAMT metric to obtain the optimal assignment.
Indeed, take the scenario in Figure \eqref{Fig:ExamMotivateOSPMAT0} for example,
the OSPAMT assignments are the optimal assignments while the OSPAT metric which only considers one-to-one assignment gives the suboptimal assignments when compared to the optimal assignments. Furthermore, allowing more than one track assigned to a track at any time index also allows the OSPAMT assignment to achieve the optimal assignment such as shown in the scenario in Figure \eqref{Fig:ExamMotivateOSPMAT0}. If we do not allow more than one track to be assigned to a track at any time, $[\tau'_1\leftrightarrow\tau_1,\tau'_2\leftrightarrow\emptyset]$ is
 not the optimal assignment for this scenario and is the OSPAT assignment. For the OSPAT metric, the computation of the one-to-one assignment leads to certain problems resulting in an incorrect optimal assignment (compared to the optimal assignment). It can be seen in scenarios in Figure \ref{Fig:ExamMotivateOSPMAT} (see Appendix \ref{A:Iss_Opt_Ass} for more information about the properties of the OSPAT assignment). Indeed, the OSPAT assignments for the scenario in Figures \eqref{Fig:ExamMotivateOSPMAT0}, \eqref{Fig:ExamMotivateOSPMAT1}, \eqref{Fig:ExamMotivateOSPMAT2} and \eqref{Fig:ExamMotivateOSPMAT3} are $[\tau'_1\leftrightarrow\tau_1,\tau'_2\leftrightarrow\emptyset]$, $[\tau'_1\leftrightarrow\tau_1,\tau'_2\leftrightarrow\tau_2]$, $[\tau_1\leftrightarrow\tau'_1]$
 and $[\tau_1\leftrightarrow\tau'_1,\tau'_2\leftrightarrow\tau_2]$ respectively. However, these OSPAT assignments are not the optimal assignments for the corresponding scenarios.

Allowing track indices assigned to $0$ also helps to achieve the optimal assignment. Take the scenarios in Figures \eqref{Fig:ExamMotivateOSPMAT2} and \eqref{Fig:ExamMotivateOSPMAT3} as examples, the OSPAMT assignment which considers track-to-0 assignment represents the optimal assignment ($[\tau_1\leftrightarrow\emptyset,\tau'_1\leftrightarrow\emptyset]$
 and $[\tau_1\leftrightarrow\tau'_1,\tau'_2\leftrightarrow\emptyset]$ are the OSPAMT for the scenarios in Figures \eqref{Fig:ExamMotivateOSPMAT2} and \ref{Fig:ExamMotivateOSPMAT} respectively because of $\lambda\in\mathcal{M}(L^{\omega'},L^{\omega}_0)$ such that $\lambda(1)=0$ and $\lambda(2)=0$ for the scenarios in Figures \eqref{Fig:ExamMotivateOSPMAT2} and \eqref{Fig:ExamMotivateOSPMAT} respectively). While the OSPAT assignment which does not consider track-to-0 assignment is $[\tau'_1\leftrightarrow\tau_1]$ the scenario in Figure \eqref{Fig:ExamMotivateOSPMAT2} (i.e. track $\tau'_1$ is the estimate of the truth track $\tau_1$) but not the optimal assignment.

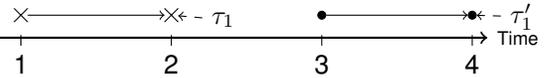
\begin{figure}[htbp!]
\centering\hspace{-1.5cm}\scalebox{1}{\hspace{-0.5cm}
\pgfdeclarelayer{background} \pgfdeclarelayer{foreground}
\pgfsetlayers{background,main,foreground}
\par
\par
\tikzset{cross/.style={cross out, draw=black, minimum size=2*(#1-\pgflinewidth), inner sep=0pt, outer sep=0pt},
cross/.default={3pt}}
\begin{tikzpicture}[y=.2cm, x=2cm,font=\sffamily,bend angle=45]
 \draw [->,thick]
 |-(4.1,0) node (xaxis) [right]
  {\scriptsize Time};
\draw (1,0) -- coordinate (x axis mid) (4.1,0);
\foreach \x in {1,...,4}
     		\draw (\x,1pt) -- (\x,-3pt)
			node[anchor=north] {\x};
\draw (1,1.5) node[cross] (x1){};
\draw (2,1.5) node[cross] (x2){};
\node[circle,draw=black, fill=black, inner sep=0pt,minimum size=3pt] (y1) at (3,1.5) {};
\node[circle,draw=black, fill=black, inner sep=0pt,minimum size=3pt] (y2) at (4,1.5){};
\draw[->,black](y1)edge (y2)
(x1)edge (x2);
 \begin{pgfonlayer}{background}
  \path (y2.east |-y2.east)+(.3,-.1) node(aa){$\tau'_1$};
  \path (x2.east|- x2.east)+(.3,-.3) node(a){$\tau_1$};
\draw[->,black,dashed](a) edge (x2)(aa)edge(y2);
\end{pgfonlayer}
\end{tikzpicture}
}
\caption{$\omega=\{\tau_1\},\omega^{\prime}=\{\tau'_1\}$ where $\tau_1$ exists from time $1$ to $2$ while $\tau'$ exists from time $3$ to $4$. Intuitively, $\tau'_1$ is false track in $\omega'$ and hence there is a missed track in $\omega'$. 
The OSPAT assignment is $[\tau_1\leftrightarrow\tau'_1]$. However, the OSPAMT assignment $[\tau_1\leftrightarrow\emptyset;\emptyset\leftrightarrow\tau'_1]$ is also the optimal assignment.}
\label{F:1_1Sepa}
\end{figure}

The property that two tracks from two finite sets of tracks cannot be assigned to each other without any common existing time helps the OSPAMT assignment to obtain the optimal assignment (see Figure \ref{F:1_1Sepa}). However, the OSPAT assignment always assigns all tracks in a set to all tracks in the other set if the two sets have the same number of tracks regardless of the existence of missing or false tracks (see Figure \ref{F:1_1Sepa}).

The optimal assignment represents the physical matching between truth tracks and tracks generated by an MTT algorithm, false tracks generated by false alarm; and missing tracks caused by low probability of detection. This is the case for OSPAMT assignment but may not be the case for the OSPAT assignment. The OSPAMT assignments shown in scenarios in Figure \ref{Fig:ExamMotivateOSPMAT} are the optimal assignments as discussed above. Indeed, tracks assigned to a truth track $i$ are estimates of the truth track $i$, tracks assigned to $0$ are false tracks, and truth tracks not assigned to tracks are missed tracks. However, the OSPAT assignment is not the optimal assignment for Figures \eqref{Fig:ExamMotivateOSPMAT0}, \eqref{Fig:ExamMotivateOSPMAT1}, and \eqref{Fig:ExamMotivateOSPMAT2}. Specifically, for the Figure \eqref{Fig:ExamMotivateOSPMAT0}, intuitively the track $\tau'_1$ is the estimate of truth track $\tau_1$ along with track $\tau'_2$ but it is a false track by the OSPAT assignment. For Figure \eqref{Fig:ExamMotivateOSPMAT1}, intuitively track $\tau'_1$ is the estimate of truth track $\tau_1$ along with track $\tau'_2$; and tracks $\tau_2$ is missed track. However, by the OSPAT assignment $\tau'_1$ is the estimate of the truth track $\tau_1$; track $\tau_2$ is the missed track. For the Figure \eqref{Fig:ExamMotivateOSPMAT2}, intuitively track $\tau'_1$ is a false track and $\tau_1$ is a missed track. However, track $\tau'_1$ is the estimate of $\tau_1$ by the OSPAT assignment.

\subsubsection{Unreliability of the OSPAMT metric at time $t$}\label{SS:UnreOSPAMT_t}

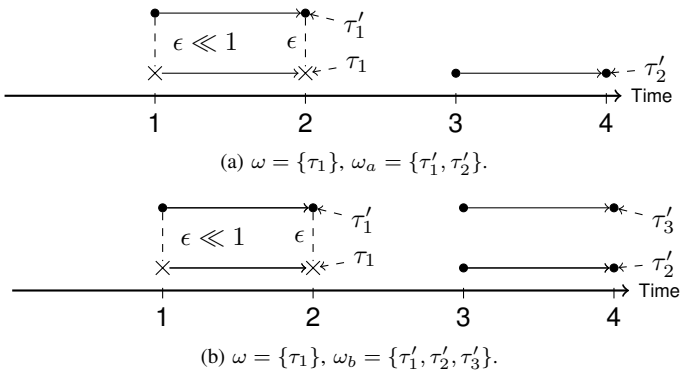
\begin{figure}[htbp!]
\centering\hspace{-1cm}
\subfloat[$\omega=\{\tau_1\}$, $\omega_a=\{\tau'_1,\tau'_2\}$.]
{\label{Fig:CompAtTime_All_1}
\centering{\hspace{-0.8cm}
\scalebox{1}
{\centering
\pgfdeclarelayer{background} \pgfdeclarelayer{foreground}
\pgfsetlayers{background,main,foreground}
\par
\par
\tikzset{cross/.style={cross out, draw=black, minimum size=2*(#1-\pgflinewidth), inner sep=0pt, outer sep=0pt},
cross/.default={3pt}}
\begin{tikzpicture}[y=.2cm, x=2cm,font=\sffamily,bend angle=45]
 \draw [->,thick]
 |-(4.1,0) node (xaxis) [right]
  {\scriptsize Time};
\draw (1,0) -- coordinate (x axis mid) (4.1,0);
\foreach \x in {1,...,4}
     		\draw (\x,1pt) -- (\x,-3pt)
			node[anchor=north] {\x};
\draw (1,1.5) node[cross] (x1){};
\draw (2,1.5) node[cross] (x2){};
\node[circle,draw=black, fill=black, inner sep=0pt,minimum size=3pt] (y1) at (1,5.5) {};
\node[circle,draw=black, fill=black, inner sep=0pt,minimum size=3pt] (y2) at (2,5.5){};
\node[circle,draw=black, fill=black, inner sep=0pt,minimum size=3pt] (2y3) at (3,1.5){};
\node[circle,draw=black, fill=black, inner sep=0pt,minimum size=3pt] (2y4) at (4,1.5){};
\draw[->,black](y1)edge (y2)(2y3)edge (2y4)
(x1)edge (x2);
\node(temdis1)  [below =.2cm of y1] {};
 \node (dis1) [right =-0.01cm of temdis1]{$\epsilon\ll 1$};
 \draw[-,thin,black,dashed](y1)edge node[swap,near start]{}
  (x1);
\node(temdis2) [below =.2cm of y2] {};
  \node (dis1) [right =-.5cm of temdis2]{$\epsilon$};
  \draw[-,thin,black,dashed](y2)edge node[swap,near start]{}
  (x2);

 \begin{pgfonlayer}{background}
 \path (2y4.east |-2y4.east)+(.3,.3) node(bb){$\tau'_2$};
 \path (y2.east |-y2.east)+(.3,-.8) node(aa){$\tau'_1$};
  \path (x2.east|- x2.east)+(.3,.6) node(a){$\tau_1$};
\draw[->,black,dashed](a) edge (x2)(aa)edge(y2)(bb) edge (2y4);
\end{pgfonlayer}
\end{tikzpicture}
}}
}\\
\vspace{0.3cm}
\centering\hspace{-1cm}
\subfloat[$\omega=\{\tau_1\}$, $\omega_b=\{\tau'_1,\tau'_2,\tau'_3\}$.
]
{\label{Fig:CompAtTime_All_2}
\centering{\hspace{-0.3cm}
\scalebox{1}
{\hspace{-.3cm}\centering
\pgfdeclarelayer{background} \pgfdeclarelayer{foreground}
\pgfsetlayers{background,main,foreground}
\par
\par
\tikzset{cross/.style={cross out, draw=black, minimum size=2*(#1-\pgflinewidth), inner sep=0pt, outer sep=0pt},
cross/.default={3pt}}
\begin{tikzpicture}[y=.2cm, x=2cm,font=\sffamily,bend angle=45]
 \draw [->,thick]
 |-(4.1,0) node (xaxis) [right]
  {\scriptsize Time};
\draw (1,0) -- coordinate (x axis mid) (4.1,0);
\foreach \x in {1,...,4}
     		\draw (\x,1pt) -- (\x,-3pt)
			node[anchor=north] {\x};
\draw (1,1.5) node[cross] (x1){};
\draw (2,1.5) node[cross] (x2){};
\node[circle,draw=black, fill=black, inner sep=0pt,minimum size=3pt] (y1) at (1,5.5) {};
\node[circle,draw=black, fill=black, inner sep=0pt,minimum size=3pt] (y2) at (2,5.5){};
\node[circle,draw=black, fill=black, inner sep=0pt,minimum size=3pt] (2y3) at (3,1.5){};
\node[circle,draw=black, fill=black, inner sep=0pt,minimum size=3pt] (2y4) at (4,1.5){};
\draw[->,black](y1)edge (y2)(2y3)edge (2y4)
(x1)edge (x2);
\node(temdis1)  [below =.2cm of y1] {};
 \node (dis1) [right =-0.01cm of temdis1]{$\epsilon\ll 1$};
 \draw[-,thin,black,dashed](y1)edge node[swap,near start]{}
  (x1);
\node(temdis2) [below =.2cm of y2] {};
  \node (dis1) [right =-.5cm of temdis2]{$\epsilon$};
  \draw[-,thin,black,dashed](y2)edge node[swap,near start]{}
  (x2);
\node[circle,draw=black, fill=black, inner sep=0pt,minimum size=3pt] (3y3) at (3,5.5){};
\node[circle,draw=black, fill=black, inner sep=0pt,minimum size=3pt] (3y4) at (4,5.5){};  
\draw[->,black](y1)edge (y2)(2y3)edge (2y4)
(x1)edge (x2)(3y3)edge(3y4);

 \begin{pgfonlayer}{background}
 \path (3y4.east |-3y4.east)+(.3,-.8) node(cc){$\tau'_3$};
 \path (2y4.east |-2y4.east)+(.3,.3) node(bb){$\tau'_2$};
 \path (y2.east |-y2.east)+(.3,-.8) node(aa){$\tau'_1$};
 \path (x2.east|- x2.east)+(.3,.6) node(a){$\tau_1$};
\draw[->,black,dashed](a) edge (x2)(aa)edge(y2)(bb) edge (2y4)(cc) edge(3y4);
\end{pgfonlayer}
\end{tikzpicture}
}}}
\caption{
Intuitively,
$\omega_a$ is a better estimate of $\omega$ than $\omega_b$ because $\omega_b$ has more false tracks than $\omega_a$. This is also the conclusion of the OSPAMT metric since
$d_{c,p}^{\Delta}(\omega,\omega_a)=\sqrt[p]{(\epsilon^p+c^p)/2}<\sqrt[p]{(\epsilon^p+2c^p)/3}=d_{c,p}^{\Delta}(\omega,\omega_b)$ even the OSPAMT distances at time $t$ (for all $t=1,\ldots,4$) between $\omega$ and $\omega_a$ and between $\omega$ and $\omega_b$ are the same (i.e. $d_{c,p}^{\Delta,t}(\omega,\omega_a)=d_{c,p}^{\Delta,t}(\omega,\omega_b)$ for all $t=1,\ldots,4$).}
\label{Fig:CompAtTime_All}
\end{figure}

The OSPAMT metric on the space of finite sets of tracks reliably evaluates the distance between two finite sets of tracks. However, the OSPAMT distance at each time index between two finite sets of tracks does not evaluate the distance between two finite sets of tracks reliably. Similarly, the OSPAT metric does not also evaluate the distance between two finite sets of tracks reliably. This is because the OSPAMT distance at each time index and the OSPAT metric are calculated in a similar way as the OSPA metric.
Indeed, the scenario in Figure \eqref{Fig:CompAtTime_All_1} is intuitively better than the other scenario in Figure \eqref{Fig:CompAtTime_All_2} because $\omega_b$ in this scenario has more false tracks than the one in Figure \eqref{Fig:CompAtTime_All_1}.
\renewcommand{\arraystretch}{1.3}

\begin{center}
\begin{table}[htpb!]%
\caption{\textbf{Distances 
of 
Figure \ref{Fig:CompAtTime_All}}}
\begin{center}
\begin{tabular}
{
|m{2.8cm}|m{0.58cm}|m{.68cm}|m{.68cm}|m{.68cm}|m{.68cm}|@{}m{0pt}@{}}\cline{1-6}
Distance&Figure&$t=1$&$t=2$&$t=3$&$t=4$\\
\cline{1-6}
\multicolumn{1}{|m{2.9cm}|}{\multirow{2}{*}
{OSPAMT, OSPAT, OSPA}} &
\multicolumn{1}{m{0.5cm}|}{\eqref{Fig:CompAtTime_All_1}} & $\epsilon$&$\epsilon$ &$c$&$c$ \\
\cline{2-6}
\multicolumn{1}{|m{0.5cm}|}{}                        &
\multicolumn{1}{m{0.5cm}|}{\eqref{Fig:CompAtTime_All_2}} & $\epsilon$&$\epsilon$ &$c$&$c$ \\ 
\cline{1-6}
\end{tabular}
\label{table:SameTimeButDiffAll}
\end{center}
\end{table}
\end{center}
\renewcommand{\arraystretch}{1}
This is the case for the OSPAMT distance ($d_{c,p}^{\Delta}(\omega,\omega_a)=\sqrt[p]{(\epsilon^p+c^p)/2}<\sqrt[p]{(\epsilon^p+2c^p)/3}=d_{c,p}^{\Delta}(\omega,\omega_b)$) but not the case for OSPAMT distance at any time $t=1,\ldots,4$ by Table \ref{table:SameTimeButDiffAll}
(i.e. $d_{c,p}^{\Delta,t}(\omega,\omega_a)=d_{c,p}^{\Delta,t}(\omega,\omega_b)$ for $t=1,\ldots,4$). By Table \ref{table:SameTimeButDiffAll}, the OSPAT distance is also not reliable in evaluating the distance between two finite sets of tracks because $\boldsymbol{\mathcal{D}}^{\alpha,t}_{c,p}(\omega,\omega_a)=d_{c,p}^{\Delta,t}(\omega,\omega_b)$ for all $t=1,\ldots,4$.
\subsection{Experiments}\label{S:Experiment}
\begin{figure}[htbp!]
\centering
\includegraphics[width=\textwidth/3-.1cm]
{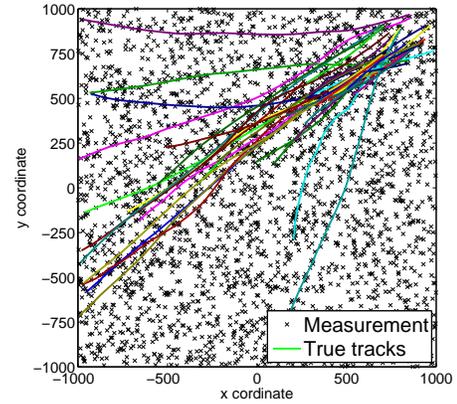}
\caption{Ground truth targets are immersed with their measurements and clutter.}
\label{FigGrdtruth}
\end{figure}

\begin{figure}[htbp!]
\centering
\includegraphics[width=\textwidth/2-0.5cm]
{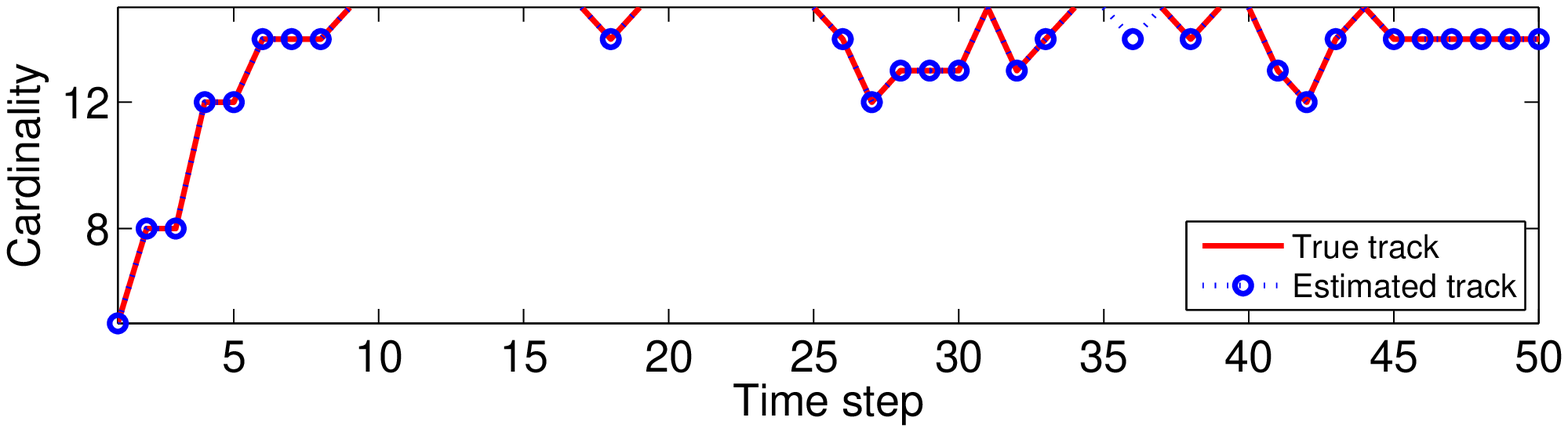}
\caption{Number of true targets and targets 
versus time for the scenario with $38$ targets.}
\label{FigScenario2}
\end{figure}

\begin{figure}[htbp!]
\centering
\includegraphics[width=\textwidth/2-0.5cm]
{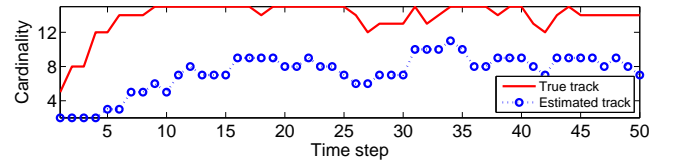}
\caption{Number of true targets and targets 
versus time for the scenario with $27$ targets.}
\label{FigScenario1}
\end{figure}

In order to illustrate application of the OSPAMT metric to evaluate an MTT algorithm, we use the same result and analysis which was published in \cite{Tuyet14OSPAMT}. In this section, the OSPAMT metric is demonstrated using the two scenarios of the multi-target tracking application in \cite{TuyetPMMHT14} where $38$ targets are plotted in Figure \ref{FigGrdtruth}. The first scenario considers the (accurately) estimated tracks ($38$ targets) of the ground truth tracks where the number of targets are estimated correctly apart from time $36$ (see Figure \ref{FigScenario2}) while the second scenario involves the poorly estimated tracks ($27$ targets) of the ground truth tracks where the number of targets are estimated poorly at any time index (see Figure \ref{FigScenario1}). In both examples, we compare the distances between the OSPA metric \cite{Schuhmacher2008}, the OSPAT metric \cite{ristic2011metric} and the proposed OSPAMT metric.

\begin{figure}[htbp!]
\centering
\includegraphics[width=\textwidth/2-0.3cm]
{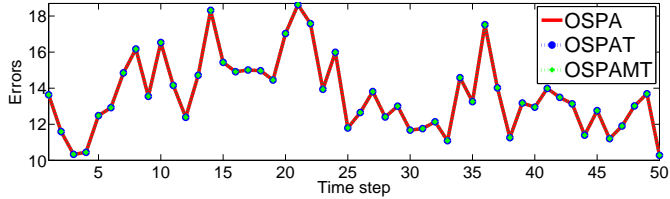}
\caption{Error versus time calculated under OSPA, OSPAT and OSPAMT metric for the scenario with $38$ targets.}
\label{Fig2_OSPA}
\end{figure}

In both scenarios, we chose $c=80$, $\Delta=\alpha=10$. The distances in these scenarios using the three metrics are calculated and plotted in Figures \ref{Fig2_OSPA} and \ref{Fig1_OSPA}. In both scenarios, the OSPA metric and OSPAT metric have the same results because there is no mislabeling in these scenarios. This is because both of these metrics calculate the minimum distance between states across two finite sets of tracks at each time index regardless of which tracks these states belong to. For the proposed OSPAMT metric, the distance between the associated tracks at each time index is calculated so when two states of a pair of associated tracks are far from each other, the distance is large. It can be seen at time $t=11$ and the period from $t=26$ to $t=41$ in Figure \ref{Fig1_OSPA}, the proposed OSPAMT metric has larger error than the other two metrics.

\begin{figure}[htbp!]
\centering
\includegraphics[width=\textwidth/2-0.3cm]
{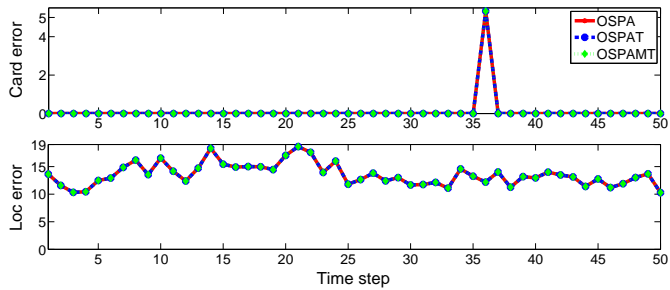}
\caption{Cardinality and localization errors 
versus time for the scenario with $38$ targets.}
\label{Simul2_CardLoc}
\end{figure}
When the tracks are estimated accurately, the three metrics have the same results (see Figures \ref{Fig2_OSPA} and \ref{Simul2_CardLoc}) because there are no missed tracks and the number of estimated tracks equals the number of ground truth tracks. However, when the number of estimated tracks is different from that of ground truth tracks, the OSPA and OSPAT metric are the same because they compute the distances of targets between two finite sets at each time without considering false tracks and missed tracks while the OSPAMT takes into account the false tracks and missed tracks (see Figures \ref{Fig1_OSPA} and \ref{Simul1_CardLoc}). As a result, at a time index where there exist missed tracks or false tracks, the OSPAMT distance is higher than the OSPA metric while the OSPAT is the same as the OSPA metric (note that there is no mislabeling in this scenario). Indeed, from Figure \ref{Simul1_CardLoc}, it can be seen that from time $7$ to $13$ the OSPAMT metric is larger in the cardinality error than the other two metrics but smaller in localization error than the other two metrics. This is because distances between all target states and the ground truth states are considered in the other two metrics even though some of the states are from false tracks while the OSPAMT metric distinguishes the false tracks, missed tracks and estimated tracks. Furthermore, in the period from time $t=26$ to time $t=41$ apart from time $t=38$, the OSPAMT may be larger in the localization error than the other two metrics and is larger in the cardinality error than the other two metrics because the OSPAMT assignment is many-to-one, i.e. some tracks are the estimates of a ground truth track (see the paragraph after \eqref{E:LocDist} for explanation).

\begin{figure}[H]
\centering
\includegraphics[width=\textwidth/2-0.3cm]
{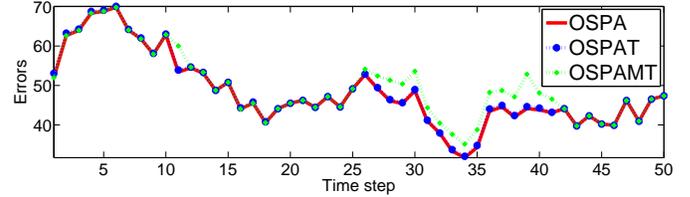}
\caption{Error versus time calculated under OSPA, OSPAT and OSPAMT metric for the scenario with $27$ targets.}
\label{Fig1_OSPA}
\end{figure}
\begin{figure}[htbp!]
\centering
\includegraphics[width=\textwidth/2-0.3cm]
{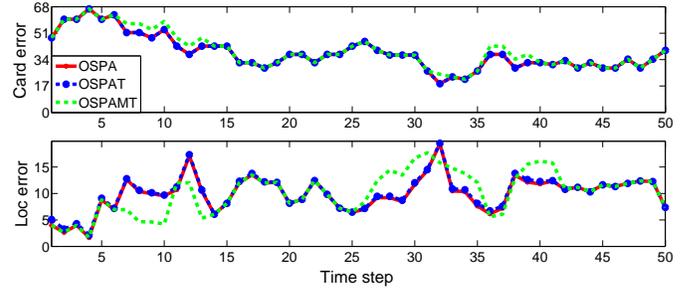}
\caption{Cardinality and localization errors 
versus time for the scenario with $27$ targets.}
\label{Simul1_CardLoc}
\end{figure}
\section{Conclusion}\label{S:Conclusion}
This paper  derives a new metric for evaluating the performance of MTT algorithms. The new  OSPAMT metric,  overcomes all identified limitations of existing MTT metrics.  The new metric accounts for the diverse error mechanisms of MTT algorithms such  missed detections (by allowing multiple tracks to be assigned to a track),  missed tracks(by not assigning some truth tracks to any track), false tracks( by assigning some tracks to $0$), and also the number of targets at any time. Moreover, with a suitable choice of $c$ and $\Delta$ the new metric yields the optimal assignment between a finite set of truth tracks and a finite set of estimated tracks obtained from an  MTT algorithm. It also produces the distance between these two sets at each time.

The proposed OSPAMT metric has three parameters $p$, $\Delta$ and $c$ $(p\in[1,\infty),c>0,0<\Delta\le c)$ which are adjustable and problem dependent. The parameters $p$ and $c$ follow the same meaningful interpretation of the OSPA metric \cite{Schuhmacher2008} as the outlier sensitivity penalty and cardinality error, and $\Delta$ penalizes the case when many tracks assigned to a truth track or the reverse. For applications, if it is important to estimate the number of tracks correctly, we choose a large value of $c$ and small value of $\Delta$. If the exact position of tracks is important and the cardinality error is disregarded, $c$ and $\Delta$ should be small. However, no matter which of these three parameters are chosen, checking the satisfaction of the assignment between estimated tracks and truth tracks is advisable.

Different choices of $c$ and $\Delta$ lead to different assignments and different distances so the choice of suitable $c$ and $\Delta$ is under further  investigation.
\bibliographystyle{IEEEtran}
\bibliography{references_new}

\begin{thebibliography}{10}
\providecommand{\url}[1]{#1}
\csname url@samestyle\endcsname
\providecommand{\newblock}{\relax}
\providecommand{\bibinfo}[2]{#2}
\providecommand{\BIBentrySTDinterwordspacing}{\spaceskip=0pt\relax}
\providecommand{\BIBentryALTinterwordstretchfactor}{4}
\providecommand{\BIBentryALTinterwordspacing}{\spaceskip=\fontdimen2\font plus
\BIBentryALTinterwordstretchfactor\fontdimen3\font minus
  \fontdimen4\font\relax}
\providecommand{\BIBforeignlanguage}[2]{{%
\expandafter\ifx\csname l@#1\endcsname\relax
\typeout{** WARNING: IEEEtran.bst: No hyphenation pattern has been}%
\typeout{** loaded for the language `#1'. Using the pattern for}%
\typeout{** the default language instead.}%
\else
\language=\csname l@#1\endcsname
\fi
#2}}
\providecommand{\BIBdecl}{\relax}
\BIBdecl

\bibitem{ristic2011metric}
B.~Ristic, B.-N. Vo, D.~Clark, and B.-T. Vo, ``A metric for performance
  evaluation of multi-target tracking algorithms,'' \emph{IEEE Transactions on
  Signal Processing}, vol.~59, no.~7, p. 3453, 2011.

\bibitem{Fridling1991}
\BIBentryALTinterwordspacing
B.~E. Fridling and O.~E. Drummond, ``Performance evaluation methods for
  multiple-target-tracking algorithms,'' vol. 1481, 1991, pp. 371--383.
  [Online]. Available: \url{http://dx.doi.org/10.1117/12.45677}
\BIBentrySTDinterwordspacing

\bibitem{Ristic99TWSEval}
B.~Ristic, ``A tool for track-while-scan algorithm evaluation,'' in
  \emph{Information, Decision and Control, 1999. IDC 99. Proceedings. 1999},
  1999, pp. 105--110.

\bibitem{rothrock2000performance}
R.~L. Rothrock and O.~E. Drummond, ``Performance metrics for multiple-sensor
  multiple-target tracking,'' in \emph{AeroSense 2000}.\hskip 1em plus 0.5em
  minus 0.4em\relax International Society for Optics and Photonics, 2000, pp.
  521--531.

\bibitem{Blackman99}
S.~Blackman and R.~Popoli, \emph{{Design and analysis of modern tracking
  systems}}.\hskip 1em plus 0.5em minus 0.4em\relax Artech House Norwood, MA,
  1999.

\bibitem{Colegrove96PerAss}
S.~Colegrove, L.~Davis, and S.~Davey, ``Performance assessment of tracking
  systems,'' in \emph{Signal Processing and Its Applications, 1996. ISSPA 96.,
  Fourth International Symposium on}, vol.~1, 1996, pp. 188--191.

\bibitem{HoffmanMahler02}
J.~Hoffman and R.~Mahler, ``{Multitarget miss distance via optimal
  assignment},'' \emph{Systems, Man and Cybernetics, Part A: Systems and
  Humans, IEEE Transactions on}, vol.~34, no.~3, pp. 327 -- 336, May 2004.

\bibitem{Schuhmacher2008}
D.~Schuhmacher, B.-T. Vo, and B.-N. Vo, ``{A Consistent Metric for Performance
  Evaluation of Multi-Object Filters},'' \emph{IEEE Transactions on Signal
  Processing}, vol.~56, no.~8, pp. 3447--3457, Aug. 2008.

\bibitem{schuhmacher2008math}
\BIBentryALTinterwordspacing
D.~Schuhmacher and A.~Xia, ``A new metric between distributions of point
  processes,'' \emph{Advances in Applied Probability}, vol.~40, pp. 651--672,
  2008. [Online]. Available:
  \url{http://adsabs.harvard.edu/abs/2007arXiv0708.2777S}
\BIBentrySTDinterwordspacing

\bibitem{Reid1979}
D.~Reid, ``{An algorithm for tracking multiple targets},'' \emph{IEEE
  Transactions on Automatic Control}, vol.~24, no.~6, pp. 843 -- 854, 1979.

\bibitem{shalom1988tracking}
Y.~Bar-Shalom and T.~Fortmann, \emph{{Tracking and data association}}.\hskip
  1em plus 0.5em minus 0.4em\relax Academic Press, 1988.

\bibitem{Blackman2004}
S.~Blackman, ``Multiple hypothesis tracking for multiple target tracking,''
  \emph{IEEE Aerospace and Electronic Systems Magazine}, vol.~19, no.~1, pp. 5
  --18, 2004.

\bibitem{BlomHans2008Closespace}
H.~A.~P. Blom, E.~Bloem, Y.~Boers, and H.~Driessen, ``Tracking closely spaced
  targets: Bayes outperformed by an approximation?'' in \emph{Information
  Fusion, 2008 11th International Conference on}, 2008, pp. 1--8.

\bibitem{TuyetPMMHT14}
T.~Vu, B.-N. Vo, and R.~Evans, ``A particle marginal metropolis-hastings
  multi-target tracker,'' \emph{Signal Processing, IEEE Transactions on},
  vol.~62, no.~15, pp. 3953--3964, Aug 2014.

\bibitem{Musicki04JIPDA}
D.~Musicki and R.~Evans, ``{Joint integrated probabilistic data association:
  JIPDA},'' \emph{IEEE Transactions on Aerospace and Electronic Systems},
  vol.~40, no.~3, pp. 1093 -- 1099, Jul. 2004.

\bibitem{Oh2009}
S.~Oh, S.~Russell, and S.~Sastry, ``{Markov Chain Monte Carlo Data Association
  for Multi-Target Tracking},'' \emph{IEEE Transactions on Automatic Control},
  vol.~54, no.~3, pp. 481--497, Mar. 2009.

\bibitem{Beard17}
M.~{Beard}, B.~T. {Vo}, and B.~{Vo}, ``Ospa(2): Using the ospa metric to
  evaluate multi-target tracking performance,'' in \emph{2017 International
  Conference on Control, Automation and Information Sciences (ICCAIS)}, Oct
  2017, pp. 86--91.

\bibitem{Tuyet14OSPAMT}
T.~Vu and R.~Evans, ``{A New Performance Metric for Multiple Target Tracking
  Based on Optimal Subpattern Assignment},'' in \emph{Proceedings of the 17th
  International Conference on Information Fusion}, Jul.7-10, Salamanca, Spain
  2014.

\end{thebibliography}
\appendices
\section{Mathematical Proof}
\subsection{Proof that $d_{c,p}^{\Delta}(\omega,\omega')$ is a metric}\label{A:OSPAMT-metric}
Fix $p\in[1,+\infty)$, $c>0$ and $0<\Delta\le c$. By \eqref{E:dis_OSPATrks},
$d_{c,p}^{\Delta}(\omega,\omega')$ satisfies the axioms of non-negativity, identity, and symmetry. Using the same argument as the proof of metric \cite{Schuhmacher2008}, we prove the triangle inequality, i.e. for $\omega,\omega',\omega^*\in\Omega$
\begin{equation}\label{E:Triagle_Ineql}
d_{c,p}^{\Delta}(\omega,\omega')\le d_{c,p}^{\Delta}(\omega,\omega^*)+d_{c,p}^{\Delta}(\omega',\omega^*).
\end{equation}

By \eqref{E:NoTar_t_Sample1_2}, denote $\mathbf{n}$, $\mathbf{n}^{\prime*}$ and $\mathbf{n}^*$ as the number of distances from time $1$ to $T$ between $\omega$ and $\omega'$, between $\omega^*$ and $\omega'\!$, and between $\omega^*$ and $\omega$ respectively. Denote $\tilde{\hat{x}}^\omega_{i,t}=\{(t,i,x)\}$ if $x\in\tau^\omega_i(t)$ for $i=1,\ldots,|\omega|$ otherwise $\tilde{\hat{x}}^\omega_{i,t}=\emptyset$ for all $t\in\mathcal{T}$. Then
$X^\omega_t=\bigcup_{i=1}^{|\omega|}\tilde{\hat{x}}^\omega_{i,t}$ and hence $X^\omega=\bigcup_{t=1}^TX^\omega_t$.
For $\hat{x}=(t,j,x)\in\mathcal{T}\times\mathds{N}\times\mathds{R}^{n_x}$ denote $\varrho(\hat{x})=t$ and $\bar{\varrho}(\hat{x})=x$. For $t\in\mathcal{T}$, let $\mathbf{n}_t$ be the maximum number of targets in $\omega$ and $\omega'$ at time $t$. Then for all
$t\in\mathcal{T}$ introduce sets $W^\omega_t=X^\omega_t\cup\{(t,1,u_1),\ldots,(t,1,u_{m^\omega})\}$ and $W^{\omega'}_t=X^{\omega'}_t\cup\{(t,1,v_1),\ldots,(t,1,v_{m^{\omega'}})\}$ where $m^\omega\!=\mathbf{n}_t-|X^\omega_t|$, $m^{\omega'}\!=\mathbf{n}_t-|X^{\omega'}_t\!|$, $u_i,v_k\in\mathds{R}^{n_x}\setminus(\bar{\varrho}(X^{\omega'}_t\cup X^{\omega}_t))$, $u_i\ne u_j,v_k\ne v_l$ for $i,j=1,\ldots,m^\omega$; $k,l=1,\ldots,m^{\omega'}\!\!$; $i\ne j$ and $k\ne l$ such that
$d(u_i,y),d(v_k,x),d(u_i,v_k)\ge c$ for $(t,r,x)\in X^\omega_t$, $(t,m,y)\in X^{\omega'}_t$.

By \eqref{E:lambdaMetric_pi2} and \eqref{E:lambdaMetric_pi1}, for $\hat{x},\hat{y}\in \mathcal{T}\times\mathds{N}\times\mathds{R}^{n_x}$, 
 denote $\hat{d}^c_\Delta(\hat{x},\hat{y})=0$ if $\varrho(\hat{x})\ne\varrho(\hat{y})$. Otherwise for $\hat{x}=(t,i,x)\in X^\omega_t,\hat{y}=(t,j,y)\in X^{\omega'}_t$
\begin{align*}
\hat{d}^c_\Delta(\hat{x},\hat{y})&=\min\{c,d(x,y)\},\qquad \hat{d}^c_\Delta(\hat{x},\hat{y})^p=c^p+\Delta^p,\\
\hat{d}^c_\Delta(\hat{x},\hat{y})^p&=\min\{c,d(x,y)\}^p+\Delta^p
\end{align*}
if at time $t$ the track $i$ is assigned to the track $j$; or $\hat{d}^c_\Delta(\hat{x},\hat{y})=c$ if the track $i$ is not assigned to the track $j$. Then \eqref{E:dis_OSPATrks} becomes
\begin{align}\label{E:lambdaMetricSimp}
 \!\!\textstyle d_{c,p}^{\Delta}
(\omega,\omega')=\textstyle\left[\frac{1}{\mathbf{n}}\textstyle\sum_{i=1}^\mathbf{n}
\hat{d}^c_\Delta(\hat{x}_i,\hat{y}_{\mu(i)})^p\right]^{1/p}
\end{align}
where $\mu\in\Pi_{\mathbf{n}}$ and $\varrho(\hat{x}_i)=\varrho(\hat{y}_{\mu(i)}))$ for $i=1,\ldots,\mathbf{n}$ where $\Pi_k$ denotes the set of permutations on $\{1,2,...,k\}$  for any positive integer $k$.

If $\omega,\omega',\omega^*\ne\emptyset$, 
\eqref{E:Triagle_Ineql} holds by using the same argument as the proof of OSPA metric in \cite{Schuhmacher2008} where $l,m,n$ in \cite{Schuhmacher2008} are $\mathbf{n}$, $\mathbf{n}^{\prime*}$ and $\mathbf{n}^*$ respectively. If one of $\omega,\omega'$ and $\omega^*$ is empty, then 
\eqref{E:Triagle_Ineql} holds by \eqref{E:dis_OSPATrks} and Remark \ref{Rm:OSPAt}.
 For other cases, 
\eqref{E:Triagle_Ineql} holds by \eqref{E:dis_OSPATrks}.
\section{OSPA metric}\label{SS:Summary_OSPA}
Let $X=\{x_1,...,x_m\}$ and $Y=\{y_1,...,y_n\}$ ($X,Y\subsetneqq\mathcal{X}$) be two finite sets and assume that $m<n$. The assignment between the $m$ points of $X$ and the $n$ points of $Y$ is  determined such that it minimizes the sum of distances, subject to the constraint that the distances are capped at a preselected maximum or cut-off value $c$. This minimum sum of distances can be interpreted as the total localization error. All the unassigned points in $Y$ are penalized with $c$ the maximum error value which is regarded as a cardinality error. The OSPA metric is the sum of the localization error and the cardinality error.


The OSPA metric $\bar{d}^{(c)}_p$ is defined as follows. Let $\bar{d}^{(c)}(x,y)=\min(c,\|x-y\|)$ for $x,y\in\mathcal{X}$, and $\Pi^m_n$ ($m<n$) denotes a collection of all one-to-one functions from a set $\{1,\ldots,m\}$ to $\{1,2,...,n\}$. Then, for $p\geq 1,c>0,$
\begin{itemize}
  \item if $m\leq n$:\vspace{-.3cm}
\begin{eqnarray*}
\!\!\!\bar{d}^{(c)}_p(X,Y)\!=\!\left[\!\frac{1}{n}\!\left(\!\displaystyle\min_{\pi\in\Pi^m_n}\!\!\sum_{i=1}^m\!\bar{d}^{(c)}
(x_i,y_{\pi(i)})^p\!+\!c^p(n-m)\!\right)\!\right]^{\frac{1}{p}}&
\end{eqnarray*}
  \item if $m>n$: $\bar{d}^{(c)}_p(X,Y)=\bar{d}^{(c)}_p(Y,X)$; and
  \item if $m=n=0$: $\bar{d}^{(c)}_p(X,Y)=\bar{d}^{(c)}_p(Y,X)=0.$
\end{itemize}
%

The OSPA distance is interpreted as a $p-$th order per-target error, comprised of a $p-$th order per-target localization error and a $p-$th order per-target cardinality error. For $1\le p<\infty$ these components are given by
\begin{itemize}
  \item if $m\leq n$:\vspace{-.3cm}
\begin{align*}
\bar{e}^{(c)}_{p,\text{loc}}(X,Y)&=\left(\frac{1}{n}\min_{\pi\in\Pi^m_n}\sum_{i=1}^m\bar{d}^{(c)}
(x_i,y_{\pi(i)})^p\right)^{\frac{1}{p}},\nonumber\\
\bar{e}^{(c)}_{p,\text{card}}(X,Y)&=\left(\frac{c^p(n-m)}{n}\right)^{\frac{1}{p}}.
\end{align*}
  \item if $m>n$:\vspace{-.3cm}
\begin{alignat*}{2}
\bar{e}^{(c)}_{p,\text{loc}}(X,Y)\!=\!\bar{e}^{(c)}_{p,\text{loc}}(Y,X),\bar{e}^{(c)}_{p,\text{card}}(X,Y)\!=\!\bar{e}^{(c)}_{p,\text{card}}(Y,X).
\end{alignat*}
\end{itemize}
\section{OSPAT metric}\label{SS:Summary_OSPAT}
In this section, we will summarize the OSPAT metric \cite{ristic2011metric}. This metric is actually the OSPA distance between two sets of labeled states at each point in time. Note that the labeled state is a vector whose components are the state of the target and the target label. The labels in these two sets are added to the states in such a way that the sum of all spatial distances between pairs of tracks across these two sets is minimum. In order to compute the OSPA distance between two sets of labeled states, the labeling error is added to the standard spatial distance. The states of the pair tracks in the OSPAT assignment are labeled the same. Since the OSPAT assignment is achieved by finding the minimum spatial distance between only pairs of tracks across two sets of tracks (i.e ignores tracks from a bigger set not paired to any track from other set), the OSPAT metric suffers some limitations as discussed in Appendix  \ref{A:Iss_OSPAT}.

Note that the representation of $\omega\in\Omega$ given here is different from the representation of a set of tracks in \cite{ristic2011metric} because the elements of singleton sets of a track in \eqref{E:Trk} are in $\mathcal{X}$ while the elements of singleton sets of a track in \cite{ristic2011metric} are in $\mathds{N}\times\mathcal{X}$. Note also that $\mathcal{X}$ is the state space while $\mathds{N}\times\mathcal{X}$ is the labeled state space. Let $\omega,\omega'\in\Omega$. The OSPAT distance between $\omega$ and $\omega'$ at each time $t\in \mathcal{T}$ is actually the OSPA distance between two sets of labeled states whose states are in $\omega$ and $\omega'$ at time $t$. The labels are added to the tracks such that they minimize the total spatial distance between the pairs of tracks across $\omega$ and $\omega'$. The idea is detailed below, following \cite{ristic2011metric}.
\subsubsection{Reorder tracks}\label{SS:Relabel}
Let $|A|$ be the number of elements in a set $A$. 
Then 
$\lambda\in\Pi^{\min\{|\omega|,|\omega'|\}}_{\max\{|\omega|,|\omega'|\}}$ represents a one-to-one  assignment between tracks in $\omega$ and tracks in $\omega'$.

Given $\lambda\in\Pi^{\min\{|\omega|,|\omega'|\}}_{\max\{|\omega|,|\omega'|\}}$ 
and $\bar{c}>0$, the distance between 
$\omega$ and $\omega'$ at time $t$, denoted by $D^{\bar{c},\lambda}_t(\omega,\omega')$, is
\begin{align}\label{E:OSPATdis_lambdat}
\left\{
\begin{array}{ll}
\sum_{j=1}^{|\omega'|}\underline{d}^{\bar{c}}(\tau^{\omega'}_j(t),\tau^\omega_{\lambda(j)}(t)), & \hbox{if $|\omega'|\le |\omega|$;} \\
\sum_{j=1}^{|\omega|}\underline{d}^{\bar{c}}(\tau^{\omega}_j(t),\tau^{\omega'}_{\lambda(j)}(t)), & \hbox{otherwise.}
\end{array}
\right.
\end{align}
where
\begin{align}\label{E:dis_OSPAT Assign}
\!\!\underline{d}^{\bar{c}}\!(\tilde{x},\tilde{y})=
\left\{\!\!
\begin{array}{ll}
0, & \!\!\hbox{if $\tilde{x}=\tilde{y}=\emptyset$;} \\
\min\{\bar{c},\|x-\!y\|_2\},\!\! &\!\! \hbox{if $\tilde{x}\!=\!\{x\},\tilde{y}\!=\!\{y\}$;} \\
\bar{c},\!\!&\!\!\hbox{otherwise.}
\end{array}
\right.\!\!
\end{align}

If one and only one of the targets $j$ or $\lambda(j)$ exists at time $t$, then the distance is $\bar{c}$. If both exist at time $t$, the distance is $\min\{\bar{c},\|x-y\|_2\}$ where $\tau^{\omega'}_j(t)=\{x\}$ and $\tau^\omega_{\lambda(j)}(t)=\{y\}$.

The distance between $\omega$ and $\omega'$ given $\lambda\in\Pi^{\min\{|\omega|,|\omega'|\}}_{\max\{|\omega|,|\omega'|\}}$ is
\begin{align}\label{E:OSPATdis_lambda}
\textstyle
D^{\bar{c},\lambda}(\omega,\omega')=\sum_{t=1}^TD^{\bar{c},\lambda}_t(\omega,\omega')
\end{align}
and the OSPAT assignment between $\omega$ and $\omega'$ is
\begin{align}\label{E:min_dis4Label}
\lambda_*^{\omega,\omega'}=\arg\min_{\lambda\in\Pi^{\min\{|\omega|,|\omega'|\}}_{\max\{|\omega|,|\omega'|\}}}
D^{\bar{c},\lambda}(\omega,\omega').
\end{align}

Then the global OSPAT distance between $\omega$ and $\omega'$ and the global OSPAT distance at time $t$ between $\omega$ and $\omega'$ are
\begin{align}
D_{Gb}^{\bar{c}}(\omega,\omega')=D^{\bar{c},\lambda_*^{\omega,\omega'}}(\omega,\omega')\label{E:min_Global2}\\
D_{Gb}^{\bar{c},t}(\omega,\omega')=D^{\bar{c},\lambda_*^{\omega,\omega'}}_t(\omega,\omega')\label{E:min_Global2_t}
\end{align}
respectively.
\subsubsection{OSPAT distance}\label{SS:OSPAT metric}
Denote $\mathcal{F}_X$ as the collection of all finite subsets of $\mathds{N}\times\mathcal{X}$. 
For any $\bar{X}\in\mathcal{F}_X$. If $\bar{X}=\{\bar{x}_1,\ldots,\bar{x}_{\bar{m}}\}$ ($\bar{m}\in\mathds{N}$), denote $\bar{X}^i=\bar{x}_i$ for $i=1,\ldots,\bar{m}$.

Let $p$ ($1\le p< \infty$) and $c$ ($c>0$) be the metric order parameter and the cutoff parameter respectively. Denote
$\bar{d}^\alpha_c(\bar{x},\bar{y})=\min\{c,\bar{d}^\alpha(\bar{x},\bar{y})\}$ where $\bar{x},\bar{y}\in \mathds{N}\times\mathcal{X}$; $\alpha\in[0,c]$ and the
base distance $\bar{d}^\alpha(\bar{x},\bar{y})$ is defined in the following subsection. Let $\bar{X},\bar{Y}\in \mathcal{F}_X$, the metric
$\hat{D}^\alpha_{c,p}$ on $\mathcal{F}_X$ is
\begin{align}\label{E:Metric-OSPAT_t}
\hat{D}^\alpha_{c,p}(\bar{X},\bar{Y})=
\left\{
\begin{array}{ll}
\bar{D}^\alpha_{c,p}(\overrightarrow{\bar{X},\bar{Y}}), & \hbox{if $|\bar{X}|\le|\bar{Y}|$;} \\
\bar{D}^\alpha_{c,p}(\overrightarrow{\bar{Y},\bar{X}}), & \hbox{otherwise}
\end{array}
\right.
\end{align}
where for $m=|\bar{X}|$, $n=|\bar{Y}|$ and assume that $m\le n$
\begin{eqnarray*}
	\bar{D}^\alpha_{c,p}(\overrightarrow{\bar{X},\bar{Y}})=
	\left[\displaystyle\dfrac{1}{n}\!\left(\!\min_{\sigma\in\Pi^m_{n}}\!
	\sum_{i=1}^{m}\bar{d}^\alpha_c(\bar{X}^i,\bar{Y}^{\sigma(i)})^p+
	(n-m)c^p\!\right)\!\right]^{\frac{1}{p}}&
\end{eqnarray*}
In order to apply the distance given in \eqref{E:Metric-OSPAT_t} to the distance between $\omega$ and $\omega'$ at each point in time, the
states of targets in $\omega$ and $\omega'$ must be added to their labels and grouped according to the time index provided that the optimal assignment $\lambda_*^{\omega,\omega'}$ in \eqref{E:min_dis4Label} exists. Without loss of generality assume that $|\omega'|\le|\omega|$. Let $\omega^\ast=\{\tau_j\in\omega:j\notin\{\lambda_*^{\omega,\omega'}(1),\ldots,\lambda_*^{\omega,\omega'}(|\omega'|)\}\}$.
Then
\begin{align}\label{E:state_time_t}
\bar{X}^{\omega'}_t=\textstyle
\bigcup_{l=1}^{|\omega'|}\tilde{\bar{x}}^l_t,\qquad
\bar{Y}^{\omega}_t=\textstyle
\left(\bigcup_{l=1}^{|\omega'|}\tilde{\bar{y}}^l_t\right)\cup\left(\bigcup_{i=1}^{|\omega^\ast|}\tilde{\bar{z}}^{i}_t\right)
\end{align}
for all $t\in \mathcal{T}$ where for $l=1,\ldots,|\omega'|$ and for $i=1,\ldots,|\omega^\ast|$
\begin{align*}
\tilde{\bar{x}}^l_t=&\{(l,x)\}\text{ if }x\in\tau^{\omega'}_l(t)
\text{ otherwise }\tilde{\bar{x}}^l_t=\emptyset,\\
\tilde{\bar{y}}^l_t=&\{(l,y)\}\text{ if }y\in\tau^{\omega}_{\lambda_*^{\omega,\omega'}(l)}(t)
\text{ otherwise }\tilde{\bar{y}}^l_t=\emptyset,\\
\tilde{\bar{z}}^{i}_t=&\{(i+|\omega'|,z)\}\text{ if }z\in\tau^{\omega^\ast}_i(t)
\text{ otherwise }\tilde{\bar{z}}^{i}_t=\emptyset.
\end{align*}
Thus the two sets $\omega$ and $\omega'$ after being equipped their labels via the OSPAT assignment becomes $\bar{\omega}$ and $\bar{\omega'}$ respectively
\begin{align*}
\bar{\omega}'=\{\bar{\tau}'_1,\ldots,\bar{\tau}'_{|\omega'|}\},\qquad
\bar{\omega}=\{\bar{\tau}_1,\ldots,\bar{\tau}_{|\omega'|+|\omega^\ast|}\}
\end{align*}
where $\bar{\tau}'_l,\bar{\tau}_l$ and $\bar{\tau}_i$ for $l=1,\ldots,|\omega'|$ and $i=1,\ldots,|\omega^\ast|$
\begin{align*}
\bar{\tau}'_l=(\tilde{\bar{x}}^l_1,\ldots,\tilde{\bar{x}}^l_T),\hspace{.2cm}
\bar{\tau}_l=(\tilde{\bar{y}}^l_1,\ldots,\tilde{\bar{y}}^l_T),\hspace{.2cm}
\bar{\tau}_{i+|\omega'|}=(\tilde{\bar{z}}^i_1,\ldots,\tilde{\bar{z}}^i_T).
\end{align*}
Hence the OSPAT metric between $\omega$ and $\omega'$ at time $t$ is the same as the OSPAT metric between $\bar{\omega}$ and $\bar{\omega'}$ at time $t$ ($t\in\mathcal{T}$)
\begin{align}\label{D:OSPAT_omega}
\boldsymbol{\mathcal{D}}^{\alpha,t}_{c,p}(\omega,\omega')=\boldsymbol{\mathcal{D}}^{\alpha,t}_{c,p}(\bar{\omega},\bar{\omega'})=\hat{D}^\alpha_{c,p}(\bar{X}^{\omega'}_t,\bar{Y}^{\omega}_t)
\end{align}
where $\hat{D}^\alpha_{c,p}(\bar{X}^{\omega'}_t,\bar{Y}^{\omega}_t)$ is given in \eqref{E:Metric-OSPAT_t}. Without loss of generality, from now on we assume $\bar{c}=c$. Note that the global OSPAT distance in \eqref{E:min_Global2_t} differs from the OSPAT metric in \eqref{D:OSPAT_omega}.
\subsubsection{Base Distance between two labeled states}\label{SS:label_Dis}
Let $\bar{x}=(j,x),\bar{y}=(k,y)\in\mathds{N}\times\mathcal{X}$. Then the base distance $\bar{d}^\alpha(\bar{x},\bar{y})$, a metric on the space $\mathds{N}\times\mathds{R}^{n_x}$, is
\begin{align}\label{E:Metric-label}
\bar{d}^\alpha(\bar{x},\bar{y})=\left(d(x,y)^{p'}+d^\alpha(j,k)^{p'}\right)^{\frac{1}{{p'}}}
\end{align}
where $1\le p'< \infty$ is the base distance order parameter; the localization distance $d(x,y)$ typically has the $p'$-norm: $d(x,y)=\|x-y\|_{p'}$; and the labeling error $d^\alpha(j,k)$ is
\begin{align}\label{E:dis-label}
d^\alpha(j,k)=\alpha\bar{\delta}(j,k).
\end{align}
Note that $\alpha\in[0,c]$ controls the penalty assigned to the labeling error.

\section{
	Limitations of OSPAT}\label{A:Iss_OSPAT}
%

In this Appendix we discuss certain limitations of the OSPAT metric presented in Appendix  \ref{SS:Summary_OSPAT}. 
Firstly, in Appendix \ref{A:Iss_Opt_Ass}, we show that OSPAT assignment \eqref{E:min_dis4Label} which helps to relabel/label tracks is unreliable. Secondly, the OSPAT metric is unreliable for performance evaluation of MTT algorithms and is shown in Appendix \ref{SS:Fail_Rep_real_Scen}. Finally, the most important issue is that the OSPAT metric \eqref{D:OSPAT_omega} is not incompatibility of the metric definition and is presented in Appendix \ref{SS:label_issue}.

\subsection{Unreliability of the OSPAT assignment}\label{A:Iss_Opt_Ass}
The OSPAT assignment plays an important role in the OSPAT metric because it is used to label tracks (or rename labels of labeled track). However, it can be inconsistent and unreliable. Take the scenarios in Figures \ref{Fig:ExamMotivateOSPMAT} and \ref{F:1_1Sepa}. The OSPAT assignments for these scenarios are not the optimal assignments as discussed in section \ref{SS:Co_OSPAMT_Ass}. Similarly, for the scenario in Figure \ref{Fig:RemainIssue}, $[\tau_2\leftrightarrow\tau'_1\!,\tau_3\leftrightarrow\tau'_2]$ is the OSPAT assignment but  clearly this is the worst among the one-to-one assignments shown in Figure \ref{Fig:RemainIssue}.
\begin{figure}[htbp!]
	\centering\hspace{-1.3cm}
	\scalebox{0.75}
	{\centering\hspace{-1.2cm}
		\pgfdeclarelayer{background} \pgfdeclarelayer{foreground}
\pgfsetlayers{background,main,foreground}
\par
\par
\tikzset{cross/.style={cross out, draw=black, minimum size=2*(#1-\pgflinewidth), inner sep=0pt, outer sep=0pt},
cross/.default={3pt}}
\begin{tikzpicture}[y=.2cm, x=2cm,font=\sffamily,bend angle=45]
 \draw [->,thick]
 |-(6.1,0) node (xaxis) [right]
  {\scriptsize Time};
\draw (1,0) -- coordinate (x axis mid) (6.1,0);
\foreach \x in {1,...,6}
     		\draw (\x,1pt) -- (\x,-3pt)
			node[anchor=north] {\x};
\draw (1,1.5) node[cross] (x1){};
\draw (2,1.5) node[cross] (x2){};
\draw (3,1.5) node[cross] (x3){};
\draw (4,1.5) node[cross] (x4){};
\draw (5,1.5) node[cross] (x5){};
\draw (6,1.5) node[cross] (x6){};
\draw (1,17) node[cross] (2x1){};
\draw (2,9.5) node[cross] (2x2){};
\draw (3,17) node[cross] (3x3){};
\draw (4,17) node[cross] (3x4){};
\draw (5,17) node[cross] (3x5){};
\node[circle,draw=black, fill=black, inner sep=0pt,minimum size=3pt] (y1) at (1,5.5) {};
\node[circle,draw=black, fill=black, inner sep=0pt,minimum size=3pt] (y2) at (2,5.5){};
\node[circle,draw=black, fill=black, inner sep=0pt,minimum size=3pt] (y3) at (3,5.5){};
\node[circle,draw=black, fill=black, inner sep=0pt,minimum size=3pt] (2y5) at (5,5.5){};
\node[circle,draw=black, fill=black, inner sep=0pt,minimum size=3pt] (2y6) at (6,5.5){};
\draw[->,black](y1)edge (y2)(y2)edge (y3)(2y5)edge (2y6)
(x1)edge (x2)(x2)edge (x3)(x3)edge (x4)(x4)edge (x5)(x5)edge(x6)(2x1)edge(2x2)(3x3)edge(3x4)(3x4)edge(3x5);
\node(temdis1)  [below =.2cm of y1] {};
 \node (dis1) [right =-0.01cm of temdis1]{$\epsilon\ll 1$};
 \draw[-,thin,black,dashed](y1)edge node[swap,near start]{}
  (x1);
\node(temdis2) [below =.2cm of y2] {};
  \node (dis1) [right =-.5cm of temdis2]{$\epsilon$};
  \draw[-,thin,black,dashed](y2)edge node[swap,near start]{}
  (x2);
\node(temdis3)  [below =.3cm of y3] {};
  \node (dis1) [right =-.5cm of temdis3] {$\epsilon$};
  \draw[-,thin,black,dashed](y3)edge node[midway]{}
  (x3);
\node(temdis5)  [below =.1cm of 2y5] {};
  \node (dis1) [right =-.7cm of temdis5] {$\epsilon$};
  \draw[-,thin,black,dashed](2y5)edge node[midway]{}
  (x5);
\node(temdis6)  [below =.3cm of 2y6] {};
  \node (dis1) [right =-.5cm of temdis6] {$\epsilon$};
  \draw[-,thin,black,dashed](2y6)edge node[midway]{}
  (x6);
\node(2temdis2)  [below =.1cm of 2x2] {};
  \node (dis1) [right =-.5cm of 2temdis2] {$\epsilon$};
  \draw[-,thin,black,dashed](2x2)edge node[midway]{}
  (y2);
\node(2temdis1)  [below =1cm of 2x1] {};
  \node (dis1) [left =-.6cm of 2temdis1] {$\beta$};
  \draw[-,thin,black,dashed](2x1)edge node[near start]{}
  (y1);
\node(2temdis3)  [below =1cm of 3x3] {};
  \node (dis1) [left =-1.8cm of 2temdis3] {$\beta$};
  \draw[-,thin,black,dashed](3x3)edge[bend left] node[near start]{}
  (y3);
  \draw[-,thin,black,dashed](3x5)edge[bend right] node[near start]{}
  (2y5);

 \begin{pgfonlayer}{background}
 \path (2y6.east |-2y6.east)+(.3,-.8) node(bb){$\tau'_2$};
  \path (y3.east |-y3.east)+(.3,1) node(aa){$\tau'_1$};
  \path (x6.east|- x6.east)+(.3,.6) node(a){$\tau_1$};
  \path (3x5.east|- 3x5.east)+(.3,-.9) node(c){$\tau_3$};
 \path (2x2.east|- 2x2.east)+(.3,-.6) node(b){$\tau_2$};
\draw[->,black,dashed](a) edge (x6)(b) edge (2x2)(c) edge (3x5)(aa)edge(y3)(bb) edge (2y6);
\end{pgfonlayer}
\end{tikzpicture}
	}
	\caption{$\omega=\{\tau_1,\tau_2,\tau_3\},\omega'=\{\tau'_1,\tau'_2\}$. Obviously, [$\tau_1\leftrightarrow(\tau'_1,\tau'_2)$] is the optimal assignment. However, the OSPAT assignment is [$\tau_2\leftrightarrow\tau'_1\!,\tau_3\leftrightarrow\tau'_2$].}
	\label{Fig:RemainIssue}
\end{figure}

The cause of the unreliability of the OSPAT is that the distance at any time $t$ in \eqref{E:OSPATdis_lambdat} is inconsistent with the number of targets in two sets of tracks at that time. 
The inconsistences happen because it does not take into account the unassigned tracks from a set with a larger number of tracks at that time. That is, the distances in \eqref{E:OSPATdis_lambdat} with different assignments have different numbers of distances. Take Figure \ref{Fig:RemainIssue} as an example. If an assignment $\lambda$ is chosen such that $\lambda(1)=1$ and $\lambda(2)=3$, then $D_6^{c,\lambda}(\omega,\omega')=\underline{d}^c(\tau^\omega_1(6),\tau^{\omega'}_1(6))+ \underline{d}^c(\tau^\omega_2(6),\tau^{\omega'}_2(6))=2c$ which is the sum of two distances where $\underline{d}^c(\cdot,\cdot)$ is defined in \eqref{E:dis_OSPAT Assign}. However, if $\lambda$ is chosen such that $\lambda(2)=1$ and $\lambda(1)=2$, then $D_6^{c,\lambda'}(\omega,\omega')=\underline{d}^c(\tau^\omega_1(6),\tau^{\omega'}_2(6))=\epsilon$ which includes only one distance. In fact, there is only one distance at time $6$ between $\omega$ and $\omega'$ because each set has only one exiting target. This inconsistency may result in the assignment [$\tau_1\!\leftrightarrow\!\tau'_2;\!\tau_2\!\leftrightarrow\!\tau'_1$] as the OSPAT assignment for Figure \ref{Fig:RemainIssue} which is not the optimal assignment.
\subsection{Unreliability of the OSPAT metric}\label{SS:Fail_Rep_real_Scen}
The unreliability of the OSPAT metric happens because it does not measure the distance between two sets of tracks. Instead, it actually measures the distance between two sets of tracks at each point in time. Thus it is incompatible for measuring two sets of tracks when the OSPAT distances at time $t$ between $\omega'$ and $\omega$, $\boldsymbol{\mathcal{D}}^{\alpha,t}_{c,p}(\omega,\omega')$ and between $\omega^{\prime\prime}$ and $\omega$, $\boldsymbol{\mathcal{D}}^{\alpha,t}_{c,p}(\omega,\omega^{\prime\prime})$, are different for all $t=1,\ldots,T$. When two estimated track is better than the other, the OSPAT metric also give the wrong conclusion which is the two estimated track $\omega'$ and $\omega^{\prime\prime}$ are equal. This is because $\boldsymbol{\mathcal{D}}^{\alpha,t}_{c,p}(\omega,\omega')=\boldsymbol{\mathcal{D}}^{\alpha,t}_{c,p}(\omega,\omega^{\prime\prime})$ for all $t=1,\ldots,T$. Indeed, consider the case when the OSPAT assignment is the optimal assignment such as the scenarios in Figure \ref{Fig:CompAtTime_All}. By Table \ref{table:SameTimeButDiffAll}, both scenarios in Figures \eqref{Fig:CompAtTime_All_1} and \eqref{Fig:CompAtTime_All_2} are the same when using the OSPAT metric but clearly the scenario in Figure \eqref{Fig:CompAtTime_All_1} is better than the scenario in Figure \eqref{Fig:CompAtTime_All_2}.
Similarly, take Figure \ref{Fig:Inc_Sol} as an example where we need to evaluate the performance of the two sets of tracks $\omega'$ and $\omega^{\prime\prime}$ with the set of the truth tracks  $\omega$ ($\omega'$ and $\omega^{\prime\prime}$ obtained from two different MTT algorithms). Clearly, $\omega^{\prime\prime}$ is always a better estimate of $\omega$ than $\omega'$. However, as discussed in Section \ref{SS:AnalysisMetrics}, the OSPAT metric gives different conclusion depending on the choice of $\alpha$.

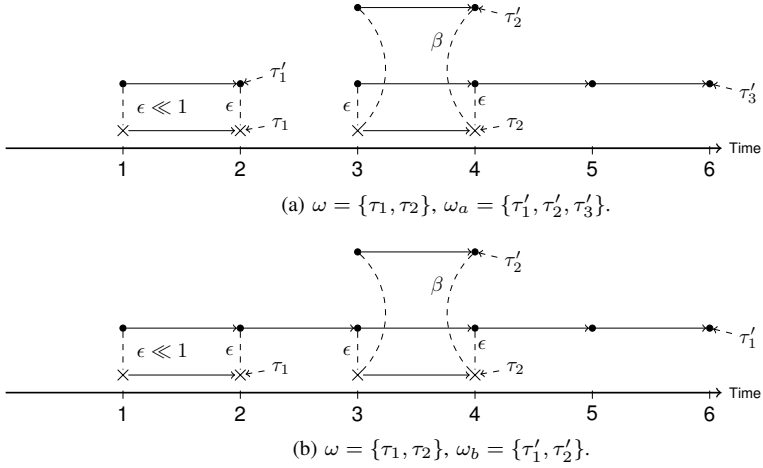
\begin{figure}[htbp!]
	\centering
	\subfloat[$\omega=\{\tau_1,\tau_2\}$, $\omega_a=\{\tau'_1,\tau'_2,\tau'_3\}$. 
	]
	{\label{Fig:EvalSmeRis_DifTuyet1}
		\centering{\hspace{-2cm}
			\scalebox{0.78}
			{\centering
				\pgfdeclarelayer{background} \pgfdeclarelayer{foreground}
\pgfsetlayers{background,main,foreground}
\par
\par
\tikzset{cross/.style={cross out, draw=black, minimum size=2*(#1-\pgflinewidth), inner sep=0pt, outer sep=0pt},
cross/.default={3pt}}
\begin{tikzpicture}[y=.2cm, x=2cm,font=\sffamily,bend angle=45]
 \draw [->,thick]
 |-(6.1,0) node (xaxis) [right]
  {\scriptsize Time};
\draw (1,0) -- coordinate (x axis mid) (6.1,0);
\foreach \x in {1,...,6}
     		\draw (\x,1pt) -- (\x,-3pt)
			node[anchor=north] {\x};
\draw (1,1.5) node[cross] (x1){};
\draw (2,1.5) node[cross] (x2){};

\draw (3,1.5) node[cross] (2x3){};
\draw (4,1.5) node[cross] (2x4){};
\node[circle,draw=black, fill=black, inner sep=0pt,minimum size=3pt] (y1) at (1,5.5) {};
\node[circle,draw=black, fill=black, inner sep=0pt,minimum size=3pt] (y2) at (2,5.5){};
\node[circle,draw=black, fill=black, inner sep=0pt,minimum size=3pt] (2y3) at (3,5.5){};
\node[circle,draw=black, fill=black, inner sep=0pt,minimum size=3pt] (2y4) at (4,5.5){};
\node[circle,draw=black, fill=black, inner sep=0pt,minimum size=3pt] (2y5) at (5,5.5){};
\node[circle,draw=black, fill=black, inner sep=0pt,minimum size=3pt] (2y6) at (6,5.5){};
\node[circle,draw=black, fill=black, inner sep=0pt,minimum size=3pt] (3y3) at (3,12){};
\node[circle,draw=black, fill=black, inner sep=0pt,minimum size=3pt] (3y4) at (4,12){};
\draw[->,black]
(y1)edge (y2)(2y3)edge(2y4)(2y4)edge (2y5)(2y5)edge (2y6)(3y3)edge (3y4)
(x1)edge (x2)(2x3)edge (2x4);
\node(temdis1)  [below =.2cm of y1] {};
 \node (dis1) [right =-0.01cm of temdis1]{$\epsilon\ll 1$};
 \draw[-,thin,black,dashed](y1)edge node[swap,near start]{}
  (x1);
\node(temdis2) [below =.2cm of y2] {};
  \node (dis1) [right =-.5cm of temdis2]{$\epsilon$};
  \draw[-,thin,black,dashed](y2)edge node[swap,near start]{}
  (x2);
\node(2temdis3)  [below =.2cm of 2y3] {};
  \node (dis1) [right =-.5cm of 2temdis3] {$\epsilon$};
  \draw[-,thin,black,dashed](2y3)edge node[midway]{}
  (2x3);
\node(temdis4)  [below =.1cm of 2y4] {};
  \node (dis1) [right =-.2cm of temdis4] {$\epsilon$};
  \draw[-,thin,black,dashed](2y4)edge node[midway]{}
  (2x4);
  \draw[-,thin,black,dashed](3y3)edge[bend left] node[midway]{}
  (2x3);
\node(3temdis5)  [below =.4cm of 3y4] {};
  \node (dis1) [right =-1cm of 3temdis5] {$\beta$};
  \draw[-,thin,black,dashed](3y4)edge[bend right] node[midway]{}
  (2x4);


 \begin{pgfonlayer}{background}
 \path (2y6.east |-2y6.east)+(.3,-.8) node(bb){$\tau'_3$};
 \path (y2.east |-y2.east)+(.3,1) node(aa){$\tau'_1$};
 \path (x2.east|- x2.east)+(.3,.6) node(a){$\tau_1$};
 \path (2x4.east|- 2x4.east)+(.3,.6) node(b){$\tau_2$};
 \path (3y4.east |-3y4.east)+(.3,-.8) node(cc){$\tau'_2$};

\draw[->,black,dashed](a) edge (x2)(b) edge (2x4)(aa)edge(y2)(bb) edge (2y6)(cc)edge(3y4);
\end{pgfonlayer}
\end{tikzpicture}
		}}
	}\\%
	\subfloat[$\omega=\{\tau_1,\tau_2\}$, $\omega_b=\{\tau'_1,\tau'_2\}$. 
	]
	{\label{Fig:EvalSmeRis_DifTuyet2}
		\centering{\hspace{-2cm}
			\scalebox{.78}
			{\centering
				\pgfdeclarelayer{background} \pgfdeclarelayer{foreground}
\pgfsetlayers{background,main,foreground}
\par
\par
\tikzset{cross/.style={cross out, draw=black, minimum size=2*(#1-\pgflinewidth), inner sep=0pt, outer sep=0pt},
cross/.default={3pt}}
\begin{tikzpicture}[y=.2cm, x=2cm,font=\sffamily,bend angle=45]
 \draw [->,thick]
 |-(6.1,0) node (xaxis) [right]
  {\scriptsize Time};
\draw (1,0) -- coordinate (x axis mid) (6.1,0);
\foreach \x in {1,...,6}
     		\draw (\x,1pt) -- (\x,-3pt)
			node[anchor=north] {\x};
\draw (1,1.5) node[cross] (x1){};
\draw (2,1.5) node[cross] (x2){};

\draw (3,1.5) node[cross] (2x3){};
\draw (4,1.5) node[cross] (2x4){};
\node[circle,draw=black, fill=black, inner sep=0pt,minimum size=3pt] (y1) at (1,5.5) {};
\node[circle,draw=black, fill=black, inner sep=0pt,minimum size=3pt] (y2) at (2,5.5){};
\node[circle,draw=black, fill=black, inner sep=0pt,minimum size=3pt] (2y3) at (3,5.5){};
\node[circle,draw=black, fill=black, inner sep=0pt,minimum size=3pt] (2y4) at (4,5.5){};
\node[circle,draw=black, fill=black, inner sep=0pt,minimum size=3pt] (2y5) at (5,5.5){};
\node[circle,draw=black, fill=black, inner sep=0pt,minimum size=3pt] (2y6) at (6,5.5){};
\node[circle,draw=black, fill=black, inner sep=0pt,minimum size=3pt] (3y3) at (3,12){};
\node[circle,draw=black, fill=black, inner sep=0pt,minimum size=3pt] (3y4) at (4,12){};
\draw[->,black]
(y1)edge (y2)(y2)edge (y3)(2y3)edge(2y4)(2y4)edge (2y5)(2y5)edge (2y6)(3y3)edge (3y4)
(x1)edge (x2)(2x3)edge (2x4);
\node(temdis1)  [below =.2cm of y1] {};
 \node (dis1) [right =-0.01cm of temdis1]{$\epsilon\ll 1$};
 \draw[-,thin,black,dashed](y1)edge node[swap,near start]{}
  (x1);
\node(temdis2) [below =.2cm of y2] {};
  \node (dis1) [right =-.5cm of temdis2]{$\epsilon$};
  \draw[-,thin,black,dashed](y2)edge node[swap,near start]{}
  (x2);
\node(2temdis3)  [below =.2cm of 2y3] {};
  \node (dis1) [right =-.5cm of 2temdis3] {$\epsilon$};
  \draw[-,thin,black,dashed](2y3)edge node[midway]{}
  (2x3);
\node(temdis4)  [below =.1cm of 2y4] {};
  \node (dis1) [right =-.2cm of temdis4] {$\epsilon$};
  \draw[-,thin,black,dashed](2y4)edge node[midway]{}
  (2x4);
  \draw[-,thin,black,dashed](3y3)edge[bend left] node[midway]{}
  (2x3);
\node(3temdis5)  [below =.4cm of 3y4] {};
  \node (dis1) [right =-1cm of 3temdis5] {$\beta$};
  \draw[-,thin,black,dashed](3y4)edge[bend right] node[midway]{}
  (2x4);


 \begin{pgfonlayer}{background}
 \path (2y6.east |-2y6.east)+(.3,-.8) node(bb){$\tau'_1$};
 \path (x2.east|- x2.east)+(.3,.6) node(a){$\tau_1$};
 \path (2x4.east|- 2x4.east)+(.3,.6) node(b){$\tau_2$};
 \path (3y4.east |-3y4.east)+(.3,-.8) node(cc){$\tau'_2$};

\draw[->,black,dashed](a) edge (x2)(b) edge (2x4)(bb) edge (2y6)(cc)edge(3y4);
\end{pgfonlayer}
\end{tikzpicture}
	}}}\\
	\caption{
		$\epsilon\ll c<\beta$. Obviously, $[\tau_1\leftrightarrow\tau'_1\!,\tau_2\leftrightarrow\tau'_3]$ and $[\tau'_1\leftrightarrow(\tau_1,\tau_2)]$ are the optimal assignments of Figures \eqref{Fig:EvalSmeRis_DifTuyet1} and \eqref{Fig:EvalSmeRis_DifTuyet2} respectively.}
	\label{Fig:EvalSmeRis_DifTuyet}
\end{figure}

A further unreliability of the OSPA metric is caused by the OSPAT assignment. This is because the OSPAT metric includes the labeling errors where the labels are added depending on the OSPAT assignment.
When the OSPAT assignment is not the optimal assignment, the OSPAT metric is unreliable. Consider Figure \ref{Fig:EvalSmeRis_DifTuyet} where the scenario in Figure \eqref{Fig:EvalSmeRis_DifTuyet1} is clearly better than the scenario in Figure \eqref{Fig:EvalSmeRis_DifTuyet2}. However, the OSPAT metric concludes that the two scenarios are the same by Table \ref{table:EvalSmeRis_DifTuyet}. This is because at time $3$ and $4$, $\tau'_3$ (from $\omega_a$) and $\tau_2$ have different labels; and labels of $\tau'_1$ (from $\omega_b$) and
$\tau_2$ are different. These labels result from the OSPAT assignment ($[\tau_1\leftrightarrow\tau'_1\!,\tau_2\leftrightarrow\tau'_2]$) for these scenarios in Figure \ref{Fig:EvalSmeRis_DifTuyet}.
\renewcommand{\arraystretch}{1.7}
\begin{center}
	\begin{table}[htpb!]%
		\caption{\textbf{OSPAT Distances at $t=3,4,5$
				of 
				Figure \ref{Fig:EvalSmeRis_DifTuyet}}}
		\begin{center}
			\begin{tabular}
				{|m{1cm}|m{0.7cm}|m{1.4cm}|m{1.75cm}|m{1.75cm}|@{}m{0pt}@{}}\cline{1-5}
				Distance&Figure&$t=3$&$t=4$&$t=5$\\
				\cline{1-5}
				\multicolumn{1}{|m{.18cm}|}{\multirow{2}{*}{OSPAT}} &
				\multicolumn{1}{m{0.5cm}|}{\eqref{Fig:EvalSmeRis_DifTuyet1}} & $\sqrt[p]{\frac{\epsilon^p+c^p}{2}}$&$\sqrt[p]{\frac{\epsilon^p+\Delta^p+c^p}{2}}$ &$\sqrt[p]{\frac{\epsilon^p+\Delta^p+c^p}{2}}$\\\cline{2-5}
				\multicolumn{1}{|m{0.5cm}|}{}                        &
				\multicolumn{1}{m{0.5cm}|}{\eqref{Fig:EvalSmeRis_DifTuyet2}} & $\sqrt[p]{\frac{\epsilon^p+c^p}{2}}$&$\sqrt[p]{\frac{\epsilon^p+\Delta^p+c^p}{2}}$ &$\sqrt[p]{\frac{\epsilon^p+\Delta^p+c^p}{2}}$
				\\ \cline{1-5}
			\end{tabular}
			\label{table:EvalSmeRis_DifTuyet}
		\end{center}
	\end{table}
\end{center}
\renewcommand{\arraystretch}{1}
\subsection{Incompatibility with the metric definition}
\label{SS:label_issue}
The OSPAT metric is not the distance measure between two sets of (labeled) tracks. Instead it is the distance between two sets of labeled states. If the states of the tracks include labels, then it is necessary to rename the labels of these states by the OSPAT assignment before grouping the labeled states according to time indices (see Appendix \ref{SS:Summary_OSPAT}). If the states of the tracks do not include labels, then we need to add the labels to these states by the OSPAT assignment before grouping the labeled states according to time indices (see Appendix \ref{SS:Summary_OSPAT}). More precisely, the OSPAT metric is the OSPA metric \cite{Schuhmacher2008} on the space of sets of labeled states at each time index and hence it involves the computation of the labeling error. The labeling error in \eqref{E:dis-label} means that the OSPAT distance in \eqref{D:OSPAT_omega} does not satisfy the triangle inequality. This is because the labels in the labeled states are not fixed but change according to the OSPAT assignment in \eqref{E:min_dis4Label}. Indeed, the distance in \eqref{D:OSPAT_omega} is not a metric although the distances in
\eqref{E:Metric-OSPAT_t} and \eqref{E:Metric-label} are metrics if $\alpha$ in \eqref{E:dis-label} is positive. If $\alpha=0$, the distances in \eqref{E:Metric-OSPAT_t}, \eqref{D:OSPAT_omega} and \eqref{E:Metric-label} violate the identity property. Indeed, let $\bar{x}=(l,x),\bar{y}=(k,x)\in\mathds{N}\times\mathcal{X}$ where $l\ne k$  and then $\bar{X}=\{\bar{x}\},\bar{Y}=\{\bar{y}\}$. Thus $\bar{x}\ne\bar{y}$ but $\bar{d}^\alpha(\bar{x},\bar{y})=0$ and hence $\hat{D}^\alpha_{c,p}(\bar{X},\bar{Y})=0$.
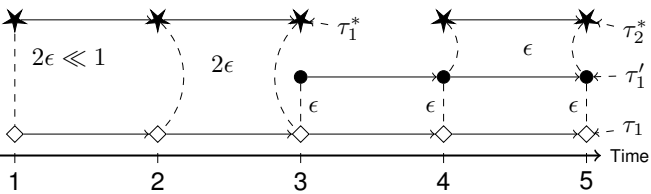
\begin{figure}[htbp!]
	\hspace{-2cm}
	\scalebox{.95}
	{\centering\hspace{-.3cm}
		\pgfdeclarelayer{background} \pgfdeclarelayer{foreground}
\pgfsetlayers{background,main,foreground}
\par
\par
\tikzset{cross/.style={cross out, draw=black, minimum size=2*(#1-\pgflinewidth), inner sep=0pt, outer sep=0pt},
cross/.default={3pt}}
\begin{tikzpicture}[y=.2cm, x=2cm,font=\sffamily,bend angle=45]
 \draw [->,thick]
 |-(5.1,0) node (xaxis) [right]
  {\scriptsize Time};
\draw (1,0) -- coordinate (x axis mid) (5.1,0);
\foreach \x in {1,...,5}
     		\draw (\x,1pt) -- (\x,-3pt)
			node[anchor=north] {\x};
\node[rectangle,draw,yscale=.7,xscale=.7, rotate=45](x1)at(1,1.5) {};
\node[rectangle,draw,yscale=.7,xscale=.7, rotate=45](x2)at(2,1.5) {};
\node[rectangle,draw,yscale=.7,xscale=.7, rotate=45](x3)at(3,1.5) {};
\node[rectangle,draw,yscale=.7,xscale=.7, rotate=45](x4)at(4,1.5) {};
\node[rectangle,draw,yscale=.7,xscale=.7, rotate=45](x5)at(5,1.5) {};
\node[star, star points=5, star point ratio=0.3, fill=black, draw] (2x1)at (1,9.5){};
\node[star, star points=5, star point ratio=0.3, fill=black, draw] (2x2)at (2,9.5){};
\node[star, star points=5, star point ratio=0.3, fill=black, draw] (2x3)at (3,9.5){};
\node[star, star points=5, star point ratio=0.3, fill=black, draw] (3x4)at (4,9.5){};
\node[star, star points=5, star point ratio=0.3, fill=black, draw] (3x5)at (5,9.5){};
\node[circle,draw=black, fill=black, inner sep=0pt,minimum size=5pt] (y3) at (3,5.5){};
\node[circle,draw=black, fill=black, inner sep=0pt,minimum size=5pt] (y4) at (4,5.5){};
\node[circle,draw=black, fill=black, inner sep=0pt,minimum size=5pt] (y5) at (5,5.5){};
\draw[->,black](y3)edge (y4)(y4)edge (y5)
(x1)edge (x2)(x2)edge (x3)(x3)edge (x4)(x4)edge (x5)(2x1)edge(2x2)(2x2)edge(2x3)(3x4)edge(3x5);
\node(temdis1)  [below =.2cm of 2x1] {};
 \node (dis1) [right =-0.01cm of temdis1]{$2\epsilon\ll 1$};
 \draw[-,thin,black,dashed](2x1)edge node[swap,near start]{}
  (x1);
  \draw[-,thin,black,dashed](2x2)edge[bend left] node[swap,near start]{}
  (x2);
\node(temdis3)  [below =.3cm of 2x3] {};
  \node (dis1) [right =-1.5cm of temdis3] {$2\epsilon$};
  \draw[-,thin,black,dashed](2x3)edge[bend right] node[midway]{}
  (x3);
\node(temdis5)  [below =.2cm of y3] {};
  \node (dis1) [left =-.5cm of temdis5] {$\epsilon$};
  \draw[-,thin,black,dashed](y3)edge node[midway]{}
  (x3);
\node(temdis6)  [below =.2cm of y4] {};
  \node (dis1) [right =-.5cm of temdis6] {$\epsilon$};
  \draw[-,thin,black,dashed](y4)edge node[midway]{}
  (x4);
\node(2temdis2)  [below =.2cm of y5] {};
  \node (dis1) [right =-.5cm of 2temdis2] {$\epsilon$};
  \draw[-,thin,black,dashed](y5)edge node[midway]{}
  (x5);
\node(2temdis1)  [below =.1cm of 3x4] {};
  \node (dis1) [left =-1.5cm of 2temdis1] {$\epsilon$};
  \draw[-,thin,black,dashed](3x4)edge[bend left] node[near start]{}
  (y4);
  \draw[-,thin,black,dashed](3x5)edge[bend right] node[near start]{}
  (y5);


 \begin{pgfonlayer}{background}
 \path (3x5.east |-3x5.east)+(.3,-.8) node(c){$\tau^*_2$};
  \path (y5.east |-y5.east)+(.3,.1) node(aa){$\tau'_1$};
  \path (x5.east|- x5.east)+(.3,.1) node(a){$\tau_1$};
 \path (2x3.east|- 2x3.east)+(.3,-.6) node(b){$\tau^*_1$};
\draw[->,black,dashed](a) edge (x5)(b) edge (2x3)(c) edge (3x5)(aa)edge(y5);
\end{pgfonlayer}
\end{tikzpicture}
	}
	\caption{$\omega=\{\tau_1\},\omega'=\{\tau'_1\}$ and $\omega^*=\{\tau^*_1,\tau^*_2\}$. By OSPAT assignment, $\tau^*_2$ is labeled $1$ (because it is assigned to $\tau'_1$), but must not be labeled $1$ (because it is not assigned to $\tau_1$).}
	\label{E:OSPAT_NoMetric}
\end{figure}

We show that the distance in \eqref{D:OSPAT_omega} violates the triangle inequality. Consider the example illustrated in Figure \ref{E:OSPAT_NoMetric} where $\omega=\{\tau_1\}$ and $\omega'=\{\tau'_1\}$ and $\omega^*=\{\tau^*_1,\tau^*_2\}$. We denote the elements in \eqref{E:state_time_t} associated with $\omega$, $\omega'$ and $\omega^*$ by $\bar{X}^{\omega}$, $\bar{Z}^{\omega'}$ and $\bar{Y}^{\omega^*}$ ($\bar{\bar{Y}}^{\omega^*}$) respectively. The OSPAT assignment assigns track $\tau^*_2$ in $\omega^*$ to $\tau'_1$ in $\omega'$ since the assignment makes distance in \eqref{E:OSPATdis_lambda} smallest among other assignment. Hence, the OSPAT assignment between $\omega'$ and $\omega^*$ makes $\tau^*_2$ to be labeled $1$ while the OSPAT assignment between $\omega$ and $\omega^*$ makes $\tau^*_2$ to be labeled $2$. In summary, for time $t=4,5$, $\lambda_*^{\omega,\omega^*}$ results in $\bar{X}^{\omega}_t=\{(1,\diamond)\}$, $\bar{Y}^{\omega^*}_t=\{(2,\star)\}$; $\lambda_*^{\omega,\omega'}$  gives $\bar{X}^{\omega}_t=\{(1,\diamond)\}$, $\bar{Z}^{\omega'}_t=\{(1,\bullet)\}$; and $\lambda_*^{\omega',\omega^*}$ results in $\bar{\bar{Y}}^{\omega^*}_t=\{(1,\star)\}$, $\bar{Z}^{\omega'}_t=\{(1,\bullet)\}$. By \eqref{D:OSPAT_omega}, $t=4,5$,
\begin{align*}
\boldsymbol{\mathcal{D}}^{\alpha,t}_{c,p}(\omega,\omega^*)&=\sqrt[p]{(2\epsilon)^p+\alpha^p},\\
\boldsymbol{\mathcal{D}}^{\alpha,t}_{c,p}(\omega,\omega')&=\boldsymbol{\mathcal{D}}^{\alpha,t2}_{c,p}(\omega',\omega^*)=\epsilon.
\end{align*}

Therefore $\boldsymbol{\mathcal{D}}^{\alpha,4}_{c,p}(\omega,\omega^*)>\boldsymbol{\mathcal{D}}^{\alpha,4}_{c,p}(\omega,\omega')+\boldsymbol{\mathcal{D}}^{\alpha,4}_{c,p}(\omega',\omega^*)$ for any $\alpha>0$ which violates the triangle inequality. Both $\bar{\bar{Y}}^{\omega^*}_t$ and $\bar{Y}^{\omega^*}_t$ represent the same physical multi-target state of $\omega^*$ at $t=4,5$ but for any $\bar{X}_t=\{(l,x)\}\subseteq\mathds{N}\times\mathcal{X}$ ($l=1,2$), we have $\hat{D}^\alpha_{c,p}(\bar{X}_t,\bar{Y}^{\omega^*}_t)\ne\hat{D}^\alpha_{c,p}(\bar{X}_t,\bar{\bar{Y}}^{\omega^*}_t)$ due to the different label assignment.

Thus, the OSPAT distance in \eqref{D:OSPAT_omega} is not a metric for any $\alpha\in[0,c]$. Furthermore, the distances in \eqref{E:Metric-OSPAT_t} and \eqref{E:Metric-label} are not metrics for $\alpha=0$.
\end{document}